\documentclass[aps, prd, a4paper, floatfix,showpacs, preprintnumbers, superscriptaddress, nofootinbib, onecolumn]{revtex4}
\usepackage{amssymb,amsmath,amsfonts,amsbsy,graphicx,rotating,bbold}
\usepackage{color,microtype}

\usepackage{tabularx}
\usepackage{eucal}
\usepackage{epsfig}
\usepackage{{geometry}}
\usepackage{float}
\usepackage{xcolor}
\usepackage{lipsum}
\usepackage{textcomp}
\usepackage{siunitx}
\usepackage{makecell}
\usepackage{lipsum}
\usepackage{slashbox}
\usepackage{hyperref}
\usepackage[labelsep=period]{caption}
\DeclareCaptionJustification{justified}{\leftskip=0pt \rightskip=0pt \parfillskip=0pt plus 1fil}
\hypersetup{
	colorlinks=true,
	linkcolor=red,
	filecolor=red,   
	citecolor= red,
	urlcolor=magenta}
\makeatletter
\let\ams@underbrace=\underbrace
\def\underbrace#1_#2{%
	\setbox0=\hbox{$\displaystyle#1$}%
	\ams@underbrace{#1}_{\parbox[t]{\the\wd0}{#2}}%
}
\makeatother

\graphicspath{{Images/}}
\usepackage{subcaption}
\definecolor{vividviolet}{rgb}{0.62, 0.0, 1.0}
\definecolor{amaranth}{rgb}{0.9, 0.17, 0.31}
\definecolor{palatinateblue}{rgb}{0.15, 0.23, 0.89}
\definecolor{brightpink}{rgb}{1.0, 0.0, 0.5}
\definecolor{cornflowerblue}{rgb}{0.39, 0.58, 0.93}
\definecolor{deepcarminepink}{rgb}{0.94, 0.19, 0.22}
\definecolor{radicalred}{rgb}{1.0, 0.21, 0.37}

\def\beq{\begin{equation}}
\def\eeq{\end{equation}}

\begin{document}

\title{ Phase space analysis and singularity classification for linearly interacting dark energy models}

\author{Muhsin Aljaf}
\email{Mohsen13.8b@gmail.com}
\affiliation{Department of Astronomy and CAS Key Laboratory for Research in Galaxies and Cosmology, University of Science and Technology of China, Hefei, Anhui 230026, China}

\author{Daniele Gregoris}
\email{danielegregoris@libero.it}
\affiliation{Center for Gravitation and Cosmology, College of Physical Science and Technology, Yangzhou University,180 Siwangting Road, Yangzhou City, Jiangsu Province 225002, China}
\affiliation{School of Aeronautics and Astronautics, Shanghai Jiao Tong University, Shanghai 200240, China}

\author{Martiros Khurshudyan}
\email{khurshudyan@yandex.com}
\email{khurshudyan@ustc.edu.cn}
\affiliation{Department of Astronomy and CAS Key Laboratory for Research in Galaxies and Cosmology, University of Science and Technology of China, Hefei, Anhui 230026, China}
\affiliation{Institut de Ciencies de lEspai (ICE-CSIC), Campus UAB, Carrer de Can Magrans, s/n 08193 Cerdanyola del Valles, Barcelona, Spain}
\affiliation{International Laboratory for Theoretical Cosmology, Tomsk State University of Control Systems and Radioelectronics (TUSUR), 634050 Tomsk, Russia}

\begin{abstract}
	In this paper,  applying the Hartman-Grobman theorem we carry out a qualitative late-time analysis of some unified dark energy-matter Friedmann cosmological
	models, where the two interact through linear energy exchanges, and the dark energy fluid obeys to the dynamical equation of state of Redlich-Kwong,
	Modified Berthelot, and Dieterici respectively. The identification of appropriate late-time attractors allows to restrict the range of validity of the free parameters of the models under investigation. In particular, we prove that the late-time attractors which support a negative deceleration parameter correspond to a de Sitter universe. We show that the strength of deviation from an ideal fluid for the dark energy does not influence the stability of the late-time attractors, as well as the values of all the cosmological parameters at equilibrium, but for the Hubble function (which represents the age of the universe). Our analysis also shows that a singularity in the effective equation of state parameter for the dark energy fluid is not possible within this class of models.
\end{abstract}

\maketitle

\section{Introduction} \label{secI}
Late-time interactions between dark energy and dark matter do not violate current observational constraints \cite{salvatelli, noint1}. In particular, it has been proposed that energy flows between the two dark components of the Universe can alleviate the  {\ttfamily"}coincidence problem{\ttfamily"}: why do we live in a special epoch of the evolution of our Universe in which the amounts of dark energy and dark matter are of the same order of magnitude? \cite{coin1,coin2,coin3}. Furthermore, they may mitigate as well the Hubble tension \cite{sunny2}, and they have been investigated in light of the 21-cm line excess at cosmic dawn by one of us \cite{cai1}.  On the other hand, the picture of the dark energy fluid as a cosmological constant term entering the Einstein equations is problematic due to many reasons: it violates the causality principle, its adiabatic speed of sound is ill-defined, there is no theoretical framework in terms of elementary particles theories which can account for its physical properties with in particular its cosmological value being different from the one estimated from vacuum energy by 120 orders of magnitude \cite{weinrev}. These considerations have been making impossible to develop appropriate technologies for a direct detection of dark energy, making the  {\ttfamily"}dark energy problem{\ttfamily"} to be considered the most urgent open question to answer in a recent survey conducted among the fifty most prominent cosmologists \cite{bolejkosurv}. In fact, the idea itself that dark energy should be a physical component of our Universe has been constantly challenged suggesting that it may indeed constitute just an interpretative aspect of the current standard model of cosmology which, for example, neglects the gravitational role that astrophysical objects like galaxies, clusters, voids, and filaments have on the large scale evolution of the Universe \cite{inhomo1,inhomo2,buch4,buch3}.

For addressing the limits of the description of the dark energy as a cosmological constant, a number of fluid models based on  non-ideal equations of state have been proposed and tested against observational datasets with the twofold purpose of accounting for its thermodynamical properties and modifying the value of its adiabatic speed of sound in such a way that it will not violate the causality principle \cite{nonideal1,nonideal2,nonideal3,nonideal4,nonideal5,nonideal6}. Moreover, a time evolution of the equation of state for dark energy does not violate current observations \cite{plank,panteon,bao}, and various possibilities for its parametric reconstruction have been investigated \cite{parametric1,parametric2,parametric3}. Exact solutions to the field equations become scarce when the picture of dark energy with a cosmological constant is abandoned, and thus dynamical system techniques constitute a valuable tool for predicting the late-time state of the evolution of the models. For example, they  have been employed in Galileon gravity \cite{qual1}, Horava-Lifshitz gravity \cite{qual2}, modified teleparallel models \cite{qual3}, Einstein-Aether model \cite{qual4}, inflationary models  \cite{cai3}, for the study of the evolution of instabilities in a Friedmann universe \cite{buch1}, and by one of us to various modified Chaplygin gas scenarios \cite{mar1} in a Friedmann cosmology including interacting cases \cite{mar2}, ghost energies regimes \cite{mar3}, and frameworks involving evolving both $G$ and $\Lambda$ \cite {mar4}. 

As a subsequent step, dynamical system techniques are usually adopted as well for studying the singularity properties of the cosmological model in hand \cite{coley1,coley2,coley3,sergei2}. In fact, the results known under the name of \textit{singularity theorems} prove that under certain conditions on the matter content of the Universe, a singular state characterized by the divergence of certain physical quantities can be reached even in a finite amount of time during the evolution \cite{tsing1,tsing2,tsing3,tsing4}. However, the mathematical demonstration of such results are usually based on the modeling of dark energy with a cosmological constant, and the roles of nonideal fluids have not been fully accounted for.

In this paper we investigate the asymptotic late-time evolution of a Friedmann universe whose matter content is a mixture of dark energy and pressure-less dark matter adopting three descriptions proposed in \cite{capo}, but allowing also for interactions between these two constituents modeled as a linear flow of dark energy, or dark matter, or of their sum, respectively. We will adopt dynamical system techniques which allow us to compute the attractor equilibrium points in terms of elementary functions showing explicitly that they can account for an accelerated expanding Universe, and how this physical requirement permits to restrict the range of validity of the free parameters entering our models. The singularity type of these physical attractors will be presented. Then, the nine cases (three possible modelings for dark energy times three possible modelings for dark interactions) will be compared and contrasted with each other in light of their late-time properties. 

The plan of the work is as follows: In section (\ref{secII}), we introduce three different cosmological interacting models between dark energy and dark matter with three different nonideal equations of state for dark energy and provide a formalism to transform their Friedmann equations into an  autonomous dynamical system. In section (\ref{secIII}) we perform the phase-space analysis for each interacting model and provide a summary of the obtained results commenting on the patterns which can be recognized.  Section (\ref{secIV}) classifies the possible singularities arising in these models by discussing the corresponding cosmological implications. Finally, in section (\ref{secV}) we summarize our results and make our conclusion.
In appendix (\ref{appendix I}) a brief review will be provided about the approach of dynamical system theory, and  the dynamical autonomous equations for each model that we consider will be derived. Moreover, appendix (\ref{appendix II}) is devoted for the study of transition between decelerating and accelerating epochs for each of our model.

\section{Basic  equations of our cosmological model} \label{secII}

In this section we will introduce the notations that we will follow  throughout the paper and review the basic properties of the Friedmann cosmology under investigation.
We adopt  units such that $8 \pi G=1=c$. $\rho_{\rm de}$ denotes the dark energy density, $\rho_{\rm dm}$ the dark matter  density, $p_{\rm de}$ the dark energy pressure, while dark matter is assumed to be pressure-less. $e$ is the number of Nepero. Let $H=\frac{\dot a}{a}$ be the Hubble function, $H_0$ its present value, $a$ the scale factor of the universe and $a_0$ its present value. The dimensionless energy-matter parameters on which we will base our dynamical system analysis are:
\beq
\label{matterp}
x=\frac{\rho_{\rm de}}{ 3H^2}\,, \quad y=\frac{p_{\rm de}}{ 3H^2}\,, \quad z=\frac{\rho_{\rm dm}}{ 3H^2}\,.
\eeq
An over dot stands for differentiation with respect to the coordinate time $t$,   and  $Q$ will be the interaction term between dark matter and dark energy. The parameter for the effective dark energy equation of state is constrained by current astrophysical observations as \cite{plank,panteon,bao}:
\beq
-\frac{6}{5}=-1.2 \leq w=\frac{y}{x} < 0.
\eeq

\noindent We assume that the geometrical properties of the Universe can be accounted for by the flat Friedmann metric \cite{exact}
\beq
ds^2=g_{\mu\nu} dx^\mu dx^\nu=-dt^2 +a^2(t)(dr^2 + r^2 d\theta^2 + r^2 \sin^2 \theta d\phi^2) \,.
\eeq
The matter content is described by the stress-energy tensor $T_{\mu \nu}=(\rho +p) u_\mu u_\nu +p g_{\mu\nu} $, where $u^\mu=\delta_t^\mu$ is the four-velocity of the reference observer. Introducing the Einstein tensor $G_{\mu\nu}$, the  equations governing the dynamics of the model are the Einstein field equations $G_{\mu\nu}=T_{\mu\nu}$ and the Bianchi identities $T^{\mu\nu}{}_{;\nu}=0$, in which a semicolon  denotes a covariant differentiation.
Thus,  the basic equations can be reduced to the Friedmann equation, the acceleration equation and the equations for the conservation of dark energy and dark matter:
\begin{eqnarray}\label{4}
H^2  &=&\left(  \frac{\dot a}{a}  \right)^2  \,=\,  \frac{\rho_{\rm de} + \rho_{\rm dm} }3\label{hb1} \\
\label{5}
\dot H +H^2 &=& \frac{\ddot a}{a}  \,=\,  -\frac{\rho_{\rm de} + \rho_{\rm dm} + 3p_{\rm de}}6 \\
\label{6}
\dot \rho_{\rm de} &=& -3H (\rho_{\rm de} + p_{\rm de}) +Q\\
\label{7}
\dot \rho_{\rm dm} &=& -3H \rho_{\rm dm}  -Q.
\end{eqnarray}
 In terms of the dimensionless matter parameters (\ref{matterp}), the Friedmann equation  can be cast into the constraint
\beq
\label{cons}
x+z=1 \,,
\eeq
and we can observe that its form does not depend on the modeling of the interaction term coupling dark energy and dark matter that we can choose. Furthermore, physical requirements impose the energy densities to be non-negative, and thus we have the bounded variables $0 \leq x \leq 1$ and $0 \leq z \leq 1$ for which dynamical system techniques can be used \cite{coley1}, while $ H $ is the appropriate monotonic function for the Friedmann model. The particular cases $x=1$ and $z=1$ correspond to a dark energy and a dark matter dominated universe respectively.
The deceleration parameter reads as \cite{peebles,gravitation}:
\beq
q=-1-\frac{\dot H}{H^2}=\frac{1}{2}(1+3y),
\eeq
which must be negative for cosmological meaningful models \cite{plank,panteon,bao}.  Now we need to derive also the remaining evolution equations in terms of appropriate variables for exploiting the techniques provided by the theory of dynamical systems.
For this purpose, let a prime denote differentiation with respect to the number of $e$-folds of the Universe  $N= \ln a$ \cite{linde}. We compute the  equations of the model  using the chain rule
\beq
{\mathcal X}  ' = \frac{d {\mathcal X}} {dN} = \frac{d {\mathcal X}} {dt} \cdot \frac{d t} {da} \cdot  \frac{d a} {dN}=\frac{\dot {\mathcal X}}{H}.
\eeq
The evolution eq. of the Hubble function does not depend on the modeling of the interaction term and it reads as:
\begin{eqnarray} 
H' &=& -\frac{3H}{2} (1+y)\,.
\end{eqnarray}
The evolution equations for the matter parameters become
\begin{eqnarray} 
x' &=&  \frac{1}{H} \left[\frac{\dot \rho_{\rm de} }{3 H^2} - \frac{2}{3}\rho_{\rm de} \frac{\dot H}{H^3}\right] =   3y (x-1) +\frac{Q }{3 H^3} \\
z' &=&  \frac{1}{H} \left[\frac{\dot \rho_{\rm dm} }{3 H^2} - \frac{2}{3}\rho_{\rm dm} \frac{\dot H}{H^3}\right] =  3y (1-x)    -\frac{Q }{3 H^3} \,,
\end{eqnarray}
from which we can see that the compatibility condition $x'=-z'$ which follows from the Friedmann equation (\ref{cons}) automatically applies implying  that we can develop the dynamical system analysis in terms of $x$ or $z$ without any difference \cite{coley1}.
In our analysis we will focus our attention on the following three phenomenological types of interactions between dark matter and dark energy \cite{rint1, rint2, rint3, rint4, rint5, rint6, rint7}:
\begin{eqnarray}
\label{int1}
Q_{1} &=& 3Hb \rho_{\rm de} + \gamma \dot \rho_{\rm de} \\
\label{int2}
Q_{2} &=& 3Hb \rho_{\rm dm} + \gamma \dot \rho_{\rm dm} \\
\label{int3}
Q_{3} &=& 3Hb (\rho_{\rm de}+\rho_{\rm dm}) + \gamma (\dot \rho_{\rm de} + \dot \rho_{\rm dm} )\,,
\end{eqnarray}
with the physical constraints on the free parameters  $-1 \leq b \leq 1$ and $-1 \leq \gamma \leq 1$. By considering both positive and negative signs in the model parameters, we allow both the dark energy to flow into dark matter and viceversa; moreover the strength of such flows are proportional to the amounts of dark energy, or dark matter, or a combination of the two, respectively, for the three cases. We choose to consider these types for the interaction terms in the dark sector because we assume that the interaction represents a small contribution in the whole evolution of the energy budget of the Universe, and in fact these terms can be interpreted as a first order Taylor  expansion. Moreover, they are based on the assumption that the propagator of the dark particles is energy dependent. These choices have been shown to be a valid tool for alleviating the coincidence problem \cite{binwang}. Letting  $y=y(H,x)$ be the equation of state for the dark energy fluid we can derive the second differential equation entering the dynamical system:
\begin{eqnarray} 
\label{evoly}
y' &=&  \frac{1}{H} \left[\frac{\partial y}{\partial H} \dot H + \frac{\partial y}{\partial x} \dot x \right] = -\frac{3H}{2} (1+y)  \frac{\partial y}{\partial H}   + \frac{\partial y}{\partial x}  x' \,.
\end{eqnarray}
The two equations which will constitute our dynamical systems are thus provided by ($x'$, $y'$) and the equilibrium points for which  ($x'=0$, $y'=0$) will be ($x_{\rm eq}$, $y_{\rm eq}$). We stress the fact that the Hubble function $H$ is not a good variable for applying the algorithm of the theory of dynamical systems since it is not dimensionless and it is not necessarily bounded: it must be eliminated from the dynamical equations by inverting the equation of state for the dark energy as $H=H(x,y)$. Then, the dynamical system analysis will let us estimate this cosmological parameter as $H_{\rm eq}=H(x_{\rm eq}, \, y_{\rm eq})$. In the current paper we will consider different equations of state  for dark energy which are known under the name of Redlich-Kwong \cite{reos1}, Modified Berthelot \cite{reos3}, and Dieterici \cite{reos2} which read respectively as:

\begin{eqnarray}
p(\rho)&=& \frac{1- (\sqrt{2} -1) \alpha \rho}{ 1- (1-\sqrt{2}) \alpha \rho}\beta \rho \label{Redlich}\\
p(\rho)&=& \frac{\beta \rho}{1+ \alpha \rho}\label{modified} \\
p(\rho)&=& \frac{\beta \rho e^{2(1- \alpha \rho)}}{2- \alpha \rho} \,.\label{Dieterici}
\end{eqnarray}
We are interested in such equations of state for the modeling of the dark energy fluid because of their applicability in cosmology \cite{capo}. We refer the reader to the appendix of such paper for a review of their microscopic foundation and of their main thermodynamical properties. Here we just want to mention that the positive parameter $\alpha$ quantifies the deviations from a perfect fluid behavior expressed in units of the present day critical density $3 H^2_0$  because in all the above cases we reduce to a linear equation of state $p \sim \rho$ in the limit $\alpha \to 0$. On the other hand,  the parameter $\beta$, which must be negative for accounting for a dark-energy-like fluid (as easily recognized from the limit at small $\alpha$), can be removed in favor of cosmologically meaningful quantities as
\begin{eqnarray}
\beta&=& \frac{(2 q_0 -1) [1-(1-\sqrt{2}) x_0 ] }{  3(1-z_0) [1- (\sqrt{2} -1) x_0]}  \\
\beta&=&    \frac{ (2 q_0 -1) (1+x_0)}{ 3(1-z_0)}  \\
\beta&=&  \frac{ (2 q_0 -1) (2-x_0) e^{2(x_0 -1)}}{ 3 (1-z_0)}  \,,
\end{eqnarray}
respectively, where $q_0$ denotes the present-day value of the deceleration parameter $q=-\frac{\ddot a a}{\dot a^2}$ and $x_0=\alpha x({ \bar z}=0)$, and $z_0=z({\bar z}=0)$, ${ \bar z}$  denoting the redshift.  These equations of state have been proposed as improved and more realistic approaches to fluidodynamics than the Van der Waals equation of state still accounting for the internal forces acting between the molecules constituting a gas. They have been used for modeling liquid-vapor phase transitions with the most important difference from the Van der Waals theory being that the attractive force term has been allowed to be temperature-dependent avoiding the oscillations in the isotherm curves below the critical temperature, and they constitute an analytical alternative to the use of virial expansions \cite{chemistry1,chemistry2}. For applying dynamical system techniques in cosmology, these equations of state must be recast as:

\begin{eqnarray}
\label{eos1}
y&=& \frac{1- 3(\sqrt{2} -1) \alpha H^2 x}{ 1- 3(1-\sqrt{2}) \alpha H^2 x}\beta x \\
\label{eos2}
y&=& \frac{\beta x}{1+3H^2 \alpha x} \\
\label{eos3}
y&=& \frac{\beta x e^{2(1-3 \alpha H^2 x)}}{2-3H^2 \alpha x}.
\end{eqnarray}
From the equations of state (\ref{eos1})-(\ref{eos2})-(\ref{eos3}) we can compute respectively:
\begin{eqnarray}
H &=& \pm \sqrt{   \frac{ (\beta x -y) (1+\sqrt{2})}{ 3\alpha x (\beta x +y)}   }\\
H &=& \pm \sqrt{   \frac{ \beta x -y }{ 3 \alpha xy}   }\\
H &=& \pm \sqrt{   \frac{ W\left(- \frac{2 \beta x}  {e^2 y} \right) +4}{6 \alpha x}},
\end{eqnarray}
in which the double sign corresponds to the cases of expanding or contracting universes respectively, and where $W({\mathcal X})$ denotes the Lambert function which is the inverse  of  ${\mathcal X} e^{\mathcal X}$. Once evaluated at the equilibrium point, the requirement of having a positive argument for the square root will impose a further restrictions on the numerical values of the free parameters of the models we are considering. We note that in the evolution equation for $y$ below,  $H$ will appear always squared, so that its sign does not affect our analysis.

\section{Phase space analysis} \label{secIII}

In this section we perform phase space analysis, then reporting the values of the equilibrium points, exhibiting the restrictions they impose on the parameters of the models $\alpha$, $\beta$, $b$, and $\gamma$, and discussing their stability and whether they are cosmologically acceptable for the nine combinations equation of state for the dark energy - interaction term. Together with  a basic review on dynamical system theory, the autonomous dynamical systems equation for each of these combinations have been derived in the appendix (\ref{appendix I}).

\subsection{ Redlich-Kwong - Interaction term $Q_1$  }

The dynamical system to consider is formed by the two equations (\ref{eqx1}) and (\ref{eqy1a}). In this case we get three mathematical equilibria $(x_{\rm eq}, \, y_{\rm eq})$  which read as:
\beq
\label{eq11}
\left( \frac{1}{1-b}, \, -1 \right)\,, \qquad  \left( \frac{\gamma -b -\beta}{\beta( \gamma-1)}, \, \frac{b+\beta -\gamma}{\gamma -1}\right)\,, \qquad   \left( \frac{b-\beta -\gamma}{\beta( \gamma-1)}, \, \frac{b -\beta -\gamma}{\gamma -1}\right)\,.
\eeq
However only the first point is relevant for cosmology because the  second and third imply a divergent and a vanishing $H_{\rm eq}$ respectively. For such a physical attractor we get:
\beq
(H_{\rm eq}, \, z_{\rm eq}, \, q_{\rm eq}, \, w_{\rm eq} )\,=\,\left( \sqrt{ \frac{(b-\beta-1) (1+ \sqrt{2} )(b-1)}{3 \alpha (b +\beta -1)}}, \, \frac{b}{b-1}, \,  -1, \, b-1 \right).
\eeq
From the values of such physical quantities we can restrict $-\frac {1}{5} \leq b \leq 0$, where the case $b=0$ corresponds to a single fluid pure dark energy pictured as a cosmological constant universe. Moreover the relation $(b-\beta-1)(b +\beta -1) < 0$ must hold. Since the second factor is always negative for our choice of the model parameters, we can further restrict $\beta <b-1$.

The Jacobian matrix associated to the dynamical system when specified to such equilibrium point reads as:
\beq
J=\begin{pmatrix} \frac{3(1-b)}{\gamma -1 } & 3\left(  \frac{1}{1-b}+ \frac{1}{\gamma -1}   \right) \\
	\frac{3[ (b-1)^2 +2 (b-1) \beta -\beta^2](1-b)}{2\beta (\gamma -1)}  &  3 \frac{(b-1)^2 +2 (b-\gamma) \beta -\beta^2}{2 \beta (\gamma -1)}
\end{pmatrix}\,,
\eeq
implying
\begin{eqnarray}
{\rm Det J} & =& 9 \frac{\beta^2 - (b-1)^2}{ 2\beta (\gamma -1)}    \\
{\rm Tr J} & =&    3 \frac{(b-1)^2 +2 (1-\gamma) \beta -\beta^2}{ 2\beta (\gamma -1)}  \\
\frac{({\rm Tr J})^2}{4} - {\rm Det J} &=&  9 \frac{   [  (b-1)^2  +2 (\gamma -1) \beta - \beta^2 ]^2 }{16 \beta^2 (\gamma -1)^2} \,.
\end{eqnarray}
Thus, taking into account the range for the free parameters of the model, we get ${\rm Det J}>0$, which means that our equilibrium point is a a node (spirals are not possible because $\frac{({\rm Tr J})^2}{4} - {\rm Det J} >0$), and we can understand that it is stable because ${\rm Tr J}<3 \frac{2 (1-\gamma) \beta} { 2\beta (\gamma -1)}\leq -3 < 0$. An example of the phase portrait for such a system is displayed in fig (\ref{figura1}), panel (a), for the numerical choice of the free parameters as $b=-0.1$, $\gamma=0.5$, $\beta=-1.2$, and $\alpha=1.5$; the equilibrium point is denoted with a circle.

\subsection{ Modified Berthelot - Interaction term $Q_1$ }

The dynamical system to consider is formed by the two equations (\ref{eqx1}) and (\ref{eqy1b}).
In this case of the two mathematical attractors $(x_{\rm eq}, \, y_{\rm eq})$:
\beq
\label{eq21}
\left( \frac{1}{1-b}, \, -1 \right)\,, \qquad  \left( \frac{b- \beta -\gamma}{\beta( \gamma-1)}, \, \frac{b-\beta -\gamma}{\gamma -1}\right)\,,
\eeq
we must exclude the second because it delivers a zero Hubble function at equilibrium.
For the physically relevant equilibrium we get:
\beq
(H_{\rm eq}, \, z_{\rm eq}, \, q_{\rm eq}, \, w_{\rm eq}) \,=\, \left( \sqrt{ \frac{b-\beta -1}{3 \alpha }}, \, \frac{b}{b-1}, \,  -1, \, b-1 \right)\,
\eeq
From the values of such physical quantities we must restrict $-\frac {1}{5} \leq b \leq 0$, where the case $b=0$ corresponds to a single fluid pure dark energy pictured as a cosmological constant universe. Moreover, the relation $(b-\beta-1)> 0$ must hold providing $\beta <b-1$.

The Jacobian matrix associated to the dynamical system when specified to such equilibrium point reads as:
\beq
J=\begin{pmatrix} \frac{3(1-b)}{\gamma -1 } & 3\left(  \frac{1}{1-b}+ \frac{1}{\gamma -1}   \right)  \\
	\frac{3  (1-b)^3}{  \beta (\gamma -1)}  &  3 \frac{ (1- \gamma) \beta +(b-1)^2   }{  \beta (\gamma -1)}
\end{pmatrix}\,,
\eeq
implying
\begin{eqnarray}
{\rm Det J} & =&  \frac{ 9 (1-b) (b-\beta -1)  }{    \beta (\gamma -1)}    \\
{\rm Tr J} & =&    3 \frac{(2-b-\gamma)\beta +(b-1)^2 }{  \beta (\gamma -1)}  \\
\frac{({\rm Tr J})^2}{4} - {\rm Det J} &=&  9 \frac{   [  (\gamma -b) \beta + (b-1)^2 ]^2 }{4 \beta^2 (\gamma -1)^2}.
\end{eqnarray}
Thus, taking into account the range for the free parameters of the model, we get ${\rm Det J}>0$, which means that our equilibrium point is a a node (spirals are not possible because $\frac{({\rm Tr J})^2}{4} - {\rm Det J} >0$), and we can understand that it is stable because ${\rm Tr J}< 3 \frac{(2-b-\gamma)\beta +\beta^2 }{  \beta (\gamma -1)} = 3 \frac{2-b-\gamma +\beta }{  \gamma -1}= 3 \frac{-2+b+\gamma -\beta }{ 1- \gamma }<-3 <0$.
An example of the phase portrait for such a system is displayed in fig (\ref{figura1}), panel (b), for the numerical choice of the free parameters as $b=-0.1$, $\gamma=0.5$, $\beta=-1.2$, and $\alpha=1.5$; the equilibrium point is denoted with a circle.

\subsection{ Dieterici - Interaction term $Q_1$ }

\begin{figure}[H]
	\centering
	\begin{subfigure}[b]{.24\linewidth}
		\centering
		\includegraphics[width=.99\textwidth]{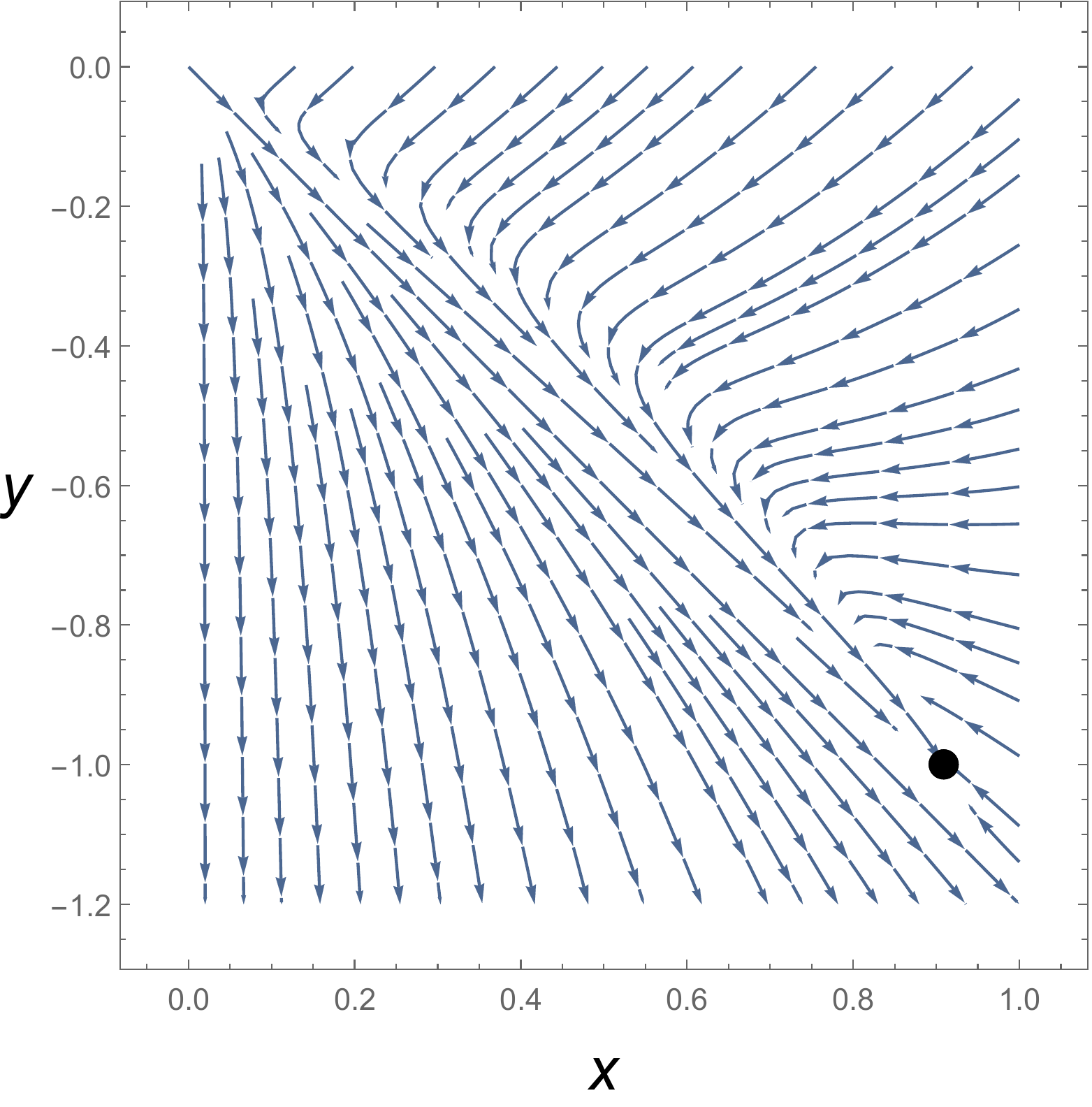}
			\caption{}\label{fig:1a}
         \end{subfigure}
     	\begin{subfigure}[b]{.24\linewidth}
     	\centering
     	\includegraphics[width=.99\textwidth]{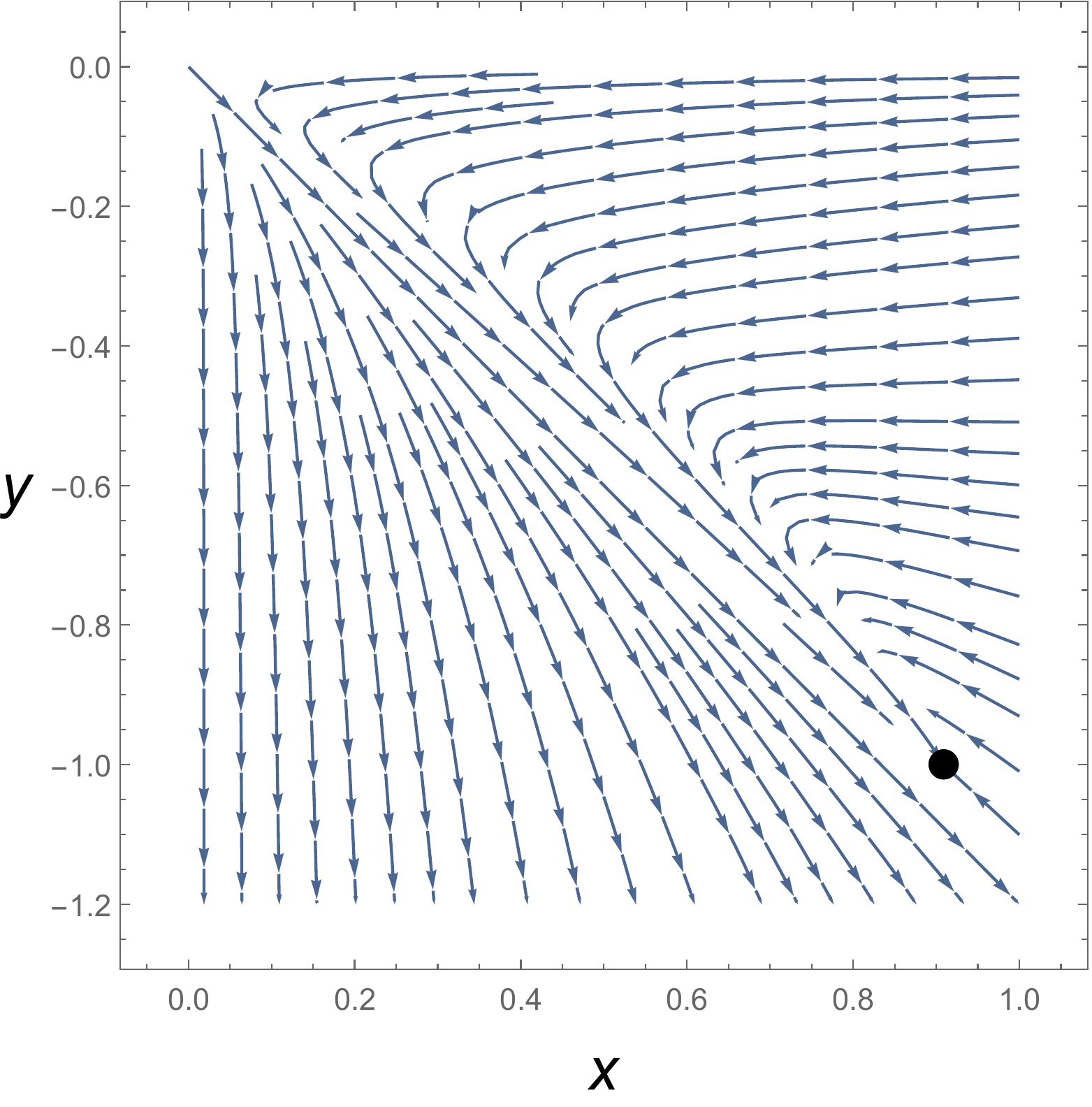}
     	\caption{}\label{fig:1b}
     \end{subfigure}
  	\begin{subfigure}[b]{.24\linewidth}
 	\centering
 	\includegraphics[width=.99\textwidth]{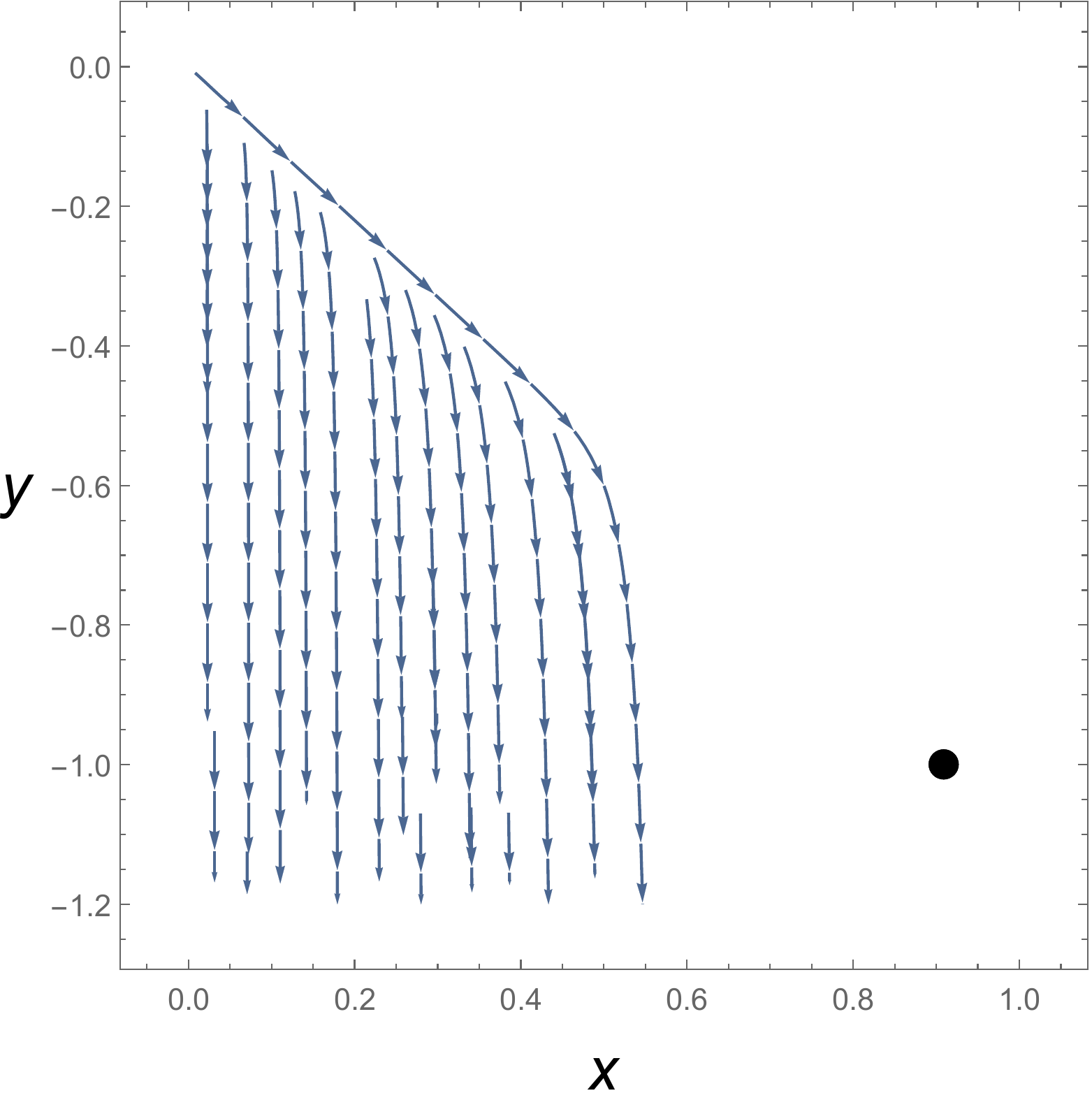}
 	\caption{}\label{fig:1c}
 \end{subfigure}
 	
	\caption{An example of the phase portrait for the dynamical system accounting for the interaction term (\ref{int1}), and equation of state (\ref{eos1}) in panel (a), (\ref{eos2}) in panel (b), (\ref{eos3}) in panel (c), for the numerical choice of the free parameters as $b=-0.1$, $\gamma=0.5$, $\beta=-1.2$, and $\alpha=1.5$; the equilibrium point is denoted with a circle. We note that the evolution of the system for case (c) stops when the dynamical equations are not anylonger defined, showing the part of the phase plane which can be effectively explored by this system.}
	\label{figura1}
\end{figure}

The dynamical system to consider is formed by the two equations (\ref{eqx1}) and (\ref{eqy1c}).
In this case we get three solutions $(x_{\rm eq}, \, y_{\rm eq})$  which read as:
\beq
\label{eq31}
\left(\frac{1}{1-b}, \, -1 \right), \qquad \left( \frac{2\beta+(\gamma -b) e}{2\beta (1-\gamma)} , \, \frac{2\beta e^{-1}-b+\gamma }{1-\gamma}  \right), \qquad \left( \frac{2 (b  -\gamma )  e^{-2} -\beta}{\beta (\gamma -1)}  ,\, \frac{2 b  -2\gamma   -\beta e^2}{2 (\gamma -1)} \right).
\eeq
The third must be excluded because it implies $H_{\rm eq}=0$.
Moreover:
\begin{eqnarray}
(H_{\rm eq}, \, z_{\rm eq}, \, q_{\rm eq}, \, w_{\rm eq}) &=& \left( \sqrt{\frac{\Big[W \left(\frac{2\beta}{e^2(1-b)} \right)+4 \Big](1-b)}{6 \alpha}}  , \, \frac{b}{b-1} , \, -1,\, b-1 \right) \\
(H_{\rm eq}, \, z_{\rm eq}, \, q_{\rm eq}, \, w_{\rm eq}) &=& \left( \sqrt{\frac{\beta (\gamma -1) }{\alpha [(b -\gamma ) e -2 \beta] }} ,\, \frac{e(b-\gamma) -2\beta \gamma}{2 \beta (1-\gamma)}, \, \frac{3(b-\gamma-2 \beta e^{-1})+\gamma -1}{2(\gamma -1)}, \, \frac{2 \beta}{e} \right),
\end{eqnarray}
for the first and second attractors respectively. We note that the second equilibrium point does not fulfill the physical requirements. In fact $x_{\rm eq} >0$ requires $e(b-\gamma) -2\beta \gamma >0$, which in turn inserted into $q_{\rm eq} <0$ would require $\gamma>1$. For the first attractor we must restrict the values of the model parameters as $-\frac {1}{5} \leq b \leq 0$, where the case $b=0$ corresponds to a single fluid pure dark energy as a cosmological constant universe, and $ \beta<2(b-1) e^{-2}$. The Jacobian matrix for the dynamical system specified to the first attractor is
\beq
J_1=\begin{pmatrix} \frac{3(1-b)}{\gamma -1 } & 3\left(  \frac{1}{1-b}+ \frac{1}{\gamma -1}   \right)  \\
	\frac{3 [W(\psi) +2]^2 (b-1)^2}{   (\gamma -1) W(\psi)}  &  3 \frac{ (b-1) W^2 (\psi)+(4b +\gamma -5) W(\psi)+4(b-1)  }{  (1-\gamma) W(\psi)}
\end{pmatrix}\,, \qquad \psi=\frac{ 2\beta}{(1-b) e^2}
\eeq
which implies
\begin{eqnarray}
{\rm Det J_1} & =&  \frac{ 9 (b-1) [W(\psi) +4] [W(\psi)+1  ]}{    (\gamma -1) W(\psi)}    \\
{\rm Tr J_1} & =&    3 \frac{(1-b) [W^2(\psi) +4] + (6-\gamma-5b) W(\psi) }{  (\gamma -1) W(\psi)}  \\
\frac{({\rm Tr J_1})^2}{4} - {\rm Det J_1} &=&  9 \frac{   [  (b-1) (W^2(\psi)+4) +(5b -\gamma -4) W(\psi)  ]^2 }{ 4(\gamma -1)^2 W^2(\psi)}.
\end{eqnarray}
Thus, taking into account the range for the free parameters of the model, and noticing that 
\begin{eqnarray}
&& {\rm Det J_1}=\frac{54 \alpha H_{\rm eq}^2 [1+W(\psi)]}{(1-\gamma)W(\psi)} 
\end{eqnarray}
we get ${\rm Det J}<0$, which means that our equilibrium point is a saddle. An example of the phase portrait for such a system is displayed in fig (\ref{figura1}), panel (c), for the numerical choice of the free parameters as $b=-0.1$, $\gamma=0.5$, $\beta=-1.2$, and $\alpha=1.5$; the equilibrium point is denoted with a circle. We note that in this case the evolution of the system stops when the dynamical equations lose meaning, that is when the argument of the Lambert W function becomes smaller than $-1/e$.

\subsection{ Redlich-Kwong - Interaction term $Q_2$ }

The dynamical system to consider is formed by the two equations (\ref{eqx2}) and (\ref{eqy2a}).
In this case we get five solutions $(x_{\rm eq}, \, y_{\rm eq})$  which read as:
\beq
\label{eq12}
(1, \, -1), \qquad (1, \, \beta), \qquad (1, \, -\beta), \qquad \left( \frac{\gamma - b}{\beta (1+\gamma)}, \, \frac{b -\gamma}{1+\gamma} \right),  \qquad \left( \frac{b -\gamma}{\beta (1+\gamma)}, \, \frac{b -\gamma}{1+\gamma}\right).
\eeq
We notice that the third and fourth equilibria  deliver an ill-defined $H_{\rm eq}$, while the second and fifth implies a zero $H_{\rm eq}$. Thus we can consider only the first point as an appropriate attractor for cosmology, for which we get:
\beq
(H_{\rm eq}, \, z_{\rm eq}, \, q_{\rm eq}, \, w_{\rm eq})\,=\,\left( \sqrt{ \frac{(1+\beta) (1+\sqrt{2})}{3 \alpha (\beta -1)}}, \, 0, \, -1, \, -1   \right)\,,
\eeq
which corresponds to a single fluid universe dominated by dark energy in form of a cosmological constant. The model parameters must be further restricted requiring $\beta<-1$.

The Jacobian matrix associated to the dynamical system when specified to such equilibrium point reads as:
\beq
J=\begin{pmatrix}- \frac{3(b+1)}{\gamma +1 } & 0   \\
	\frac{3 (\beta^2 +2 \beta -1) (1+b)}{2\beta (\gamma +1)}  &  3 \frac{ (\gamma+1) \beta^2 -\gamma -1 }{2 \beta (\gamma +1)}
\end{pmatrix}\,,
\eeq
implying
\begin{eqnarray}
{\rm Det J} & =& 9 \frac{(1-\beta^2) (1+b)}{ 2\beta (\gamma +1)}    \\
{\rm Tr J} & =&    3 \frac{  (\gamma+1) (\beta^2 -1) -2 (b+1) \beta      }{ 2\beta (\gamma +1)}  \\
\frac{({\rm Tr J})^2}{4} - {\rm Det J} &=&  9 \frac{   (\beta^2 \gamma+2 b\beta+\beta^2+2\beta-\gamma-1)^2 }{16 \beta^2 (\gamma +1)^2}\,.
\end{eqnarray}
Thus, taking into account the range for the free parameters of the model, we get ${\rm Det J}>0$, which means that our equilibrium point is a node (spirals are not possible because $\frac{({\rm Tr J})^2}{4} - {\rm Det J} >0$), and we can understand that it is stable because ${\rm Tr J}<0$.
An example of the phase portrait for such a system is displayed in fig (\ref{figura2}), panel (a), for the numerical choice of the free parameters as $b=-0.1$, $\gamma=0.5$, $\beta=-1.5$, and $\alpha=1.5$; the equilibrium point is denoted with a circle.

\subsection{ Modified Berthelot - Interaction term $Q_2$ }

The dynamical system to consider is formed by the two equations (\ref{eqx2}) and (\ref{eqy2b}).
In this case we get four attractors  $(x_{\rm eq}, \, y_{\rm eq})$  which correspond to:
\beq
(1, \, 0), \qquad  \left( \frac{b -\gamma }{\beta (1+\gamma)}, \, \frac{b -\gamma}{1+\gamma} \right),  \qquad  (1, \, -1), \qquad (1, \, \beta).
\eeq
The first equilibrium point is not physical because it delivers an ill-defined $H_{\rm eq}$, and we must exclude also the second and fourth solutions because they imply a zero $H_{\rm eq}$. For the third solution, which is the only one relevant for cosmology, we get:
\beq
(H_{\rm eq}, \, z_{\rm eq}, \, q_{\rm eq}, \, w_{\rm eq}) \,=\, \left(\sqrt{-\frac{1+\beta}{3 \alpha}}, \, 0, \, -1, \, -1 \right) ,
\eeq
which corresponds to a single fluid universe dominated by dark energy in form of a cosmological constant. The model parameters must be further restricted requiring $\beta<-1$.
The Jacobian matrix associated to the dynamical system when specified to such equilibrium point reads as:
\beq
J=\begin{pmatrix}- 3   \frac{b+1}{1+\gamma}   & 0  \\
	-3 \frac{1+b }{\beta (\gamma +1 ) }  & -3 \frac{\beta +1} {\beta}
\end{pmatrix}\,,
\eeq
which implies
\begin{eqnarray}
{\rm Det J} & =& \frac{ 9 (b+1) (\beta +1)     }{ \beta ( \gamma +1)  }    \\
{\rm Tr J} & =& - 3 \frac{ (b +1)\beta +(\gamma +1 )(1+\beta)     }{\beta (\gamma +1)}  \\
\frac{({\rm Tr J})^2}{4} - {\rm Det J} &=&  \frac{9 [b\beta -1 -(1+\beta) \gamma]^2 }{4 \beta ^2 (1+\gamma)^2} \,.
\end{eqnarray}
Thus, taking into account the range for the free parameters of the model, we get ${\rm Det J}>0$, which means that our equilibrium point is a node (spirals are not possible because $\frac{({\rm Tr J})^2}{4} - {\rm Det J} >0$), and we can understand that it is stable because ${\rm Tr J}<0$. An example of the phase portrait for such a system is displayed in fig (\ref{figura2}), panel (b), for the numerical choice of the free parameters as $b=-0.1$, $\gamma=0.5$, $\beta=-1.5$, and $\alpha=1.5$; the equilibrium point is denoted with a circle.

\subsection{ Dieterici - Interaction term $Q_2$ }

The dynamical system to consider is formed by the two equations (\ref{eqx2}) and (\ref{eqy2c}).
In this case we get five mathematical equilibria $(x_{\rm eq}, \, y_{\rm eq})$:
\beq
\
\left(1,\, \frac{2\beta}{e} \right), \qquad \left(1,\, \frac{\beta e^2}{2} \right), \qquad (1,\, -1), \qquad \left( \frac{2(b-\gamma)}{\beta(1+\gamma)e^2}, \, \frac{b-\gamma}{\gamma+1} \right), \qquad \left(\frac{(b-\gamma)e}{2\beta(1+\gamma)},\, \frac{b-\gamma}{1+\gamma} \right)
\eeq
of which we must exclude the second and fourth because they deliver a zero $H_{\rm eq}$. 
For the three cosmologically relevant solutions we get:
\begin{eqnarray}
(H_{\rm eq}, \, z_{\rm eq}, \, q_{\rm eq}, \, w_{\rm eq})&=&\left(\sqrt{ \frac{1}{2\alpha}  }, \, 0, \, \frac{1}{2}+\frac{3\beta}{e} , \, \frac{2\beta}{e}  \right) \\
(H_{\rm eq}, \, z_{\rm eq}, \, q_{\rm eq}, \, w_{\rm eq})&=&\left( \sqrt{\frac{W(2e^{-2}\beta)+4}{6\alpha}}, \, 0, \, -1, \,-1 \right) \\
(H_{\rm eq}, \, z_{\rm eq},\, q_{\rm eq}, \, w_{\rm eq})&=&\left(\sqrt{\frac{\beta (1+\gamma)}{\alpha (b-\gamma) e}}, \, \frac{e(\gamma -b) +2\beta (1+\gamma)}{2\beta (1+\gamma)} , \, \frac{1+3b-2\gamma}{2(\gamma +1)} , \, \frac{2\beta}{e} \right).
\end{eqnarray}
respectively.
For the first attractor, which pictures a dark-matter-free universe, we must restrict $ -\frac{3e}{5} \leq \beta  <- \frac{e}{6}$; the particular case $\beta=-\frac{e}{2}$ corresponds to a cosmological constant. The second attractor pictures as well a dark-matter-free universe in which dark energy is pictured with a cosmological constant, and for it we must restrict $\beta \geq -\frac{e}{2}$ for having a well-defined Hubble function. The third equilibrium point requires $-\frac{3e}{5} \leq \beta\leq \frac{e(b-\gamma)}{2(1+\gamma)}$ where the lower limit comes from requiring a negative deceleration parameter and a negative parameter  $ w_{\rm eq} $ for the dark energy equation of state, while the upper limit comes from requiring a positive dark matter abundance parameter: moreover a negative deceleration parameter requires $b<\frac{2\gamma -1}{3}$; then all the other physical requirement are automatically satisfied. 

\begin{figure}[ht]
	
	\centering
\begin{subfigure}[b]{.24\linewidth}
	\centering
\includegraphics[scale=0.3, angle=0]{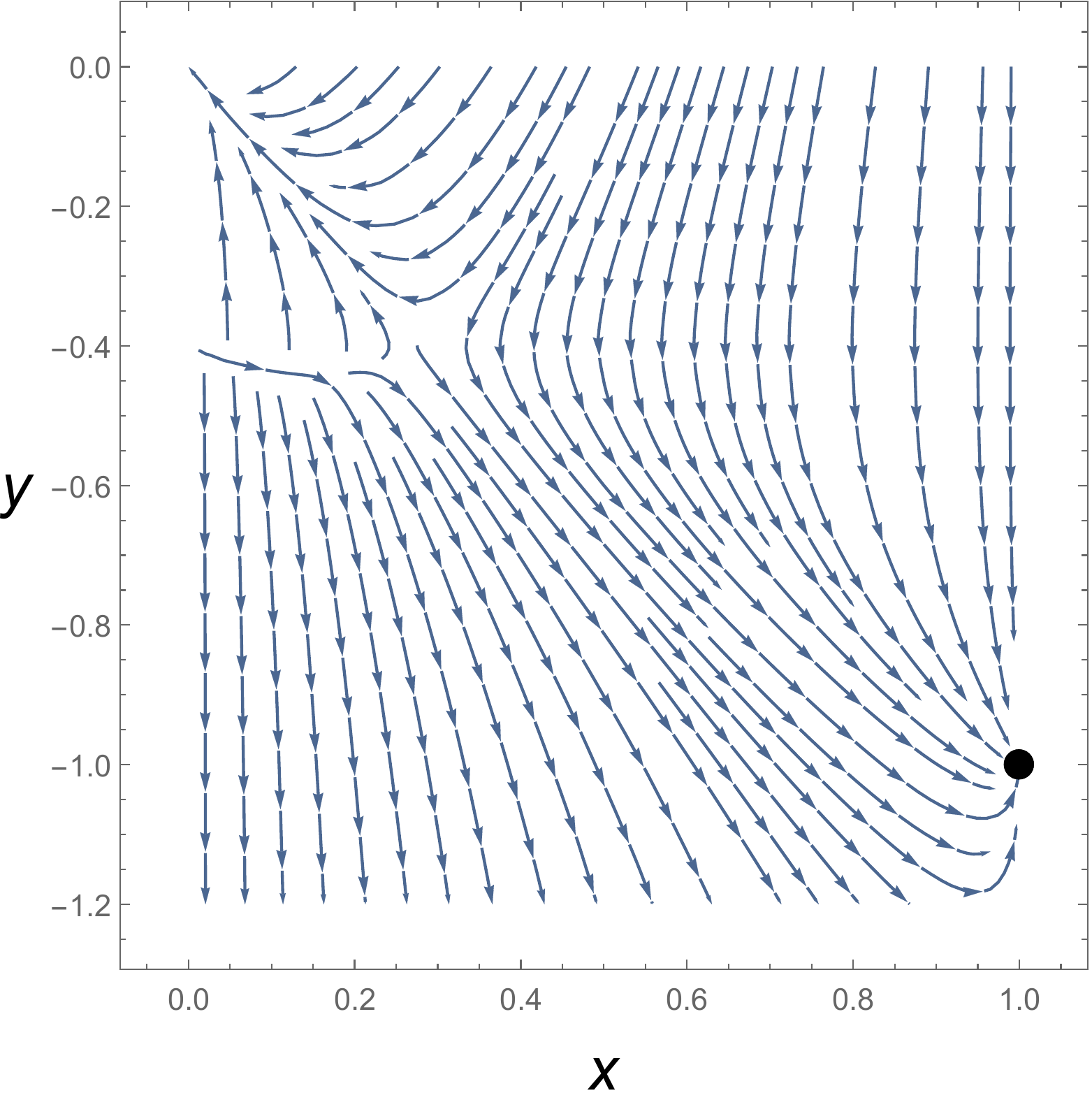}
	\caption{}
\end{subfigure}
	\centering
\begin{subfigure}[b]{.24\linewidth}
	\centering
\includegraphics[scale=0.3, angle=0]{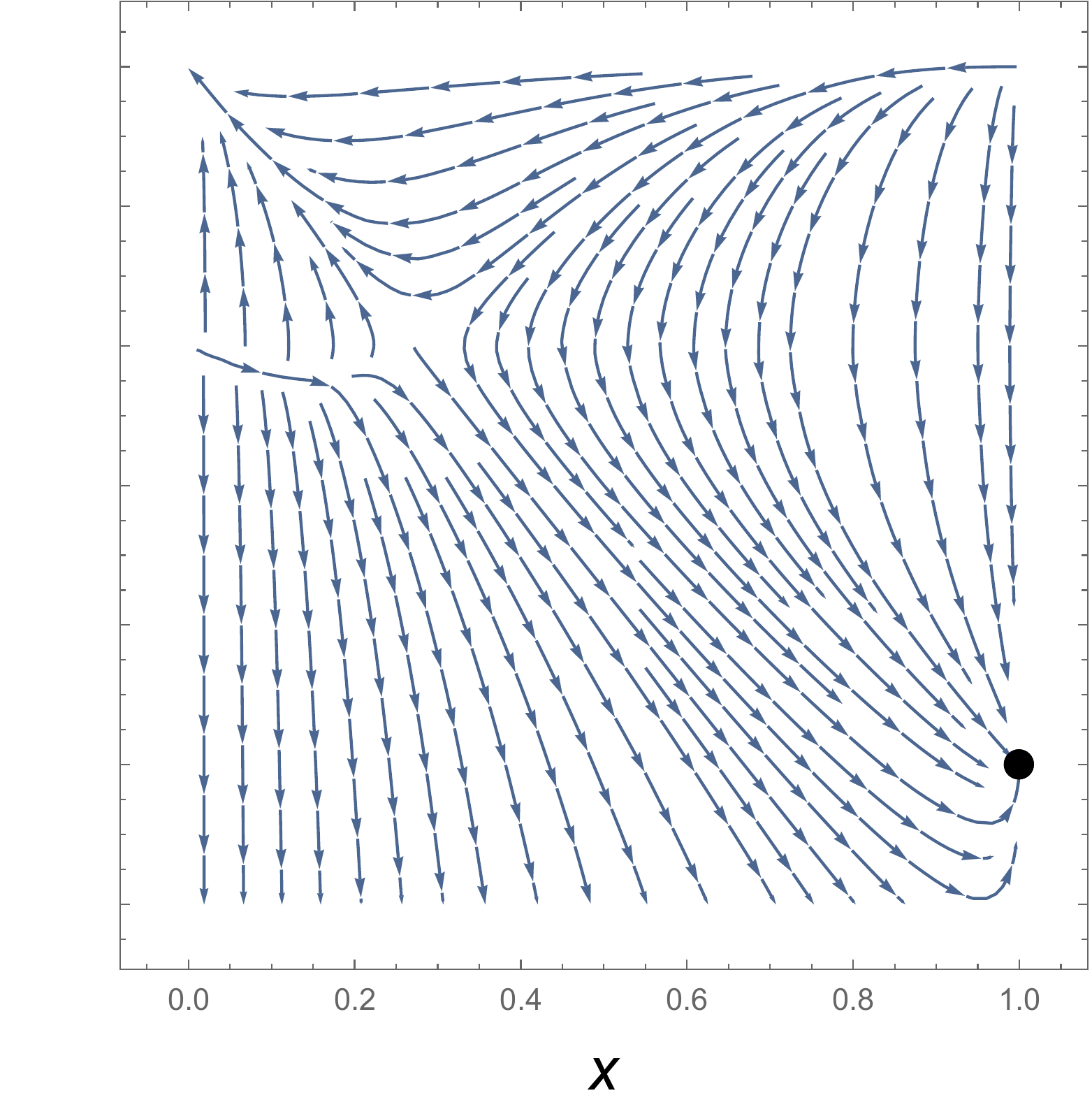}
	\caption{}
\end{subfigure}
	\centering
\begin{subfigure}[b]{.24\linewidth}
	\centering
\includegraphics[scale=0.3]{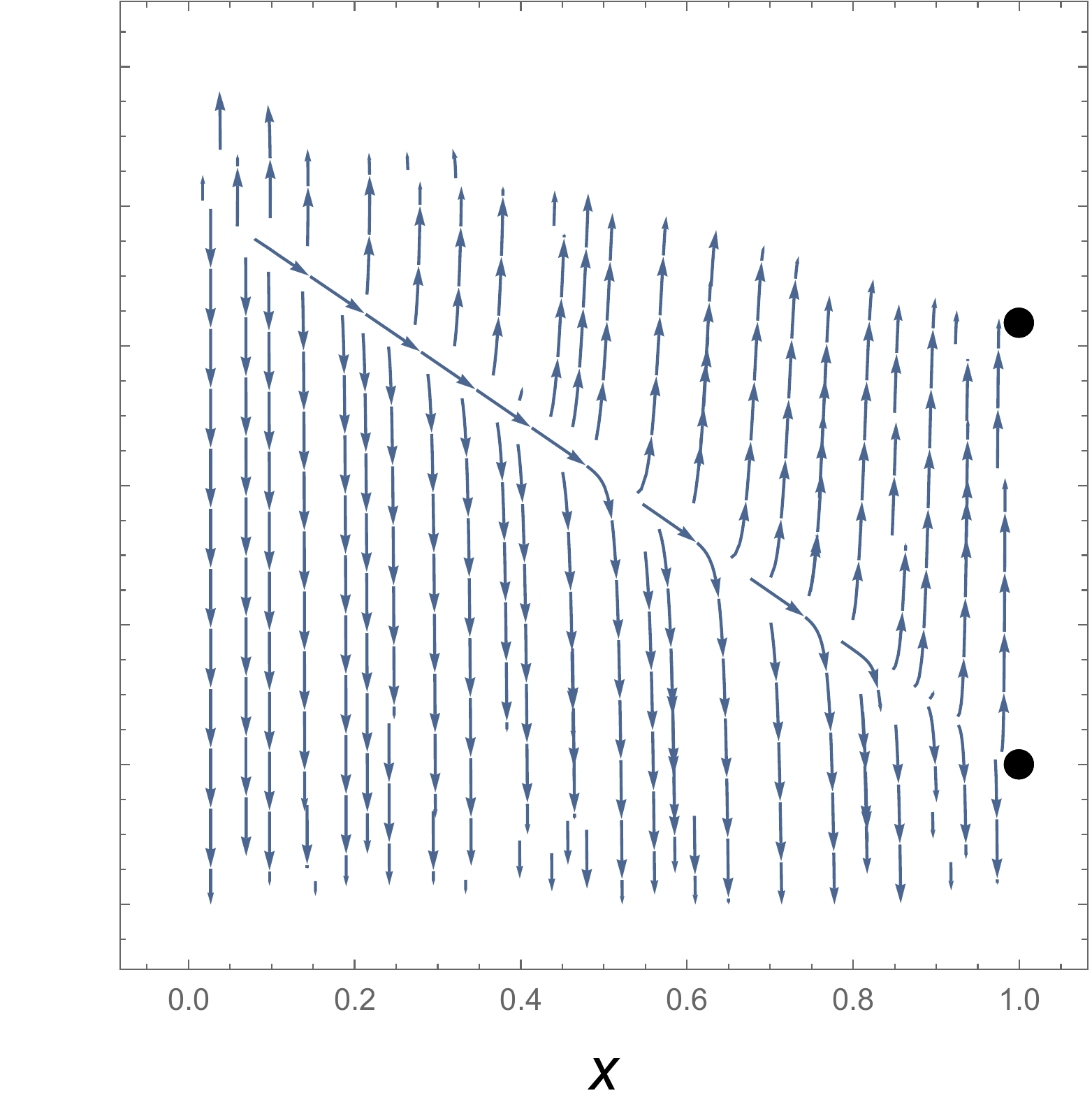}
	\caption{}
\end{subfigure}
\begin{subfigure}[b]{.24\linewidth}
	\centering
\includegraphics[scale=0.3]{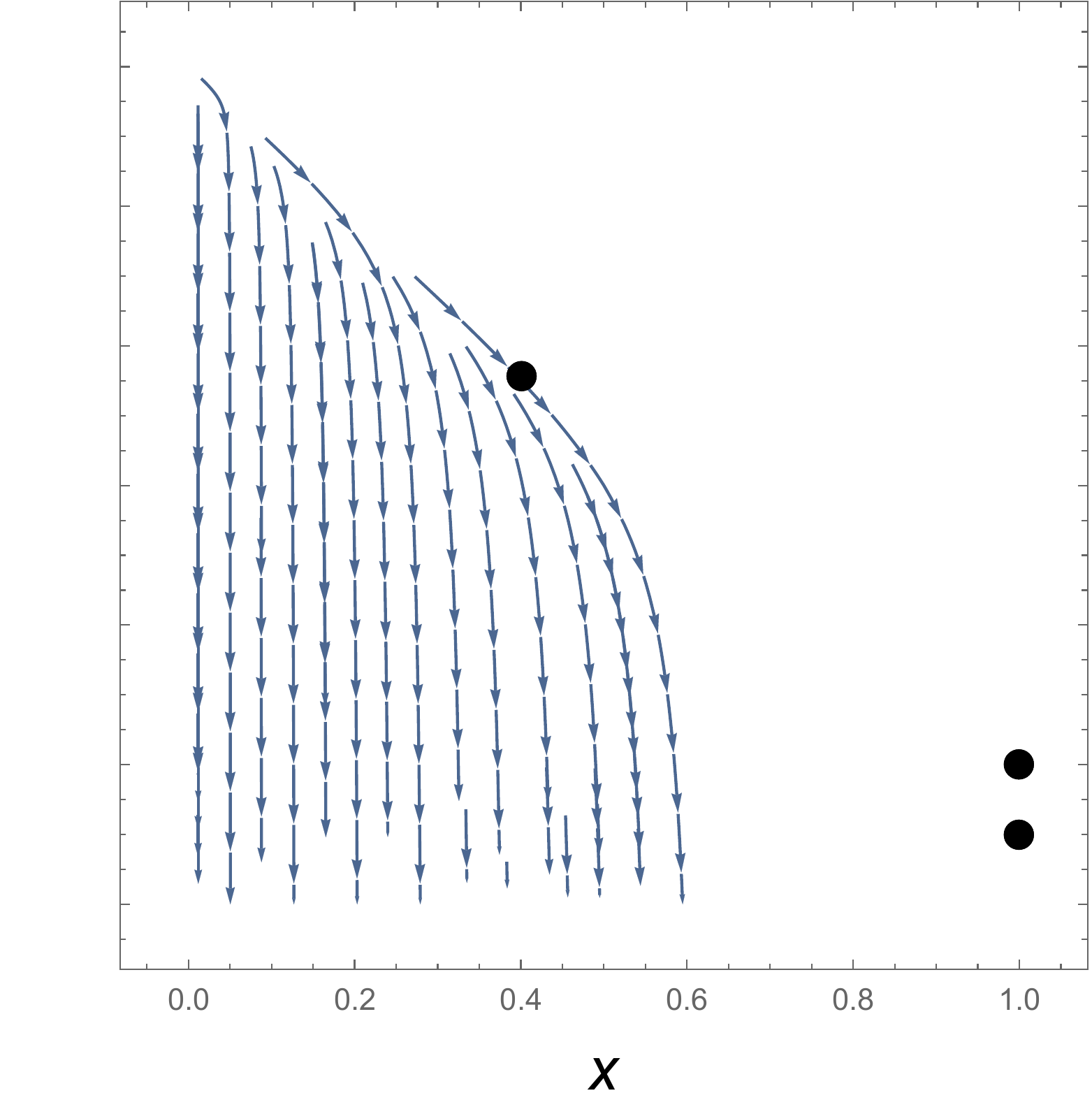}
	\caption{}
\end{subfigure}\\
\begin{subfigure}[b]{.24\linewidth}
	\centering
\includegraphics[scale=0.3]{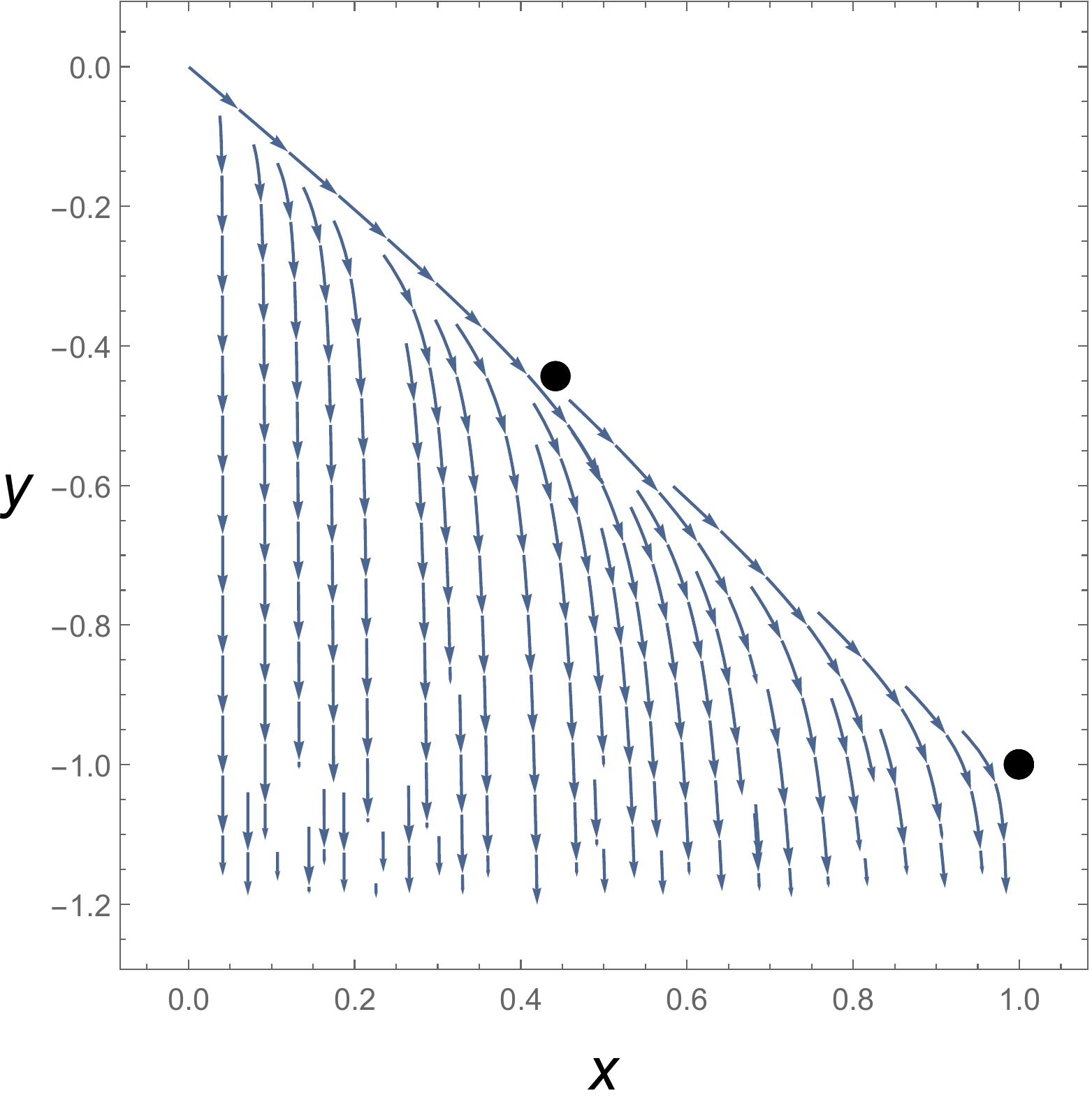}
	\caption{}
\end{subfigure}
\begin{subfigure}[b]{.24\linewidth}
	\centering
\includegraphics[scale=0.3]{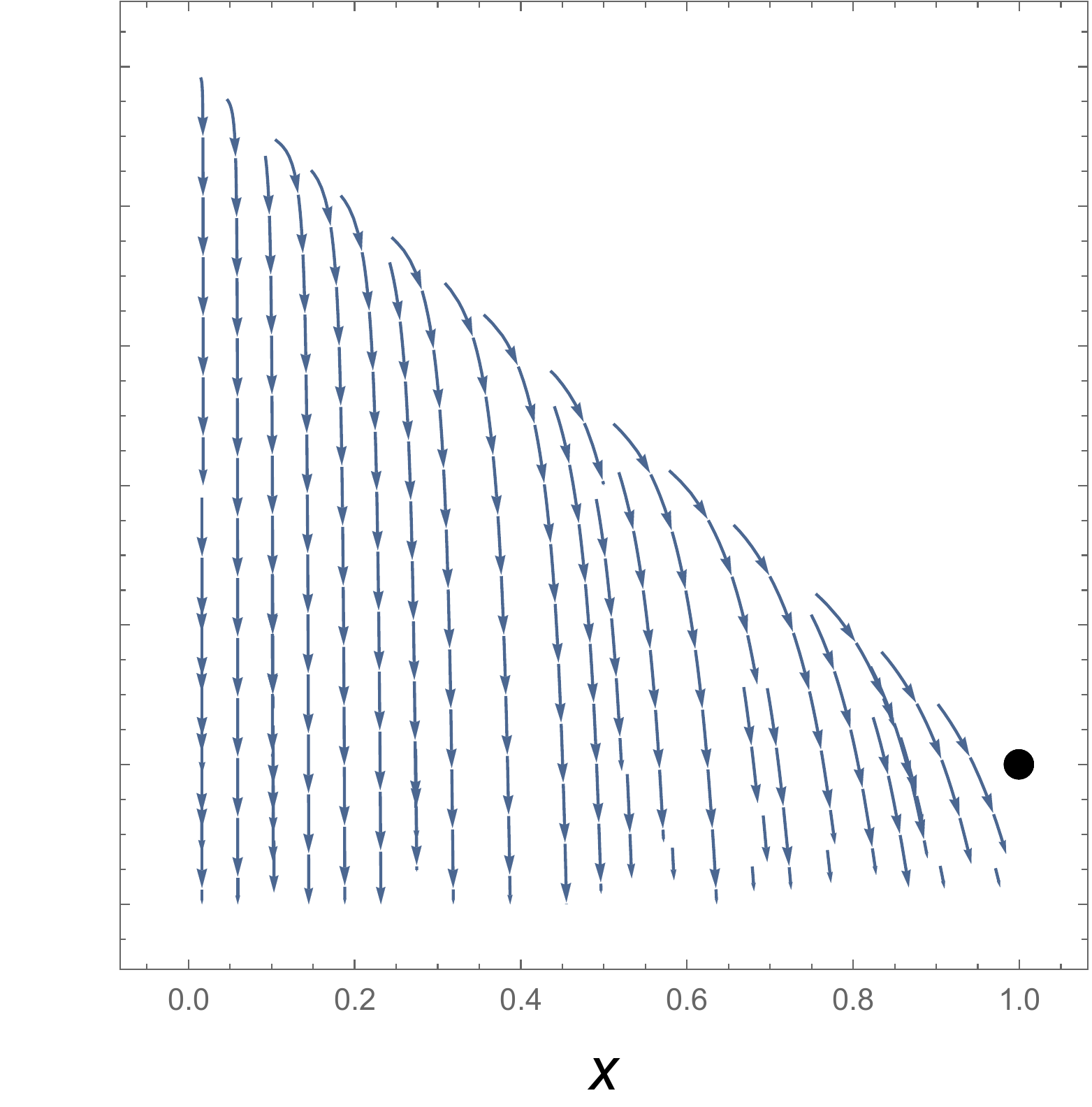}
	\caption{}
\end{subfigure}
\begin{subfigure}[b]{.24\linewidth}
	\centering
\includegraphics[scale=0.3]{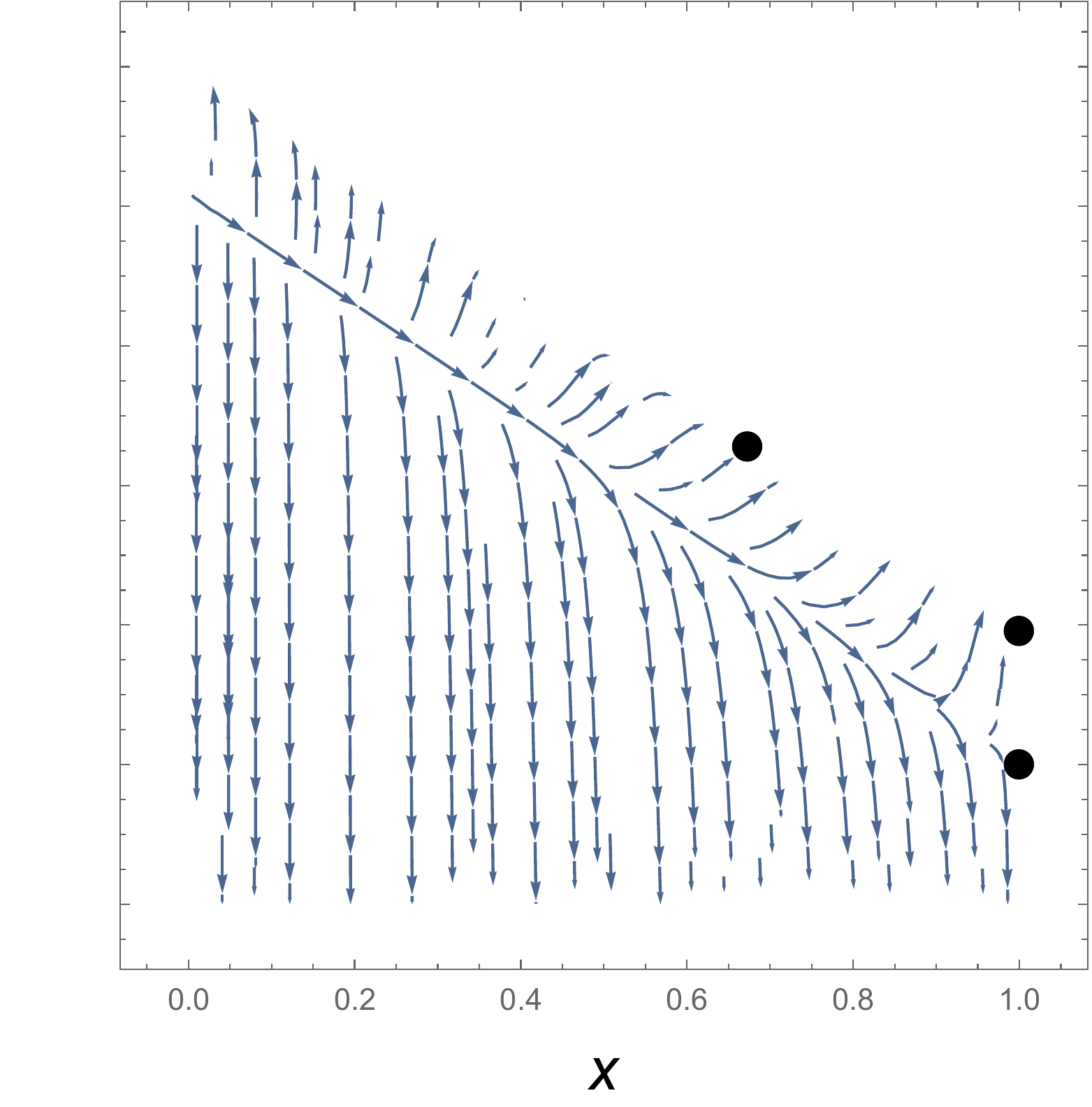}
	\caption{}
\end{subfigure}
\caption{An example of the phase portrait for the dynamical system accounting for the interaction term (\ref{int2}), and equation of state (\ref{eos1}) in panel (a), (\ref{eos2}) in panel (b), for the numerical choice of the free parameters as $b=-0.1$, $\gamma=0.5$, $\beta=-1.5$, and $\alpha=1.5$; the equilibrium point is denoted with a circle. Panels (c)-(d)-(e)-(f)-(g) display the phase space for the system for the same interaction term, but with equation of state (\ref{eos3}) for the choices of the free parameters as ($\gamma=0.1$, $b=0.5$, $\beta=-0.5$), ($\gamma=-0.1$, $b=-0.5$, $\beta=-1.5$), ($\gamma=-0.1$, $b=-0.5$, $\beta=-e/2$), ($\gamma=-0.99$, $b=-0.3$, $\beta=-e/2$), and ($\gamma=0.1$, $b=-0.5$, $\beta=-1.1$) respectively showing a rich behavior and possible bifuractions among the various attractors. Note that in this latter case the evolution equation can be studied as long as the Lambert W function is well defined, thus showing if the system can effectively reach these attractors.
}
	\label{figura2}
\end{figure}

We note that the second attractor constitutes a limit of the first attractor for the particular choice $\beta=-\frac{e}{2}$, which consequently delivers a bifurcation.  A bifurcation between the first and third attractor is possible when the condition $\frac{2\beta (1+\gamma)}{(b-\gamma)e}=1$ holds. The three attractors reduce to a unique equilibrium point for the choice ($\beta=-e/2$, $b=-1$). The Jacobian matrices are ill-defined in the first and third equilibria, while in the second it reads:
\beq
J=\begin{pmatrix}- 3   \frac{b+1}{1+\gamma}   & 0  \\
	-3 \frac{(1+b)(W(2e^{-2}\beta)+2)^2 }{ (\gamma +1 ) W(2e^{-2}\beta) }  & -3 \frac{(4+W(2e^{-2}\beta))(1+W(2e^{-2}\beta))} {W(2e^{-2}\beta)}
\end{pmatrix}\,,
\eeq
which implies
\begin{eqnarray}
{\rm Det J} & =&  \frac{9(b+1)(4+W(2e^{-2}\beta))(1+W(2e^{-2}\beta))}{(1+\gamma)W(2e^{-2}\beta)}  \,.
\end{eqnarray}
Being ${\rm Det J} <0$, this equilibrium point is a saddle.
Panels (c)-(d)-(e)-(f)-(g) of figure (\ref{figura2}) display the phase space for this cosmological system for the choices of the free parameters as ($\gamma=0.1$, $b=0.5$, $\beta=-0.5$), ($\gamma=-0.1$, $b=-0.5$, $\beta=-1.5$), ($\gamma=-0.1$, $b=-0.5$, $\beta=-e/2$), ($\gamma=-0.99$, $b=-0.3$, $\beta=-e/2$), and ($\gamma=0.1$, $b=-0.5$, $\beta=-1.1$) respectively showing a rich behavior and possible bifuractions among the various attractors. Note that in this latter case the evolution equations can be studied as long as the Lambert $ W $ function is well defined, thus showing if the system can effectively reach these attractors.

\subsection{ Redlich-Kwong - Interaction term $Q_3$ }

\begin{figure}[H]
	\begin{center}
		{\includegraphics[scale=0.3, angle=0, width=7.0cm]{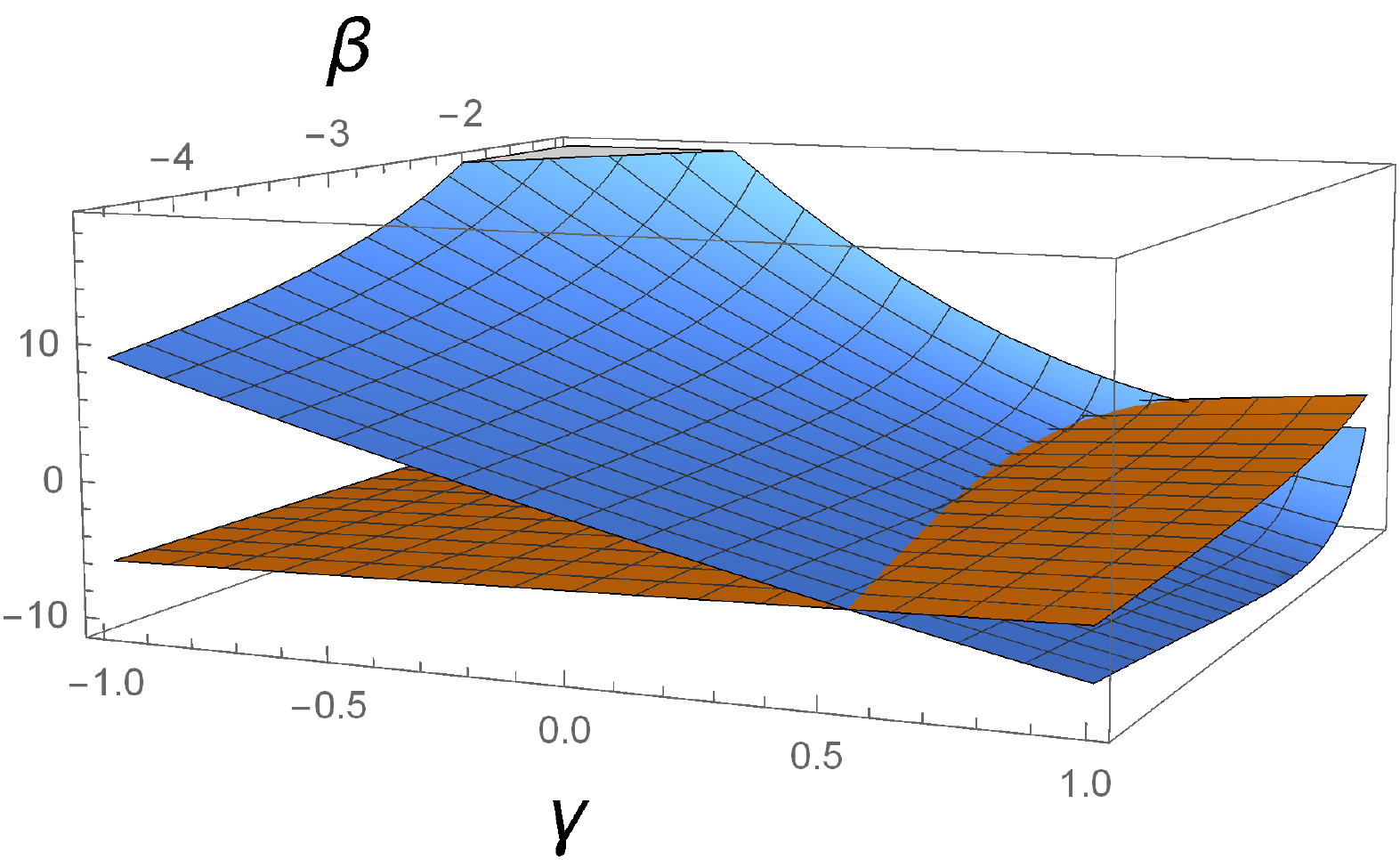}}
	\end{center}
	\caption{
		This figure shows the behavior of ${\rm Tr J}$, and of $\frac{({\rm Tr J})^2}{4} - {\rm Det J}$ given by eqs (\ref{peq1}) and (\ref{peq2}), as functions of $\gamma$ and $\beta$ for the choice $b=-0.1$ with the latter being on the top for $\gamma=-1$. 
	}
	\label{param1}
\end{figure}

The dynamical system to consider is formed by the two equations (\ref{eqx3}) and (\ref{eqy3a}).
In this case we get five mathematical equilibria $(x_{\rm eq}, \, y_{\rm eq})$:
\begin{eqnarray}
\label{eq13}
&&\left(b+1,\, -1 \right), \\
&& \left(\frac{(1+\gamma)\sqrt{\beta[\beta (1+\gamma)^2 +4(b-\gamma)]}-\beta (1+\gamma)^2+2 (\gamma -b)}{\sqrt{\beta[\beta (1+\gamma)^2 +4(b-\gamma)]}-\beta (1+\gamma)},\, \frac{\sqrt{\beta [\beta (1+\gamma)^2 +4(b-\gamma)  ]}-\beta (1+\gamma)}{2} \right), \nonumber\\
&& \qquad \left(\frac{(1+\gamma)\sqrt{\beta[\beta (1+\gamma)^2 +4(b-\gamma)]}+\beta (1+\gamma)^2-2 (\gamma -b)}{\sqrt{\beta[\beta (1+\gamma)^2 +4(b-\gamma)]}+\beta (1+\gamma)},\, -\frac{\sqrt{\beta [\beta (1+\gamma)^2 +4(b-\gamma)  ]}+\beta (1+\gamma)}{2} \right) \nonumber\\
&&\left(\frac{(1+\gamma)\sqrt{\beta[\beta (1+\gamma)^2 -4(b-\gamma)]}+\beta (1+\gamma)^2-2 (\gamma -b)}{\sqrt{\beta[\beta (1+\gamma)^2 -4(b-\gamma)]}+\beta (1+\gamma)},\, \frac{\sqrt{\beta [\beta (1+\gamma)^2 -4(b-\gamma)  ]}+\beta (1+\gamma)}{2} \right), \nonumber\\
&& \left(\frac{(1+\gamma)\sqrt{\beta[\beta (1+\gamma)^2 -4(b-\gamma)]}-\beta (1+\gamma)^2-2 (\gamma -b)}{\sqrt{\beta[\beta (1+\gamma)^2 -4(b-\gamma)]}-\beta (1+\gamma)},\, \frac{-\sqrt{\beta [\beta (1+\gamma)^2 -4(b-\gamma)  ]}+\beta (1+\gamma)}{2} \right), \nonumber
\end{eqnarray}
of which we can keep only the first since the others deliver an ill-defined (second and third), or zero (fourth and fifth) $H_{\rm eq}$. For the cosmologically relevant solution we get:
\beq
(H_{\rm eq}, \, z_{\rm eq}, \, q_{\rm eq}, \, w_{\rm eq})\,=\, \left( \sqrt{\frac{[\beta(1+b) +1] (1+\sqrt{2})}{3 \alpha (1+b) [\beta(1+b) -1]}} , \, -b , \, -1, \, -\frac{1}{1+b} \right)        .
\eeq
The restrictions to apply to the free parameters of the model are $-\frac{1}{6} \leq b \leq 0$ (from the values of $x_{\rm eq}$, $z_{\rm eq}$, and $w_{\rm eq}$), and $[\beta(1+b) +1] \cdot [\beta(1+b) -1] >0$, which implies $\beta<-\frac{1}{1+b}$, from the value of $H_{\rm eq}$. The particular case $b=0$ corresponds to the case of a dark-matter-free universe in which dark energy is pictured as a cosmological constant. The Jacobian matrix for the dynamical system specified to this attractor is
\beq
J=\begin{pmatrix}- 3      & 3(b-\gamma)  \\
	3 \frac{\beta^2(1+b)^2 +2(1+b)\beta -1 }{2\beta (b +1 )^2 }  & 3 \frac{(b+1)^2 (1+\gamma) \beta^2 -2 (1+b)\beta (b-\gamma)-\gamma-1} {2\beta (1+b)^2}
\end{pmatrix}\,,
\eeq
which implies
\begin{eqnarray}
{\rm Det J} & =& \frac{ 9 [1-\beta^2(1+b)^2]     }{2 \beta ( b +1)  }    \\
\label{peq1}
{\rm Tr J} & =&  3 \frac{(b+1)^2 (1+\gamma)\beta^2 -2(2b -\gamma +1)(1+b)\beta -\gamma-1   }{2\beta (b +1)^2}  \\
\label{peq2}
\frac{({\rm Tr J})^2}{4} - {\rm Det J} &=&  \frac{9 \left[ (1+\gamma)^2 (1+b)^2 \beta^2 +2\left(2b-\gamma+1\right)^2(1+b)\beta -(1+\gamma)^2 \right][(1+b)^2 \beta^2 +2 (1+b)\beta -1] }{16 \beta ^2 (1+b)^4}\,.\nonumber\\
\end{eqnarray}
Since ${\rm Det J}>0$ this equilibrium point cannot be a saddle, but only a node or a spiral. We note that the value of the parameter $\alpha$ does not affect this analysis, whose outcome consequently depends on the relation between $\gamma$ and the other two remaining free parameters. Fig. (\ref{param1}) depicts the behavior of ${\rm Tr J}$, and of $\frac{({\rm Tr J})^2}{4} - {\rm Det J}$ as functions of $\gamma$ and $\beta$ for the choice $b=-0.1$ with the latter being on the top for $\gamma=-1$. An example of the phase portrait for such a system is displayed in fig (\ref{figura3}), panels (a)-(b)-(c), for the numerical choices of the free parameters as ($b=-0.1$, $\gamma=0.9$, $\beta=-1.2$,  $\alpha=1.5$), ($b=-0.1$, $\gamma=-0.9$, $\beta=-1.2$,  $\alpha=1.5$), ($b=-0.1$, $\gamma=1.0$, $\beta=-1.12$,  $\alpha=1.5$) respectively; the equilibrium points are denoted with a circle, and are respectively a stable spiral, a stable node, and a unstable node for the previously mentioned choices of the free parameters.

\subsection{ Modified Berthelot - Interaction term $Q_3$ }

\begin{figure}[H]
	\begin{center}
		{\includegraphics[scale=0.3, angle=0, width=7.0cm]{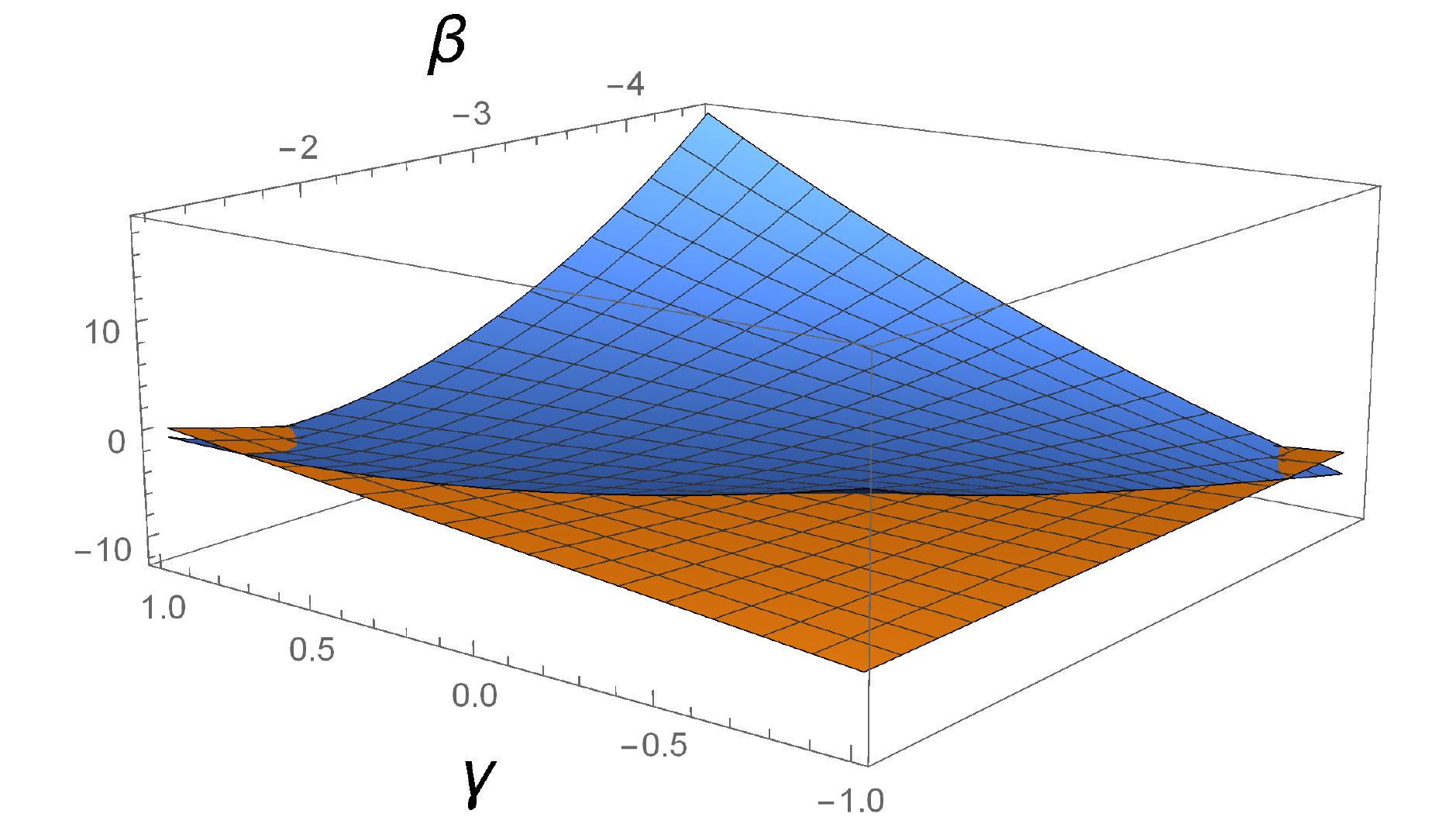}}
	\end{center}
	\caption{
		This figure represents the behavior of ${\rm Tr J}$, and of $\frac{({\rm Tr J})^2}{4} - {\rm Det J}$ given by eqs (\ref{peq3}) and (\ref{peq4}), as functions of $\gamma$ and $\beta$ for the choice $b=-0.1$ with the latter being on the top for $\gamma=-1$. 
	}
	\label{param2}
\end{figure}

The dynamical system to consider is formed by the two equations (\ref{eqx3}) and (\ref{eqy3b}).
In this case we get three mathematical equilibria $(x_{\rm eq}, \, y_{\rm eq})$:
\begin{eqnarray}
&& (b+1, \, -1) \\
&& \left( \frac{(1+\gamma) \sqrt{\beta[\beta(1+\gamma)^2 +4(\gamma -b)]}+\beta(1+\gamma)^2 +2(\gamma -b)}{\sqrt{\beta[\beta(1+\gamma)^2 +4(\gamma -b)]}+\beta(1+\gamma)}   ,\, \sqrt{\beta \left(\frac{\beta(1+\gamma)^2}{4}-b +\gamma\right)}+\frac{\beta(1+\gamma)}{2} \right) \nonumber\\ 
&&  \left( \frac{(1+\gamma) \sqrt{\beta[\beta(1+\gamma)^2 +4(\gamma -b)]}-\beta(1+\gamma)^2 -2(\gamma -b)}{\sqrt{\beta[\beta(1+\gamma)^2 +4(\gamma -b)]}-\beta(1+\gamma)}   ,\, \sqrt{\beta \left(\frac{\beta(1+\gamma)^2 }{4}-b+\gamma\right)}+\frac{\beta(1+\gamma)}{2} \right) \,, \nonumber
\end{eqnarray}
of which we can keep only the first since the others deliver a zero  $H_{\rm eq}$. For the cosmologically relevant solution we get:
\beq
(H_{\rm eq}, \, z_{\rm eq}, \, q_{\rm eq}, \, w_{\rm eq} )\,=\, \left( \sqrt{-\frac{\beta(1+b)+1}{3 \alpha (1+b) }} , \, -b , \, -1, \, -\frac{1}{1+b} \right)        .
\eeq
The restrictions to apply to the free parameters of the model are $-\frac{1}{6} \leq b \leq 0$ (from the values of $x_{\rm eq}$, $z_{\rm eq}$, and $w_{\rm eq}$), and $\beta<-\frac{1}{1+b}$ from the value of $H_{\rm eq}$. The particular case $b=0$ corresponds to the case of a dark-matter-free universe in which dark energy is pictured as a cosmological constant. The Jacobian matrix for the dynamical system specified to this attractor is
\beq
J=\begin{pmatrix}- 3      & 3(b-\gamma)  \\
	- \frac{3 }{\beta (b +1 )^2 }  & - 3 \frac{\beta (b+1)^2 +\gamma+1} {\beta (1+b)^2}
\end{pmatrix}\,,
\eeq
which implies
\begin{eqnarray}
{\rm Det J} & =& 9\frac{ 1+(1+b)\beta     }{ \beta ( b +1)  }    \\
\label{peq3}
{\rm Tr J} & =&  -3 \frac{2\beta (b+1)^2 +\gamma+1   }{\beta (b +1)^2}  \\
\label{peq4}
\frac{({\rm Tr J})^2}{4} - {\rm Det J} &=&  9 \frac{(1+\gamma)^2 +4 (b+1)^2 (\gamma -b) \beta}{4 \beta ^2 (1+b)^4}\,.
\end{eqnarray}
Since ${\rm Det J}>0$ this equilibrium point cannot be a saddle, but only a node or a spiral. We note that the value of the parameter $\alpha$ does not affect this analysis, whose outcome consequently depends on the relation between $\gamma$ and the other two remaining free parameters. Fig. (\ref{param2}) depicts the behavior of ${\rm Tr J}$, and of $\frac{({\rm Tr J})^2}{4} - {\rm Det J}$ as functions of $\gamma$ and $\beta$ for the choice $b=-0.1$ with the latter being on the top for $\gamma=-1$. An example of the phase portrait for such a system is displayed in fig (\ref{figura3}), panels (d)-(e)-(f), for the numerical choices of the free parameters as ($b=-0.1$, $\gamma=-0.5$, $\beta=-1.2$,  $\alpha=1.5$), ($b=-0.1$, $\gamma=-0.9$, $\beta=-1.2$,  $\alpha=1.5$), ($b=-0.1$, $\gamma=1.0$, $\beta=-1.12$,  $\alpha=1.5$) respectively; the equilibrium points are denoted with a circle, and are respectively a stable spiral, a stable node, and a unstable node for the previously mentioned choices of the free parameters.

\subsection{ Dieterici - Interaction term $Q_3$ }

\begin{figure}[H]
	\begin{center}
		{\includegraphics[scale=0.3, angle=0, width=7.0cm]{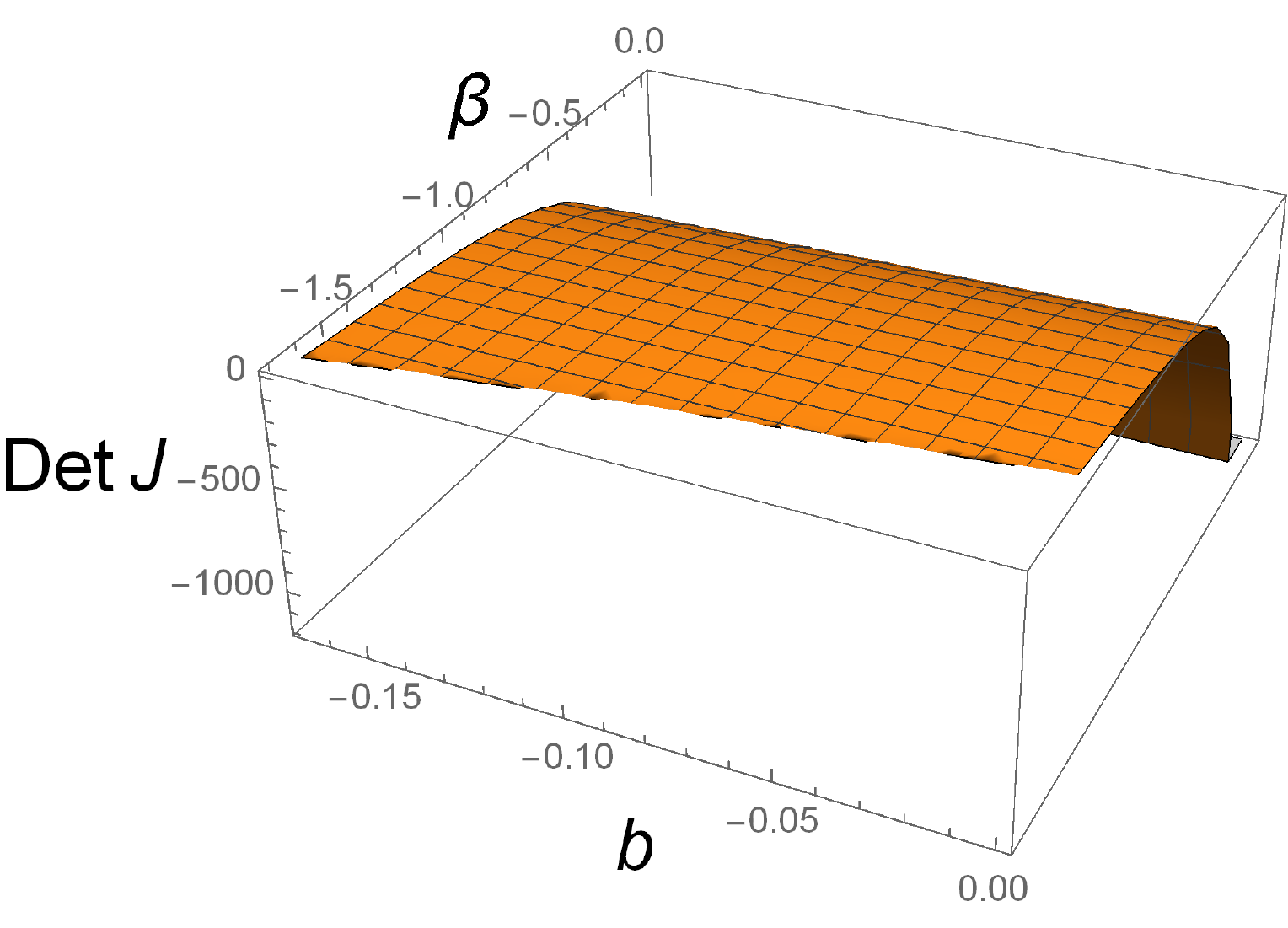}}
	\end{center}
	\caption{This figure 
		represents the behavior of ${\rm Det J}$ given by eq. (\ref{eqplot}) as function of $b$ and $\beta$ showing that it is negative for the range of parameters we are interested in. 
	}
	\label{detplot}
\end{figure}

\begin{figure}[H]
	\centering
	\begin{subfigure}[b]{.24\linewidth}
		\centering
	\includegraphics[scale=0.3, angle=0]{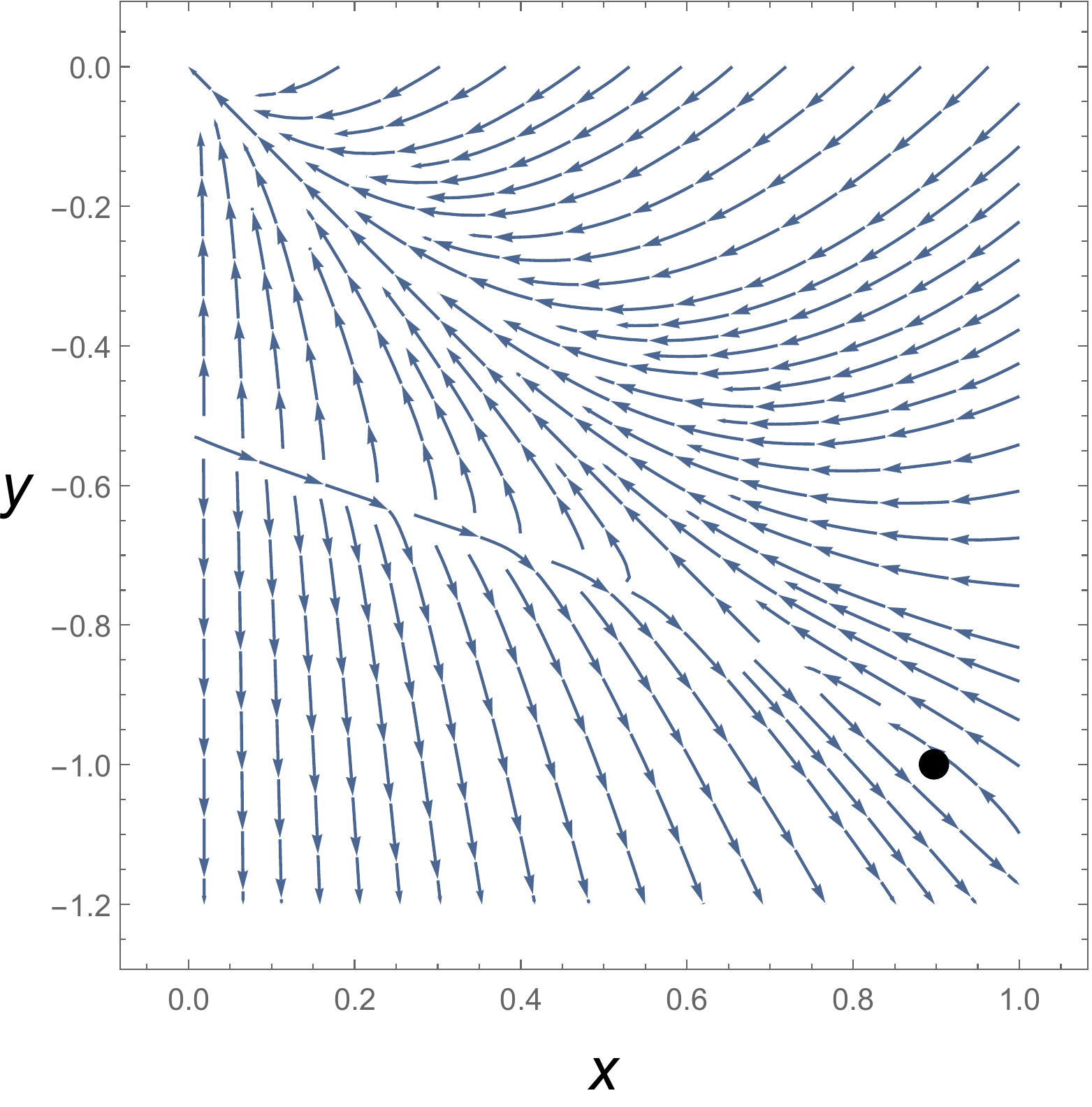}
		\caption{}
	\end{subfigure}
	\begin{subfigure}[b]{.24\linewidth}
	\centering
	\includegraphics[scale=0.3, angle=0]{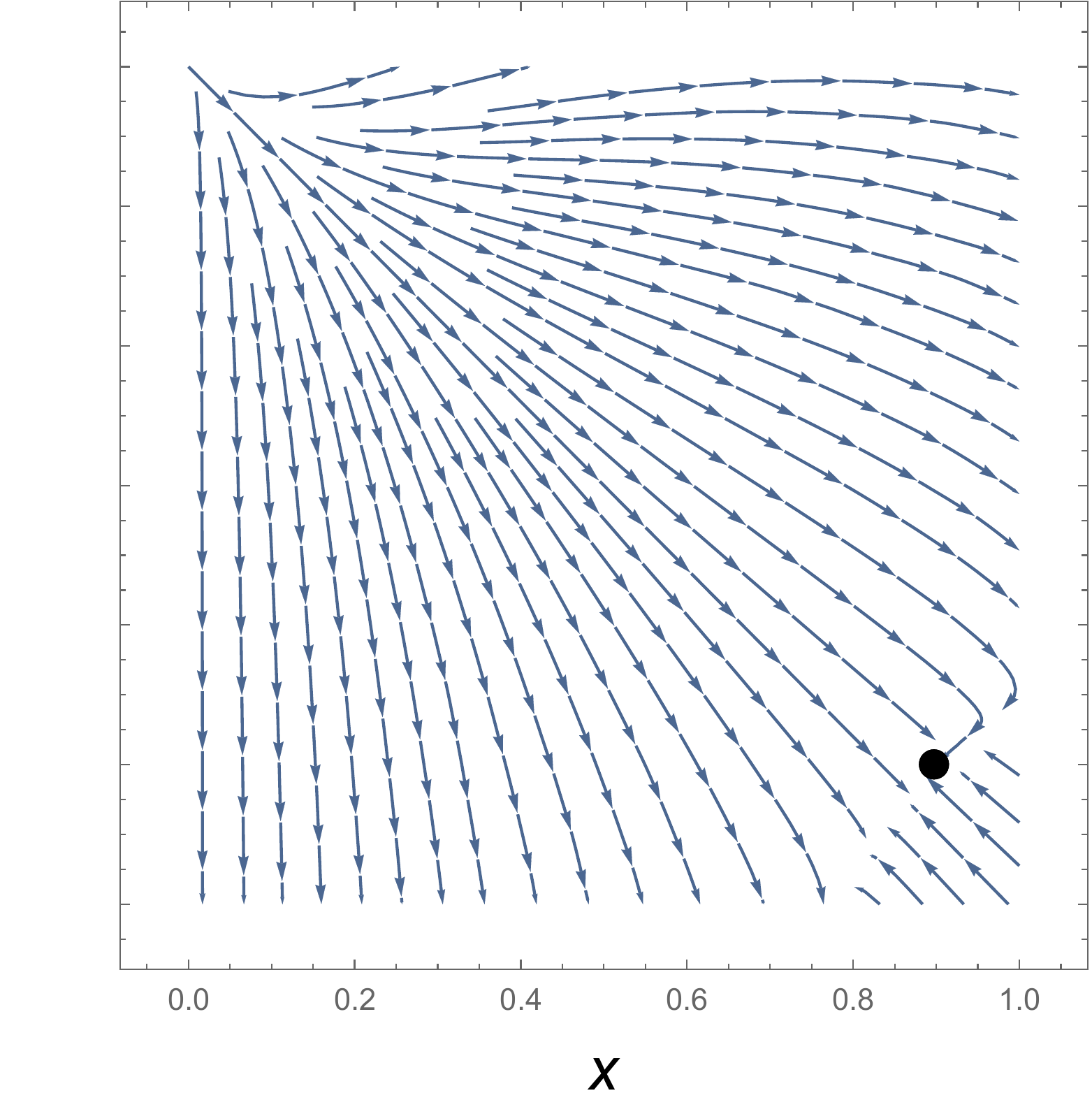}
	\caption{}
     \end{subfigure}
      \begin{subfigure}[b]{.24\linewidth}
	\centering
	\includegraphics[scale=0.3]{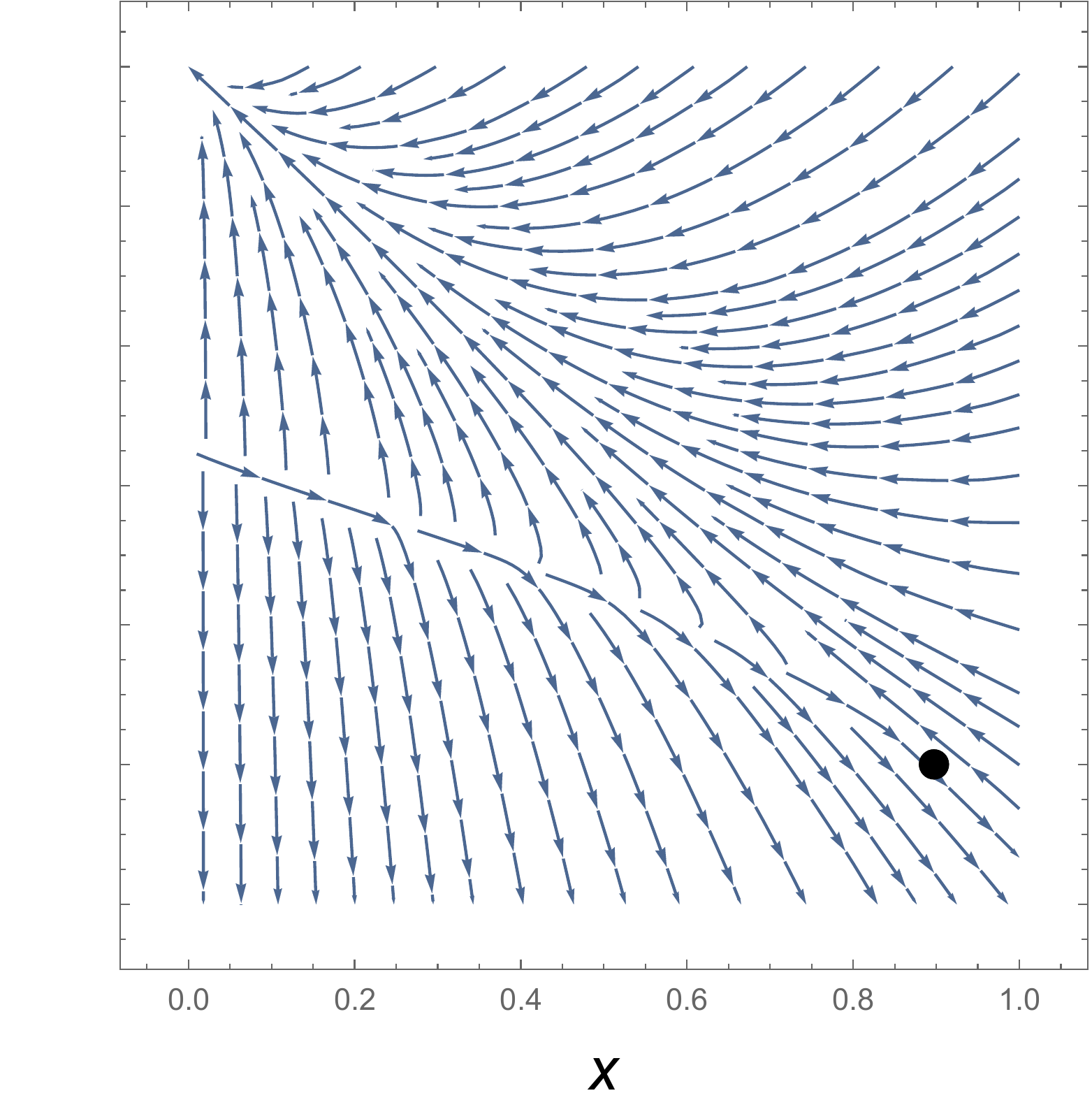}
	\caption{}
     \end{subfigure}
   \begin{subfigure}[b]{.24\linewidth}
	\centering
	\includegraphics[scale=0.3]{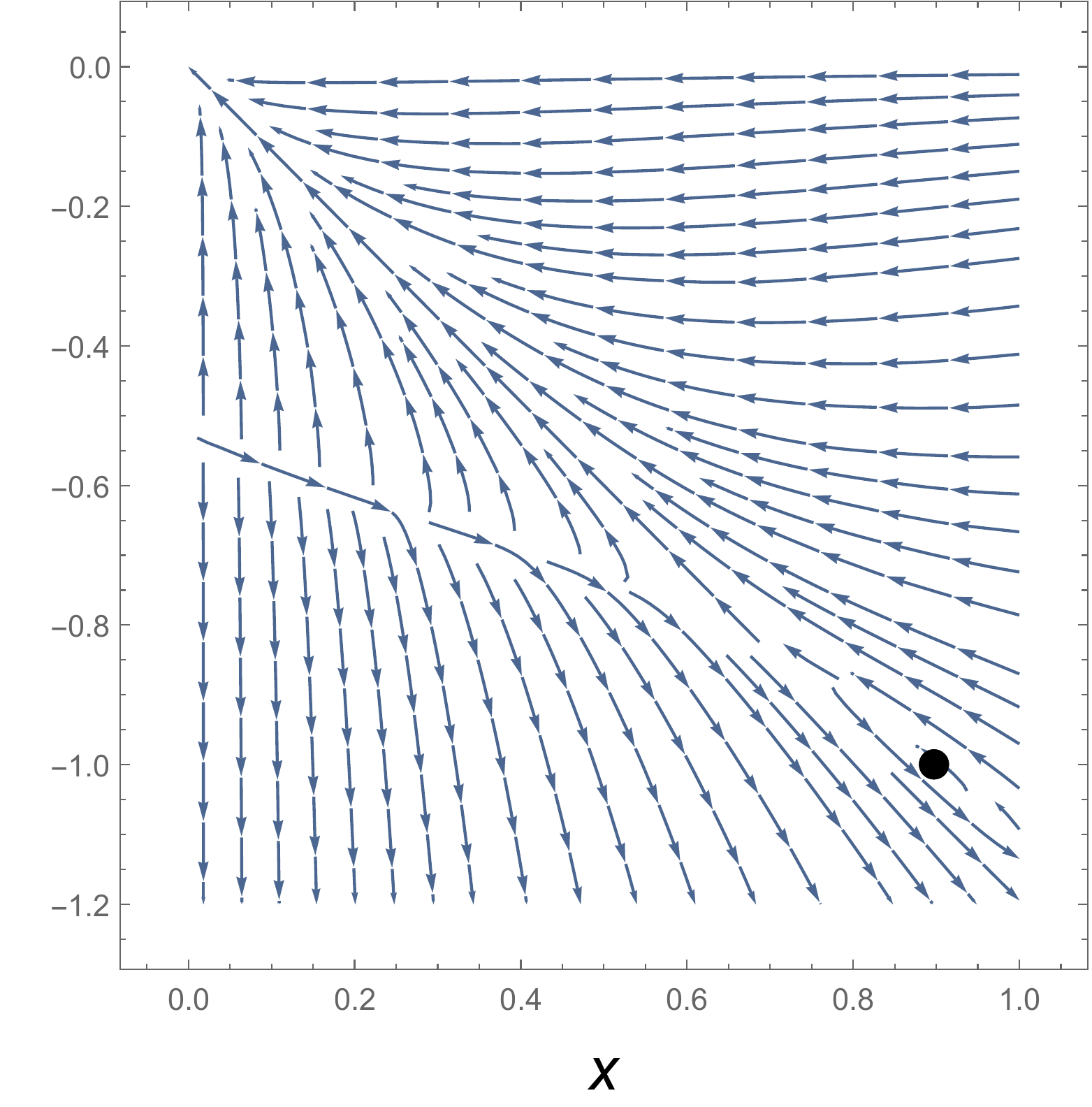}
	\caption{}
\end{subfigure}\\
\begin{subfigure}[b]{.24\linewidth}
	\centering
	\includegraphics[scale=0.3]{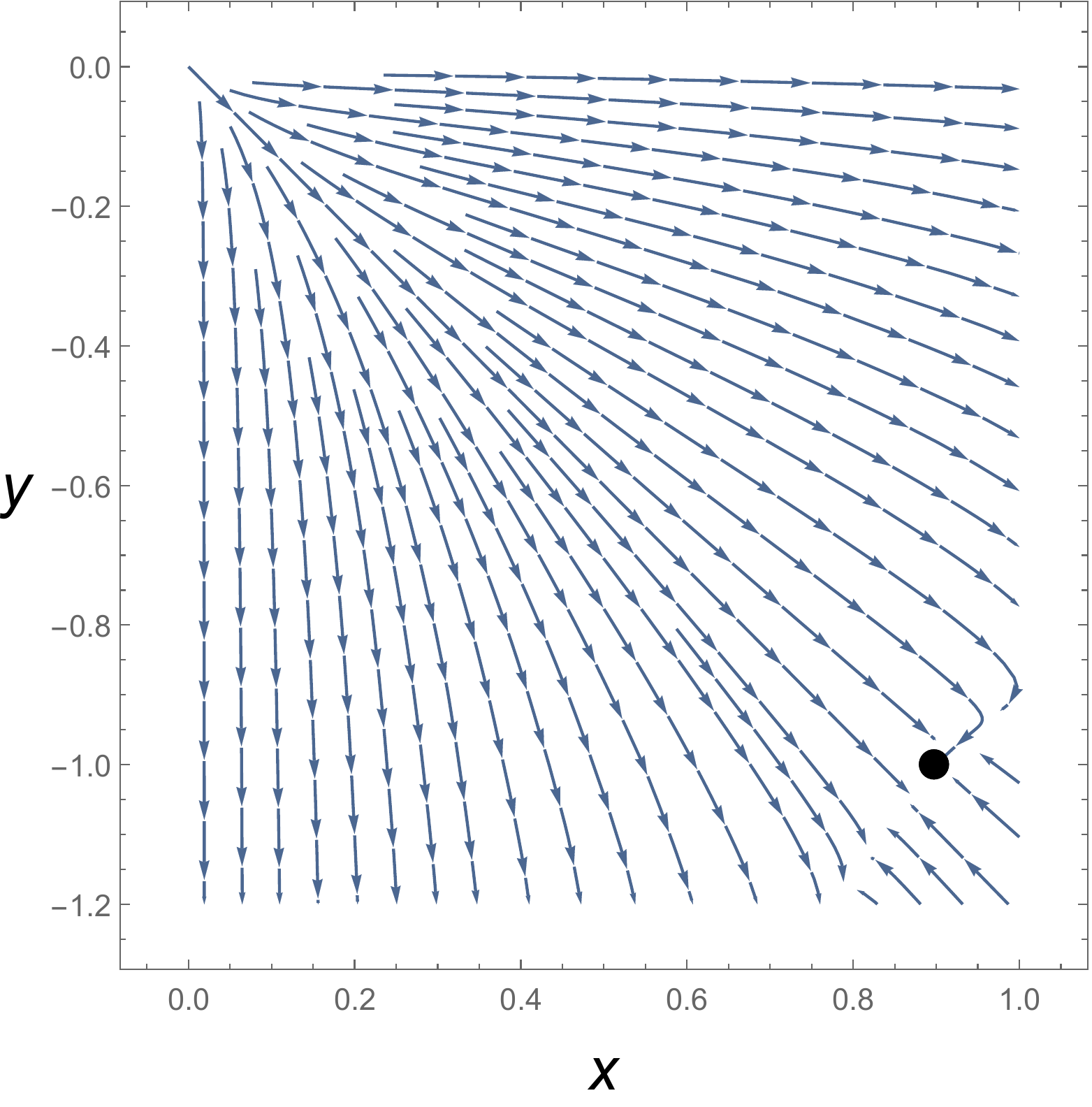}
	\caption{}
\end{subfigure}
\begin{subfigure}[b]{.24\linewidth}
	\centering
	\includegraphics[scale=0.3]{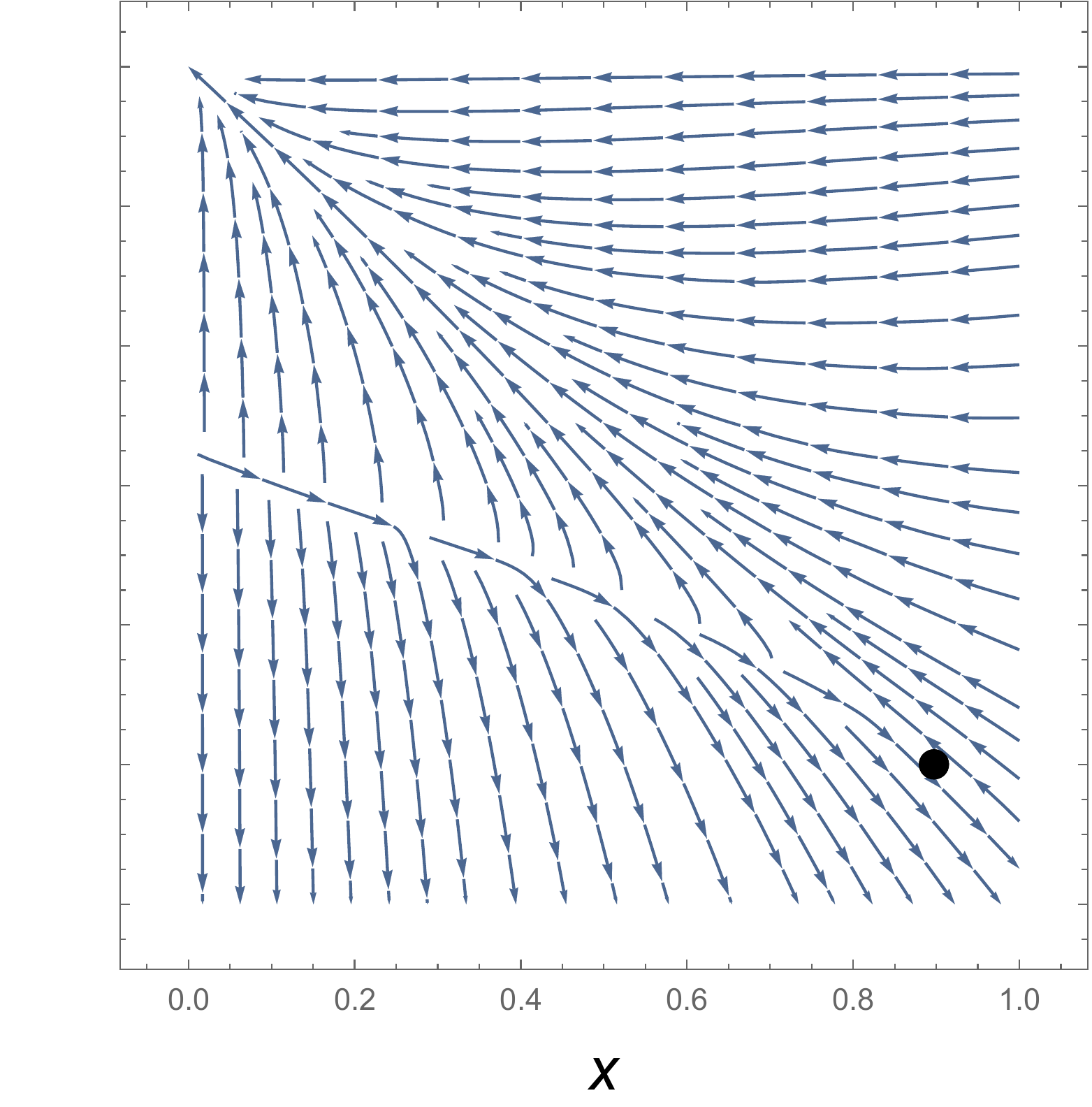}
	\caption{}
\end{subfigure}
\begin{subfigure}[b]{.24\linewidth}
	\centering
	\includegraphics[scale=0.3]{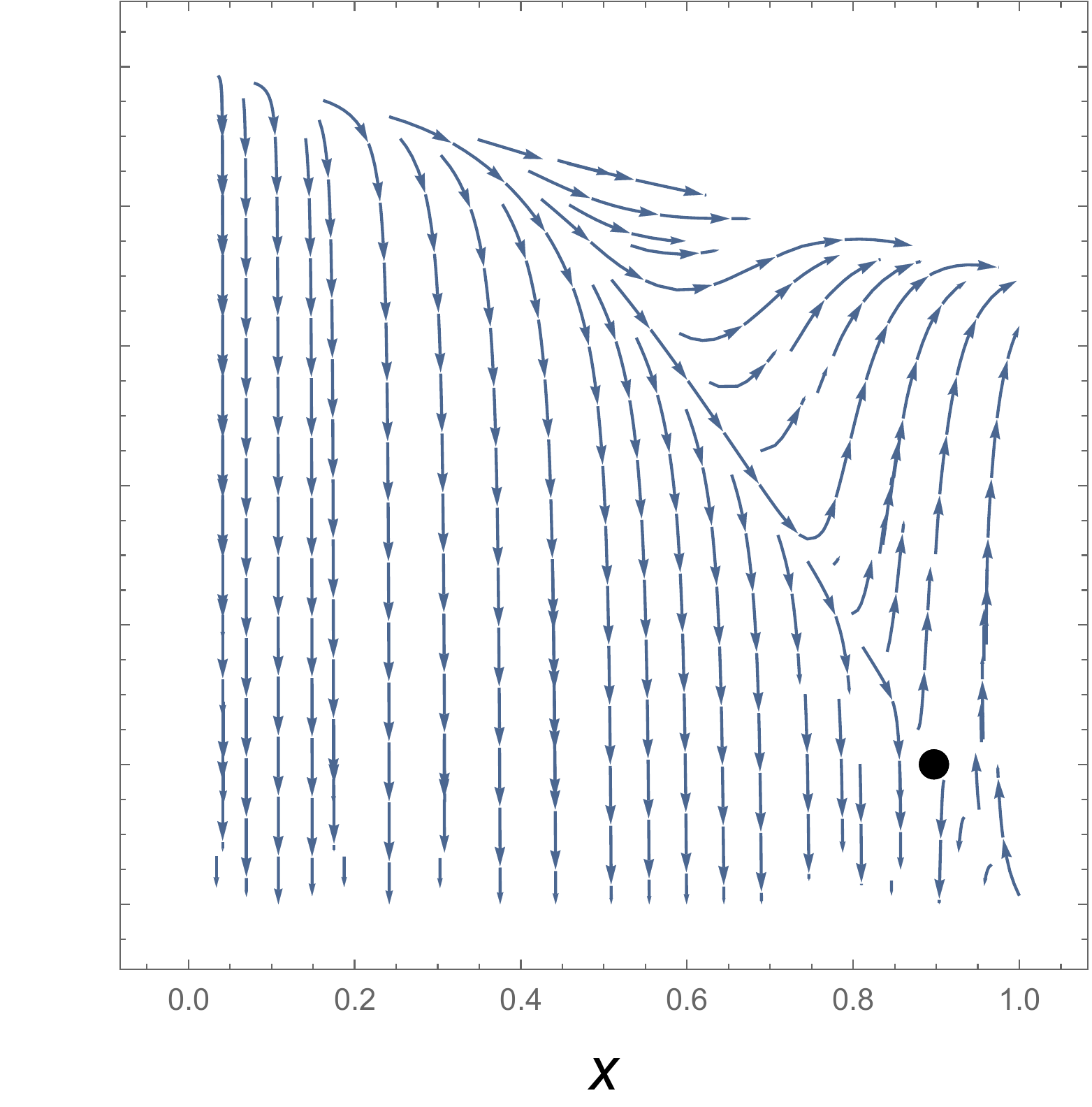}
	\caption{}
     \end{subfigure}

\caption{ An example of the phase portrait for the dynamical system accounting for the interaction term (\ref{int3}), and equation of state (\ref{eos1}) in panel (a)-(b)-(c), (\ref{eos2}) in panel (d)-(e)-(f), and (\ref{eos3}) in panel (g), for the numerical choices of the free parameters as ($b=-0.1$, $\gamma=0.9$, $\beta=-1.2$, and $\alpha=1.5$), ($b=-0.1$, $\gamma=-0.9$, $\beta=-1.2$, and $\alpha=1.5$), ($b=-0.1$, $\gamma=1.0$, $\beta=-1.12$, and $\alpha=1.5$), and ($b=-0.1$, $\gamma=0.9$, $\beta=-0.4$,  $\alpha=1.5$),  respectively; the equilibrium points are denoted with a circle, and are respectively a stable spiral, a stable node, and a unstable node, and a spiral for the previously mentioned choices of the free parameters.}
\label{figura3}

\end{figure}

The dynamical system to consider is formed by the two equations (\ref{eqx3}) and (\ref{eqy3c}).
In this case we get five mathematical equilibria $(x_{\rm eq}, \, y_{\rm eq})$:

\begin{eqnarray}
&& (b+1, \, -1) \\
&& \left( \frac{\beta (\gamma+1)^2+e (\gamma-b)+\sqrt{\beta(\beta (\gamma+1)^2+2e(\gamma-b))}(\gamma+1)} {\beta\gamma+\sqrt{\beta(\beta (\gamma+1)^2+2e(\gamma-b))}+\beta} ,\, \frac{\beta\gamma+\sqrt{\beta(\beta (\gamma+1)^2+2e(\gamma-b))}+\beta}{e} \right) \nonumber\\ 
&&  \left( \frac{\beta (\gamma+1)^2+e (\gamma-b)-\sqrt{\beta(\beta (\gamma+1)^2+2e(\gamma-b))}(\gamma+1)} {\beta\gamma+\sqrt{\beta(\beta (\gamma+1)^2+2e(\gamma-b))}+\beta} ,\, \frac{\beta\gamma-\sqrt{\beta(\beta (\gamma+1)^2+2e(\gamma-b))}+\beta}{e} \right) \nonumber \\
&& \left( \frac{e^2 \beta(1+\gamma)^2+e\sqrt{\beta(e^2 \beta (\gamma+1)^2+8(\gamma-b))}(\gamma+1)+4(\gamma-b)}{e^2 \beta (\gamma+1)+e \sqrt{\beta(e^2 \beta (\gamma+1)^2+8(\gamma-b))}} , \, \frac{e^2 \beta (\gamma+1)+e \sqrt{\beta(e^2 \beta (\gamma+1)^2+8(\gamma-b))}} {4} \right) \nonumber\\
&& \left( \frac{e^2 \beta(1+\gamma)^2-e\sqrt{\beta(e^2 \beta (\gamma+1)^2+8(\gamma-b))}(\gamma+1)+4(\gamma-b)}{e^2 \beta (\gamma+1)-e \sqrt{\beta(e^2 \beta (\gamma+1)^2+8(\gamma-b))}} , \, \frac{e^2 \beta (\gamma+1)-e \sqrt{\beta(e^2 \beta (\gamma+1)^2+8(\gamma-b))}} {4} \right)\,, \nonumber
\end{eqnarray}
of which we can keep only the first since the others deliver a zero or an ill-defined $H_{\rm eq}$. For the cosmologically relevant solution we get:
\beq
(H_{\rm eq}, \, z_{\rm eq}, \, q_{\rm eq}, \, w_{\rm eq} )\,=\, \left( \sqrt{\frac{W(2(b+1)\beta e^{-2}) +4)}{6\alpha(b+1)}} , \, -b , \, -1, \, -\frac{1}{1+b} \right)        .
\eeq
The restrictions to apply to the free parameters of the model are $-\frac{1}{6} \leq b \leq 0$ (from the values of $x_{\rm eq}$, $z_{\rm eq}$, and $w_{\rm eq}$), and $ -\frac{e}{2(b+1)}<\beta<-\frac{2}{e^2 (1+b)}$ from the value of $H_{\rm eq}$. The Jacobian matrix for the dynamical system specified to this attractor is
\beq
J=\begin{pmatrix}- 3      & 3(b-\gamma)  \\
	- 3\frac{[W(2(b+1)\beta e^{-2}) +2]^2 }{ (b +1 ) W(2(b+1)\beta e^{-2}) }  & - 3 \frac{(\gamma+1)[W^2(2(b+1)\beta e^{-2})+4]+(b+4\gamma +5)W(2(b+1)\beta e^{-2})} {(1+b) W(2(b+1)\beta e^{-2})}
\end{pmatrix}\,,
\eeq
which implies
\begin{eqnarray}
\label{eqplot}
{\rm Det J} & =& 9\frac{ W^2(2(b+1)\beta e^{-2}) +5W(2(b+1)\beta e^{-2})+4   }{ W(2(b+1)\beta e^{-2})  }.    
\end{eqnarray}
Fig. (\ref{detplot}) shows that ${\rm Det J} <0$ implying that our equilibrium point is a saddle. An example of the phase portrait for such a system is displayed in fig (\ref{figura3}), panel (g), for the numerical choices of the free parameters as ($b=-0.1$, $\gamma=0.9$, $\beta=-0.4$,  $\alpha=1.5$), and the equilibrium point, which is a saddle, is denoted with a circle.

\subsection{ Summary }

Table (\ref{table1}) summarizes the nine cases given by the combinations of equation of state for the dark energy and the type of interaction between dark matter and dark energy. Each cell reports the values assumed by the physical quantities $(x_{\rm eq}, \, y_{\rm eq}, \, H_{\rm eq}, \, z_{\rm eq}, \, q_{\rm eq}, \, w_{\rm eq} )$ at equilibrium, the type of the attractor, and the restrictions on the numerical values of the free parameters of the model  following from  the requirements $0 \leq x_{\rm eq} \leq 1$, $0 \leq z_{\rm eq} \leq 1$, $-\frac{6}{5}=-1.2 \leq w_{\rm eq} =\frac{y_{\rm eq}}{x_{\rm eq}} < 0$, $q_{\rm eq}=\frac{1}{2}(1+ 3y_{\rm eq}) <0$, well defined real and non-zero $H_{\rm eq}$ (for satisfying the Friedmann equation), and the general requirements $-1 \leq b \leq 1$, $-1 \leq \gamma \leq 1$, $\alpha >0$, $\beta<0$.

\begin{center}
	\begin{table}[H]
		\centering
	\begin{tabular}{ |c|c|c|c|c| }
		\hline
		\footnotesize
		\backslashbox{Interaction type}{EoS} &  Redlich-Kwong & Modified Berthelot & Dieterici  \\
		\hline
		$3Hb \rho_{\rm de} + \gamma \dot \rho_{\rm de}$    &   \makecell[l]{$\Big( \frac{1}{1-b}, \, -1, $\\$\sqrt{ \frac{(b-\beta-1) (1+ \sqrt{2} )(b-1)}{3 \alpha (b +\beta -1)}}, \, \frac{b}{b-1},$\\$  -1, \, b-1 \Big)$ \\ Stable node  \\ $-\frac15 \leq b\leq 0$ \\ $\beta<b-1$}  &  \makecell[l]{$\Big( \frac{1}{1-b}, \, -1, $\\$\sqrt{ \frac{b-\beta -1}{3 \alpha }}, \, \frac{b}{b-1},$\\$ \,  -1, \, b-1 \Big)$ \\ Stable node  \\ $-\frac15 \leq b\leq 0$ \\ $\beta<b-1$}  &  \makecell[l]{$\Big( \frac{1}{1-b}, \, -1 ,$\\$\sqrt{\frac{\Big[W \left(\frac{2\beta}{e^2(1-b)} \right)+4 \Big](1-b)}{6 \alpha}}  , \, \frac{b}{b-1} ,$\\$ \, -1,\, b-1 \Big)$ \\ Saddle  \\ $-\frac15 \leq b\leq 0$ \\ $\beta < 2(b-1)e^{-2} $}      \\
		\hline
		$3Hb \rho_{\rm dm} + \gamma \dot \rho_{\rm dm}$    &  \makecell[l]{$\Big( 1, \, -1, $\\$ \sqrt{ \frac{(1+\beta) (1+\sqrt{2})}{3 \alpha (\beta -1)}}, \, 0,$\\$ \, -1, \, -1 \Big)$ \\ Stable node   \\ $\beta<-1$}        & \makecell[l]{$\Big( 1, \, -1,$\\$ \sqrt{-\frac{1+\beta}{3 \alpha}}, \, 0, $\\$ -1, \, -1 \Big)$ \\ Stable node   \\ $\beta<-1$}  & Bifurcations here            \\
		\hline
		\makecell[l]{$3Hb (\rho_{\rm de}+\rho_{\rm dm}) + \gamma (\dot \rho_{\rm de} + \dot \rho_{\rm dm} )$} &     \makecell[l]{$\Big( b+1, \, -1 $\\$ \sqrt{\frac{[\beta(1+b) +1] (1+\sqrt{2})}{3 \alpha (1+b) [\beta(1+b) -1]}} , \, -b , $\\$ -1, \, -\frac{1}{1+b} \Big)$  \\ Stable spiral, \\or stable node, \\ or unstable node  \\ $-\frac16 \leq b\leq 0$ \\ $\beta<-\frac{1}{1+b}$} &   \makecell[l]{$\Big( b+1, \, -1, $\\$ \sqrt{\frac{\beta(1+b) +1 }{3 \alpha (1+b) }} , \, -b , $\\$ -1, \, -\frac{1}{1+b} \Big)$ \\ Stable spiral, \\or stable node, \\ or unstable node  \\ $-\frac16 \leq b\leq 0$ \\ $\beta<-\frac{1}{1+b}$}   &  \makecell[l]{$\Big( b+1, \, -1, $\\$ \sqrt{\frac{W(2(b+1)\beta e^{-2}) +4)}{6\alpha(b+1)}} , \, -b , $\\$ -1, \, -\frac{1}{1+b} \Big)$ \\ Saddle  \\ $-\frac16 \leq b\leq 0$ \\ $-\frac{e}{2(1+b)}<\beta<-\frac{2}{e^2(1+b)}$}               \\
		\hline
	\end{tabular}
	\caption{This table summarizes the values of the cosmological quantities  at the equilibrium for the nine dark energy - dark matter interacting systems we have investigated.}
	\label{table1}
\end{table}
\end{center}

To summarize, our analysis about linearly interacting dark matter - nonideal dark energy fluids in the linearized regime of the dynamical evolution allows us to establish some recurrent physical patterns among the different modelings for dark energy.
\begin{enumerate}
	\item The strength of interaction (that is the deviation from an ideal gas behavior $p \propto \rho$) inside the dark energy fluid quantified by the parameter $\alpha$ does not affect the stability nature of the late-time attractors. 
	\item Similarly, $\alpha$ enters only the value of the Hubble function $H_{\rm eq}$, and not the other physical quantities at the equilibrium like the deceleration parameter, the matter abundance, and the effective parameter of the dark energy equation of state.
	\item In the light of the two remarks above, the numerical value of  $\alpha$ is not restricted by any late-time cosmological requirement.
	\item When an interacting term linear in the dark matter is assumed, the final states are the same as the limiting $b \to 0$ case when the interaction is linear in the dark energy (i.e. that in each table the first row reduces to the second for $b=0$).
	\item Fixing the type of dark energy - dark matter interaction and varying the equation of state for dark energy brings to final states which differ only for the value of the Hubble constant, that is the age of the universe,  with the same other cosmological quantities.
	\item The limit of a diverging $\alpha \to \infty$ delivers a zero cosmological constant  for all the combinations eos - interaction type.
	\item Only the Dieterici equation of state supports bifurcations between the late-time attractors.
	\item The late-time attractors that we have found supporting a negative deceleration parameter correspond to a de Sitter universe.
	\item Our analysis has identified as well some equilibria characterized by the vanishing of the Hubble function. These points correspond to a Minkowski universe. In fact,  from (\ref{4})  we understand that a zero Hubble function necesserely implies absence of energy density for both dark energy and dark matter. Then, from the equations of state considered in this paper (\ref{Redlich})-(\ref{modified})-(\ref{Dieterici}) we obtain that a zero energy density for dark energy implies a zero dark energy pressure. Therefore, from the conservation equations (\ref{6})-(\ref{7}), taking into account our modeling of the interaction term (\ref{int1})-(\ref{int2})-(\ref{int3}), we can check explicitly that this equilibrium point is a Minkowski universe. Interestingly, this would not be the case when dark energy is modeled according to the Modified Chaplygin Gas \cite{mcg2} with $p=A+B \rho^\alpha$, or adopting the Generalized Chaplygin Gas \cite{nonideal1} with $p=\beta \rho^\alpha$ (in this latter case with negative $\alpha$). In fact, a zero dark energy density delivers a non-zero dark energy pressure, which in turn provides a non-zero $\dot H$ through (\ref{5}), and a non-zero Einstein curvature tensor contrary to the case of a Minkowski universe. The dynamical variables ($H$, $\dot H$) seem more suited for investigating the stability of this latter equilibrium point rather than ($x$, $y$) \cite{mcg3}. 
\end{enumerate}

\begin{center}
	\begin{table}[H]
		\footnotesize
		\centering
		\begin{tabular}{|c|c|c|c|c|}
			\hline
			\backslashbox{Interaction type}{EoS} &  Redlich-Kwong & Modified Berthelot & Dieterici  \\
			\hline
			$3Hb \rho_{\rm de} + \gamma \dot \rho_{\rm de}$    & \makecell[l]{$ \frac{(b-\beta-1) (1+ \sqrt{2} )}{ \alpha (1-b -\beta )}, \, \frac{(b-\beta-1) (1+ \sqrt{2} )(b-1)}{ \alpha (b +\beta -1)}, $\\$\frac{b(b-\beta-1) (1+ \sqrt{2} )}{ \alpha (b+\beta-1 )}$}   & \makecell[l]{$ \frac{(b-\beta-1)}{\alpha(1-b)}, \, \frac{b-\beta -1}{ \alpha }, $\\$\frac{b(b-\beta-1)}{\alpha(b-1)} $ }  &  \makecell[l]{$\frac{\Big[W \left(\frac{2\beta}{e^2(1-b)} \right)+4 \Big]}{2 \alpha}, \, $\\$ \frac{\Big[W \left(\frac{2\beta}{e^2(1-b)} \right)+4 \Big](1-b)}{2 \alpha} ,$\\$\frac{-b\Big[W \left(\frac{2\beta}{e^2(1-b)} \right)+4 \Big]}{2 \alpha} $}    \\
			\hline
			$3Hb \rho_{\rm de} + \gamma \dot \rho_{\rm de}$    & \makecell[l]{$  \frac{(1+\beta) (1+\sqrt{2})}{ \alpha (\beta -1)}, \,  \frac{(1+\beta) (1+\sqrt{2})}{ \alpha (\beta -1)}, \, 0$ }   &\makecell[l]{$ -\frac{1+\beta}{ \alpha}, \, -\frac{1+\beta}{ \alpha},\, 0 $ }& \makecell[l]{$ \Big( \frac{3}{ 2\alpha }, \,  \frac{3 \beta }{ \alpha e}, \, 0$ \Big)\\$\Big(  \frac{W(2 e^{-2}\beta)+4}{ 2\alpha }, \,  \frac{W(2 e^{-2} \beta) +4}{2 \alpha }, \, 0$ \Big)\\$\Big(  \frac{3\beta(1+\gamma)}{ \alpha }, \,  \frac{3 \beta }{ \alpha e}, \, 3\frac{e(\gamma-b)+2\beta(1+\gamma)}{2\alpha(b-\gamma)e}$ \Big)}          \\
			\hline
			\makecell[l]{$3Hb (\rho_{\rm de}+\rho_{\rm dm}) + \gamma (\dot \rho_{\rm de} + \dot \rho_{\rm dm} )$} &  \makecell[l]{$\frac{[\beta(1+b) +1] (1+\sqrt{2})}{ \alpha  [\beta(1+b) -1]}, \,$\\$ \frac{[\beta(1+b) +1] (1+\sqrt{2})}{ \alpha (1+b) [\beta(1+b) -1]} $\\$- \frac{b[\beta(1+b) +1] (1+\sqrt{2})}{3 \alpha (1+b) [\beta(1+b) -1]} $}  &   \makecell[l]{$ \frac{[\beta(1+b) +1] }{ \alpha  [\beta(1+b) -1]}, \,$\\$ \frac{[\beta(1+b) +1] }{ \alpha (1+b) [\beta(1+b) -1]} $\\$ \frac{b[\beta(1+b) +1] }{3 \alpha (1+b) [1-\beta(1+b) ]} $}    &      \makecell[l]{$\frac{W(2(b+1)\beta e^{-2}) +4)}{2\alpha}, \, $\\$\frac{W(2(b+1)\beta e^{-2}) +4)}{2\alpha(b+1)}, $\\$ -\frac{b W(2(b+1)\beta e^{-2}) +4)}{2\alpha}  $}      \\
			\hline
		\end{tabular}
		\caption{ This table summarizes the values of the dark energy density $\rho_{ de}$, the absolute value of the dark energy pressure $|p_{de}|$, and of the dark matter density $\rho_{ dm}$ at equilibrium for all the interacting systems we have investigated. }	\label{table2}
	\end{table}
\end{center}

\section{Singularities} \label{secIV}
 
The type of singularities that can arise at time $t_s$ in a cosmological model according to \cite{Sergei,refstaro} can be classified into five cases:
\begin{enumerate}
\item Type I or {\it Big rip singularity}: $\lim_{t \to t_s} a(t) = \infty$, $\lim_{t \to t_s} \rho(t) = \infty$, $\lim_{t \to t_s} |p(t)| = \infty$. This is a true spacetime singularity because it corresponds to have incomplete null and timelike geodesics \cite{class1,class1a};
\item Type II or {\it Sudden singularity}: $\lim_{t \to t_s} a(t) = a_s$, $\lim_{t \to t_s} \rho(t) = \rho_s$, $\lim_{t \to t_s} |p(t)| = \infty$. This is a weak singularity and geodesics are complete \cite{sergei4, class2,class3,class3a};
\item Type III or {\it Big freeze singularity}: $\lim_{t \to t_s} a(t) = a_s$, $\lim_{t \to t_s} \rho(t) = \infty$, $\lim_{t \to t_s} |p(t)| = \infty$. This can be both a weak or strong singularity and geodesics are complete \cite{class4};
\item Type IV or {\it Generalized sudden singularity}: $\lim_{t \to t_s} a(t) = a_s$, $\lim_{t \to t_s} \rho(t) = \rho_s$, $\lim_{t \to t_s} |p(t)| = p_s$, $\lim_{t \to t_s}  H^{(i)}(t) = \infty$, $i=2,...$. This is a weak singularity and geodesics are complete \cite{sergei3, class3,class3a,class5};
\item Type V or {\it $w$ singularity}: $\lim_{t \to t_s} a(t) = a_s$, $\lim_{t \to t_s} \rho(t) = 0$, $\lim_{t \to t_s} |p(t)| = 0$, $\lim_{t \to t_s} w = \lim_{t \to t_s} \frac{p(t)}{\rho(t)}=\infty$. This is a weak singularity \cite{class6,class6a};
\end{enumerate}
	
where $a_s$, $\rho_s$ and $p_s$ are some finite constant values. First of all we will explain why certain singularities cannot arise simply looking at the equations of state for dark energy we have adopted. Then, we will classify the remaining types of singularity which can arise in our model looking at the leading terms in the Friedmann evolution equations, essentially following the same line of thinking of \cite{lead1,lead2}.

	\begin{table}[H]
		\centering
		\begin{tabular}{ |c|c|c|c|c| }
			\hline
		{\bf Type I singularity}    &&&   \\
		\hline
	\backslashbox{Interaction type}{EoS} &  Redlich-Kwong & Modified Berthelot & Dieterici  \\
	\hline
	$3Hb \rho_{\rm de} + \gamma \dot \rho_{\rm de}$    &$b-1-\beta>0$& \# & \#   \\
	\hline
	$3Hb \rho_{\rm dm} + \gamma \dot \rho_{\rm dm}$    & \#& \# &   \#          \\
	\hline
	\makecell[l]{$3Hb (\rho_{\rm de}+\rho_{\rm dm}) + \gamma (\dot \rho_{\rm de} + \dot \rho_{\rm dm} )$} & $\beta< -\frac{1}{1+b}$, $-\frac{1}{6}\leqslant b\leqslant1$ and $b<\gamma$ &\# &  \#  \\
		\hline
			{\bf Type II singularity}    &&&   \\
			\hline
			\backslashbox{Interaction type}{EoS} &  Redlich-Kwong & Modified Berthelot & Dieterici  \\
			\hline
			$3Hb \rho_{\rm de} + \gamma \dot \rho_{\rm de}$    &\#& \# &  $  \alpha\neq0  $  and  $\gamma\neq1$   \\
			\hline
			$3Hb \rho_{\rm dm} + \gamma \dot \rho_{\rm dm}$    &\#& \#&        $  \alpha\neq0  $  and $\gamma\neq-1$ \\
			\hline
			\makecell[l]{$3Hb (\rho_{\rm de}+\rho_{\rm dm}) + \gamma (\dot \rho_{\rm de} + \dot \rho_{\rm dm} )$} & \# &\#&   $  \alpha\neq0 $ and $\gamma\neq-1$  \\
			\hline
			{\bf Type III  singularity}    &&&   \\
			\hline
			\backslashbox{Interaction type}{EoS} &  Redlich-Kwong & Modified Berthelot & Dieterici  \\
			\hline
			$3Hb \rho_{\rm de} + \gamma \dot \rho_{\rm de}$    &$\gamma=1$ and $1+\beta-b\neq 0$ & \#&  \#   \\
			\hline
			$3Hb \rho_{\rm dm} + \gamma \dot \rho_{\rm dm}$    &\# & \#&    \#        \\
			\hline
			\makecell[l]{$3Hb (\rho_{\rm de}+\rho_{\rm dm}) + \gamma (\dot \rho_{\rm de} + \dot \rho_{\rm dm} )$} &  \# &\#&  \# \\
			\hline
			{\bf Type IV singularity}    &&&   \\
			\hline
			\backslashbox{Interaction type}{EoS} &  Redlich-Kwong & Modified Berthelot & Dieterici  \\
			\hline
			$3Hb \rho_{\rm de} + \gamma \dot \rho_{\rm de}$    &$\gamma=1$ & $\gamma=1$& $\gamma=1$  \\
			\hline
			$3Hb \rho_{\rm dm} + \gamma \dot \rho_{\rm dm}$    &$\gamma=-1$& $\gamma=-1$&    $\gamma=-1$      \\
			\hline
			\makecell[l]{$3Hb (\rho_{\rm de}+\rho_{\rm dm}) + \gamma (\dot \rho_{\rm de} + \dot \rho_{\rm dm} )$} & \# &\#&  \#  \\
			\hline
			{\bf Type V singularity}    &&&   \\
			\hline
			\backslashbox{Interaction type}{EoS} &  Redlich-Kwong & Modified Berthelot & Dieterici  \\
			\hline
			$3Hb \rho_{\rm de} + \gamma \dot \rho_{\rm de}$    & \hspace{2  pt} {  \#}    & \hspace{2  pt} {  \#} & \hspace{2  pt} {  \#}  \\
			\hline
			$3Hb \rho_{\rm dm} + \gamma \dot \rho_{\rm dm}$    &       \hspace{2  pt} {  \#} &    \hspace{2  pt} {  \#}&  \hspace{2  pt} {  \#}   \\
			\hline
			\makecell[l]{$3Hb (\rho_{\rm de}+\rho_{\rm dm}) + \gamma (\dot \rho_{\rm de} + \dot \rho_{\rm dm} )$} &   \hspace{2  pt} {  \#}  &  \hspace{2  pt} {  \#} & \hspace{2  pt} {  \#}  \\
			\hline
		\end{tabular}
		
		\caption{ This table summarizes the values of the parameters which allows for a type I, or II, or III, or IV, or V singularity, respectively. The simbol \# means that that type of singularity cannot occur within our range of interest for the parameters. }\label{table3}
	\end{table}

\subsection {Type V Singularity}
The equations of state for the dark energy modeling we considered, namely the Redlich-Kwong (\ref{Redlich}), the modified Berthelot (\ref{modified}), and the Dieterici (\ref{Dieterici}) exhibit the following behaviors:
\begin{eqnarray}
\label{qeos1}
&&\lim_{\rho \to 0}|p|= 0\,, \qquad \lim_{\rho \to +\infty}|p|= +\infty\,, \qquad \lim_{\rho \to 0}w=\lim_{\rho \to 0}\frac{p}{\rho}= \beta  \\
\label{qeos2}
&&\lim_{\rho \to 0}|p|= 0\,, \qquad \lim_{\rho \to +\infty}|p|= \frac{|\beta|}{\alpha}\,, \qquad \lim_{\rho \to 0}w=\lim_{\rho \to 0}\frac{p}{\rho}= \beta  \\
\label{qeos3}
&&\lim_{\rho \to 0}|p|= 0\,, \qquad \lim_{\rho \to +\infty}|p|= 0\,, \qquad \lim_{\rho \to 0}w=\lim_{\rho \to 0}\frac{p}{\rho}= \frac{\beta e^2}{2}   \,,
\end{eqnarray}
respectively. Therefore, all the models considered in this papar are type V-singularity-free because they cannot admit at the same time a zero energy density and a diverging effective equation of state parameter. We stress that this result relies only on the particular equations of state we have considered, and not on the interactions terms. Moreover, this is a qualitatively similar behavior than the one that can be obtained for a cosmological modeling which adopts the Chaplygin gas $p= -\frac{A}{\rho^n}$ as dark energy for which the pressure cannot vanish in the correspondence of a vanishing energy density.
\subsection{Type I and Type III Singularity}
From (\ref{qeos1})-(\ref{qeos2})-(\ref{qeos3}) we understand that a type I singularity cannot happen when we model dark energy in terms of the modified Berthelot and of the Dieterici equations of state because it is not possible to have a diverging pressure in correspondence of a diverging energy density, which is qualitatively the same behavior of the Chaplygin gas for positive $n$. Again, we stress that this observation fully relies only on the particular functional form of the equations of state considered and not on the choice of the interaction terms. The same conclusion holds for the type III singularity.

For the Redlich-Kwong scenario, in proximity of a possible type I or III singularity (if it arises) we can approximate the pressure as 
\beq
\label{rkh}
p\simeq \beta \rho
\eeq
which holds at high energy densities. 

\subsubsection{Redlich-Kwong - Interaction term $Q_1$}

Using (\ref{rkh}) we can approximate the evolution equations (\ref{hb1}), and (\ref{RDM})-(\ref{RDE}) nearby the singularity as

\begin{eqnarray}
\left( \frac{\dot a}{a} \right)^2&=& \frac{\rho_{\rm de}+\rho_{\rm dm}}{3} \\
\dot \rho_{\rm de} &\simeq& -\frac{3}{\gamma -1}(b-1-\beta) \rho_{\rm de} \frac{\dot a}{a} \\
\dot \rho_{\rm dm} &\simeq& -\frac{3}{\gamma -1}[(\gamma-b+\gamma\beta) \rho_{\rm de}+(\gamma-1)\rho_{\rm dm} ]\frac{\dot a}{a}\,.
\end{eqnarray}
The second equation can be integrated as
\beq
\rho_{\rm de}(t)\simeq {\mathcal C}_1 a(t)^{ \frac{3(1+\beta-b)}{\gamma-1}}  \,,
\eeq
where ${\mathcal C}_1$ is a constant of integration.
Therefore, a type I singularity may be possible if
\beq
\label{92}
\frac{3(1+\beta-b)}{\gamma-1}>0 \quad \Rightarrow \quad 1+\beta-b<0\,,
\eeq
where we used the physical restriction $-1 \leq \gamma \leq 1$, while a type III singularity may be possible if
\beq
\label{93}
\gamma=1 \quad {\rm and} \quad 1+\beta-b \neq 0\,.
\eeq
We note that in the absence of the interaction term, i.e. for $b=0=\gamma$, a type I singularity may still be possible for a phantom dark energy with $\beta<-1$, while a type III singularity cannot be realized. Then the evolution eq. for the dark matter density provides
\beq
\rho_{\rm dm}(t)\simeq a(t)^{-3}\left[{\mathcal C}_2 + \frac{{\mathcal C}_1 (\gamma+\beta\gamma-b) a(t)^{ \frac{3(\gamma+\beta-b)}{\gamma-1}}}{(b-\gamma-\beta)} \right] \,,
\eeq
where ${\mathcal C}_2$ is another constant of integration. ${\mathcal C}_1$ and ${\mathcal C}_2$ can be determined by imposing $\rho_{\rm de}(a_{0})=\rho_{\rm de_{0}}$ and  $\rho_{\rm dm}(a_{0})=\rho_{\rm dm_{0}}$ respectively. In the proximity of a type I singularity we can neglect the first term in the evolution of the energy density for the dark matter respect to the contribution of the dark energy, and write the evolution of the scale factor as
\beq
a(t)\simeq \left(\frac{3 {\mathcal C}_1 \beta}{4}  \right)^{\frac{\gamma-1}{3(b-1-\beta)}}\cdot\left[\frac{((t-{\mathcal C}_3 )(\beta+1-b))^2}{(\gamma-1)(b-\beta-\gamma)} \right]^{\frac{\gamma-1}{3(b-1-\beta)}}\,,
\eeq
where ${\mathcal C}_3$ is a constant of integration which can be determined requiring $a(t_0)=a_0$. Therefore, we can have a type I singularity, which requires a diverging scale factor at a finite time $t=t_s$ if the denominator of the fraction within the square bracket is vanishing and simultaneously the exponent is positive, or if the numerator of the fraction within the square bracket is vanishing and simultaneously the exponent is negative. In the former case we get the following requirement
\beq
b=\beta+\gamma \quad {\rm and } \quad b<1+\beta\,,
\eeq
which must be exluded in the light of (\ref{92}), while in the latter case
\beq
t_s={\mathcal C}_3 \quad {\rm and } \quad b>1+\beta\,.
\eeq
Therefore, a type I singularity can still arise in the latter case for a phantom fluid with $\beta<-1$ even without interactions. We note that the strength of interactions within the dark energy fluid quantified by the parameter $\alpha$ has not played any role in this analysis. If a type III singularity arises, we can write the evolution for the scale factor as
\beq
3 (\dot a(t))^2=\frac{{\mathcal C}_1 \beta (\gamma-1) a(t)^{\frac{1+3(\beta-b)+2\gamma}{\gamma-1}}}{b-\gamma-\beta}+\frac{{\mathcal C}_2}{a(t)}\,,
\eeq
and we can neglect the second term in its rhs in light of (\ref{93}) and using that the exponential is dominating over the polynomial. Therefore
\beq
a(t)\simeq \left( \frac{3{\mathcal C}_1 \beta(b-1-\beta)^2(t-{\mathcal C}_3)^2}{4 (\gamma-1) (b-\gamma-\beta)}\right)^{\frac{\gamma-1}{3(b-1-\beta)}} \,.
\eeq
Observing that 
\beq
\lim_{\gamma \to 1^{-}} a(t)=1\,,
\eeq
we understand that we can get a type III singularity when (\ref{93}) applies. Again the parameter $\alpha$ has not played any role because in the high-energy regime the Redlich-Kwong fluid behaves effectively like an ideal fluid.
\subsubsection{Redlich-Kwong - Interaction term $Q_2$}
The evolution  for the energy density of dark matter in this case reads as 
\begin{equation}
\rho_{ dm}(t)\simeq a(t)^{-\frac{3(b+1)}{\gamma+1}} \mathcal C_{1},
\end{equation}
where $\mathcal  C_{1} $ is  some constant of the integration. The evolution for dark energy is  
\begin{equation}
\rho _{d e}(t)\simeq a(t)^{-3 \beta-3} \mathcal C_{2}-\frac{a(t)^{-\frac{3(b+1)}{\gamma +1}}(\gamma-b )\mathcal C_{1}}{\gamma (\beta+1)+\beta-b},
\end{equation}
where $\mathcal  C_{2} $ is  another constant of integration. A type I singularity  may happen if
\begin{equation}
 \frac{3(b+1)}{\gamma +1}<0 \quad \rm{or }\quad -3(\beta+1)>0.
\end{equation}
The former condition cannot happen for $-1\leqslant\gamma\leqslant 1$ and $-1\leqslant b\leqslant1$. The latter implies $-\frac{6}{5}\leqslant\beta<-1$, that is that  dark energy must be a quintessence fluid. Moreover, we ignore the case in which the denominator is zero because the exponential term dominates over the polynomial term. Nearby the singularity we can approximate
\begin{equation}
\frac{ \dot a(t)^2}{a(t)^2} \simeq\frac{\mathcal C_{2}}{3}a(t)^{-3(1+\beta)},
\end{equation}
which can be integrated for the scale factor as
\begin{equation}
a(t)\simeq  \left(\frac{3 \mathcal C_{2} (1+\beta)^2(t-\mathcal C_{3})^2}{4}\right)^{\frac{1}{3(1+\beta)}},
\end{equation}
where $ \mathcal C_{3} $ is another constant of  integration. The scale factor in this case  is always regular at a finite time for $-\frac{6}{5}\leqslant\beta<-1$. Therefore a type  I singularity cannot occur.

A type III singularity can happen when

\beq
\label{106}
\gamma(\beta+1)+\beta-b=0 \quad {\rm or } \quad - \frac{3(b+1)}{\gamma +1}\to +\infty \,.
\eeq
The first condition implies that $\beta$=$\frac{b-\gamma}{1+\gamma}$, and therefore requires b$ < $$\gamma$  for being realized. The latter condition cannot be realized since $\frac{-3(b+1)}{\gamma+1}<0$. Thus, nearby the singularity  we can approximate 
\begin{equation}
\frac{ \dot a(t)^2}{a(t)^2}= \frac{ {\mathcal C}_1(b-\gamma) a(t)^{3\frac{(1+b)}{\gamma+1}}}{3\gamma(\beta+1)+\beta-b},\
\end{equation}
which can be integrated for the scale factor as 
\begin{equation}
a(t)\simeq \left[\frac{3(t-\mathcal C_{3})^2(1+b)^2 \mathcal C_{1}(b-\gamma)}{4(1+\gamma)^2(\gamma(1+\beta)+\beta-b)}\right]^{\frac{1}{3}\frac{1+\gamma}{(1+b)}},
\end{equation}
where $\mathcal C_{3}$ is a constant of integration. Therefore a type III singularity, characterized by $ a(t)\rightarrow a_{s} $, cannot be realized because $ \frac{1+\gamma}{3(1+b)} >0$, as already discussed, and taking into the account  vanishing  factor in the denominator which follows from (\ref{106}).
\subsubsection{Redlich-Kwong - Interaction term $Q_3$}
In this limit, we can approximate the energy conservation equations (\ref{A20}) and (\ref{A21}) with
\begin{equation}
\dot{\rho_{de}}\simeq 3H\left[(b-\gamma)\rho_{ dm}+\left(b-(\gamma+1)(1+\beta)\right)\rho_{ de}\right]
\end{equation}
\begin{equation}
\dot{\rho_{dm}}\simeq 3H\left[(\gamma-1-b)\rho_{ dm}+(\gamma\beta+\gamma-b)\rho_{ de}\right],
\end{equation}
which deliver the following time evolution for the dark energy density:
\begin{equation}
\rho_{ de}\simeq a(t)^{u} \left[\mathcal C_{1} a(t)^{v}+\mathcal C_{2} a(t)^{-v} \right]
\end{equation}
\begin{equation}
u=-3\left[1+\frac{\beta(1+\gamma)}{2}\right]
\end{equation}
\begin{equation}
v=-3\sqrt{\beta\left(\frac{\beta(1+\gamma)}{4}+\gamma-b\right)},
\end{equation}
where $ \mathcal C_{1} $ and $  \mathcal C_{2}  $ are two arbitrary constants of integration. Since neither $ u $ nor $ v $ can diverge, a type III singularity, characterized by $\lim_{a \to a_{s}} \rho_{ de}=\infty$, cannot occur in this framework.
For investigating the possible occurrence of a type I singularity, characterized by $ \lim_{a(t)\to \infty} \rho(t)=\infty $, we observe that we must require $ u+v>0 $. Since $ u+v $ will dominate over $ u-v $, it is enough to study the sign of the former. Explicitly we must require
\begin{equation}
	-3\left[1+\frac{\beta(1+\gamma)}{2}\right]+3\sqrt{\beta\left(\frac{\beta(1+\gamma)}{4}+\gamma-b\right)}>0,
\end{equation}
which after some algebraic manipulations can be recast as
\begin{equation}
	\beta<-\frac{1}{1+b}.
\end{equation}
In particular, this requires b$\neq$-1, and more im general the condition $-\frac{6}{5}\leqslant\beta<1$ will impose $-\frac{1}{6}\leqslant b\leqslant1$. Moreover, a well defined $ v $ requires 
\begin{equation}
	\beta\frac{(1+\gamma)}{4}+\gamma-b<0 \qquad \Rightarrow \qquad \beta<\frac{4(b-\gamma)}{1+\gamma} \qquad \Rightarrow \qquad b<\gamma,
\end{equation} 
where in the last step we imposed $\beta<0$. Applying the parameterization of \cite{lead1} to the energy density rather than to the Hubble function we write
\begin{equation}
	\rho_{ de} \simeq (t-t_{s})^{n}+\rho_{s}\simeq \mathcal C_{1}a(t)^{u+v},
\end{equation}
where in the last step we kept only the leading term.
A diverging energy density in the neighborhood of $ t=t_{s} $ requires $n<0$. By inverting for the scale factor we get
\begin{equation}
	a(t)\simeq \left[\left(\mathcal C_{1}(t-t_{s})^{n}+\rho_{s}\right)\right]^{\frac{1}{u+v}},
\end{equation}
and since $  u+v>0$ (under the conditions we have derived previously), we can get $a(t)\to \infty$ for $t\to t_{s}$. Therefore, a type I singularity can occur when the following conditions are met simultaneously: $\beta< -\frac{1}{1+b}$, $-\frac{1}{6}\leqslant\beta\leqslant1$ and $b<\gamma$.
\subsection{Type II Singularity}
We  may have a type II singularity in correspondence of the values of the energy densities which make the denominator of the equations of state vanish. We observe that:
\begin{eqnarray}
1- (1-\sqrt{2}) \alpha \rho &\neq& 0 \quad \forall \rho  \in (0,\infty) \\
1+ \alpha \rho &\neq& 0 \quad \forall \rho  \in (0,\infty) \\
2- \alpha \rho &=& 0 \quad {\rm for } \quad \rho=\rho_s=\frac{2}{\alpha}\,,
\end{eqnarray}
in which we used $\alpha>0$ as from \cite{capo}. Therefore, a type II singularity cannot arise in the models involving the Redlich-Kwong and the modified Berthelot equations of state, but it may arise in a framework supported by the Dieterici equation of state. Following \cite{lead1}, nearby a singularity, we can parameterize the Hubble function as 
\begin{equation}
H(t)=\left(\frac{1}{t-t_{s}}\right)^{n}+H_{s}.
\end{equation}
In a type II singularity $ H(t)$ is finite because $ \rho(t)  $ is finite and we use (\ref{4}). Therefore, n $ \leqslant $0. 
Integrating 
\begin{equation}
\frac{\dot a}{a}=\left(\frac{1}{t-t_{s}}\right)^{n}+H_{s}
\end{equation}
we obtain the evolution of the scale factor as
\begin{equation}
a(t)=\mathcal C_{1} \exp \left[H_{s} t+\frac{1}{1-n}\left(\frac{1}{t-t_{s}}\right)^{n-1}\right]
\end{equation}
where ${\mathcal C}_1$ is some constant of the integration and $ t_{s} $ is the time at the singularity.
For $ n\leqslant0 $,  $ a(t) $ is finite about $ t=t_{s} $, as we need  in a type II singularity.
Therefore, we just need to check explicitly whether we  have a finite energy density nearby the singularity for each type of interaction.
First of all, by expanding the Dieterici EOS about $ \rho=\frac{2}{\alpha} $ we get
\begin{equation}
p\simeq\frac{2 \beta}{e^{2} \alpha(2-\alpha \rho_{de})}+\frac{3 \beta}{e^{2} \alpha}+O\left(\rho_{de}-\frac{2}{\alpha}\right),
\end{equation}
in which the leading term is the first one.  

\subsubsection{Dieterici - Interaction term $Q_1$}

Keeping only the leading term in the pressure and substituting it into the energy conservation equations we get
\begin{equation}\label{118}
\dot \rho_{de}\simeq \frac{3 H}{\gamma-1} p_{d e}=\frac{6 H \beta}{(\gamma-1)  e^{2} \alpha(2-\alpha \rho_{de})},
\end{equation}
\begin{equation}
\dot \rho_{ dm} \simeq \frac{3 \gamma H}{1-\gamma} p_{d e}=\frac{6  \gamma H \beta}{(1-\gamma)  e^{2} \alpha(2-\alpha \rho_{de})}.
\end{equation}
From (\ref{118}) we get
\begin{equation}\label{119}
\rho_{d e}\simeq \frac{2}{\alpha}\left[1 \pm \sqrt{1-\frac{3 \beta}{e^{2}(\gamma-1)^{2}}\left(\mathcal C_{1}+\ln a\right)}\right],
\end{equation}
where the double sign corresponds to approaching $\rho_{s}$ from the left and from the right respectively. Therefore, for avoiding a divergence in the  energy density we must require that $\alpha\neq0$ and  $\gamma\neq 1$. In this type of singularity $\alpha$ is playing an important role (contrary to other types of singularity). In fact for an ideal fluid with $ p\simeq\beta \rho $, we would get $ p\rightarrow\infty,$ and a type II singularity would not be realized. This result is compatible with $ \rho_{s}=\frac{2}{\alpha} $ $\Rightarrow \alpha \neq0 $ (for having a finite energy density), but we get one more piece of information such that $ \gamma\neq1. $ 
By plugging this $ \rho_{d e} $ into $ \dot\rho_{dm}$, and integrating over the time, we understand that these two  conditions $  \alpha\neq0 $ and $ \gamma\neq 1 $ are enough for guaranteeing a finite $ \rho_{dm} $ as well.

\subsubsection{ Dieterici - Interaction term $Q_2$}

By repeating the same steps, the dark energy density, when the second interaction is considered, can be obtained as 
\begin{equation}
\rho_{ de}\simeq \frac{2}{\alpha}\left[1\pm \sqrt{1+\frac{3\beta}{e^2}(\ln a+\mathcal C_{1})}\right],
\end{equation}
where the double sign must be interpreted in the same way in the previous paragraph.
Similarly, as in  previous case we must require $\alpha\neq0$. The  energy density for dark matter is 
\begin{equation}
\rho_{dm}\simeq \mathcal C_{1} a(t)^ {-3\frac{b+1}{\gamma+1}}.
\end{equation}
To have a finite energy density we must required  that $\gamma \neq-1$.

\subsubsection{Dieterici - Interaction term $Q_3$}  

For the case of the  third interaction term we obtain
\begin{equation}
\rho_{de}\simeq \frac{2}{\alpha}\left[1\pm \sqrt{1+\frac{3 \beta(1+\gamma)(\mathcal C_{1}+\ln a)}{e^2}}\right],
\end{equation}
where the double sign must be interpreted in the same way in the previous paragraph.
Therefore, we must require $\alpha\neq0$, that is that the dark energy must be pictured as a nonideal fluid. 
The dark matter energy density for this case reads as
\begin{equation}
\rho_{ dm}\simeq-\frac{2\gamma}{e\alpha\beta(1+\gamma)}\sqrt{3\beta(1+\gamma
	)(\ln a+\mathcal C_{1})+e^2}+\mathcal C_{2},
\end{equation}
which can be finite when $\alpha\neq$0 and $\gamma\neq-1$, (taking into account that $\beta\neq0$ for dark energy fluid).

\subsection{Type IV Singularity}
To find out if  there is such a singularity or not, it is only necessary to check whether the second derivative of Hubble function is ill-defined, since, in this case, all the higher order derivatives will be ill-defined automatically. From the Friedman equations we can  get the first derivative for Hubble function as
\begin{equation}
\dot{H}=-\frac{1}{2}(\rho_{d e}+\rho_{d m}+p_{de}).
\end{equation}
Taking the second derivative of Hubble function and using energy conservation equations get
\begin{equation}
\ddot{H}=-\frac{3 H}{2} \underbrace{\left(\rho_{d e}+\rho_{d m}+\rho_{d m}\right)}_{In Type IV singularity this term is regular by definition.}-\frac{1}{2} \dot{p}_{d e}
\end{equation}
If a type IV singularity occurs, we must have a divergent pressure for dark energy  $\left|p_{d e}\right| \rightarrow\infty$. By looking carefully at the second term $\dot{p}_{d e}$ which reads as 
\begin{equation}
\dot{p}_{d e}=\frac{\partial p}{\partial \rho} \cdot \dot{\rho},
\end{equation}
We can see that it can be divergent when $\left|\frac{\partial p}{\partial \rho }\right| \rightarrow \infty $ or $\dot{\rho}\rightarrow\infty$.
Note that the $\dot{\rho}$ depends on the type of the interaction. For each equation of state we  calculate
\begin{itemize}
	\item For Redlich-Kowng equation of state: 
	\begin{equation}
	\frac{\partial p}{\partial \rho}=\frac{\beta(\alpha \rho-1)[(2 \sqrt{2}-3) \alpha \rho-1]}{[1-(1-\sqrt{2}) \alpha \rho]^{2}}.
	\end{equation}
	Therefore, a type IV singularity may happen if $\rho=\frac{1}{(1-\sqrt{2}) \alpha}$. However in this case we get $\rho<0 $. So there is no type IV singularity.
	\item For Modified Betherlot equation of state:
\begin{equation}
	\frac{\partial p}{\partial \rho}=\frac{\beta}{(1+\alpha \rho)^{2}}.
\end{equation}
	This quantity is always regular for $\alpha>0$. Thus a type IV singularity does not happen in this case.
	\item  For Dieterici equation of state:
	\begin{equation}
	\frac{\partial p}{\partial \rho}=\frac{\beta e^{(1-\alpha \rho)}\left(\alpha^{2} \rho^{2}-2 \alpha \rho+2\right)}{(2-\alpha \rho)^{2}}.
	\end{equation}
	In this case we may have a type IV singularity if $\rho$=2/$\alpha$. However in this case we get  $|p_{de}| \to \infty $. Thus there is no type IV singularity.

	Therefore, we can have type IV singularity when
	$\to\left\{\begin{array}{l}{\gamma=1 \text {  }  \text { for 1st interaction }} \\ {\gamma=-1 \text { for 2nd interaction } \text {  }} \\ {\text {  }}  { \text {no possibilities for 3rd interaction. }}\end{array}\right.$
\end{itemize}

\section{Conclusions} \label{secV}
The theoretical construction of a cosmological model is an interplay between the following three independent aspects: a gravitational theory, the symmetry group of the manifold, a modeling for the matter content. It is important to clarify the specific role of each three of them on driving the evolution of the Universe. For example, the effects of a Chaplygin gas can be accounted for also by an appropriate $f(R)$ theory \cite{chmod}, or by an effective inhomogeneous cosmological model \cite{buch2}, and similarly the effects of dark matter can be mimicked by a non-local gravitational theory \cite{nloc}. These claims triggered a line of research trying to break the possible degeneracy between physically and conceptually different frameworks \cite{dege1,dege2,dege3}.  In particular, we are interested in understanding how this interplay affects two basic aspects of the cosmological model under investigation: the existence of late-time equilibrium solutions characterized by a negative deceleration parameter, and the types of singularities which can arise. As the starting point, in this manuscript, we explicitly checked the existence of accelerating attractors in some models based on general relativity, the flat Friedmann metric, and some non-ideal dark energy fluids (already proposed in the literature) linearly interacting with dark matter. In our models a phase transition between decelerating and accelerating universes is possible (see appendix \ref{appendix II}). We showed that the uniqueness of the accelerating late-time state depends on the modeling of the dark energy because in one case bifurcations are allowed. In our class of models, all the late-time attractors which support a negative deceleration parameter correspond to a de Sitter universe. Then, we explained that certain types of singularity cannot arise just looking at the specific equations of state we considered, without the need of any information about the theory of gravity and the metric element we adopted. Moreover, we showed that the deviations from an ideal fluid for the dark energy do not affect the existence of other types of singularity. Similarly, we showed that this quantity affects the value of the Hubble function at equilibrium (which is the age of the universe), but not of the other cosmological parameters. Furthermore the request of a negative deceleration parameter has allowed us to establish a set of constraints among the free parameters of the models under investigation. In the forthcoming works we consider appropriate to investigate whether the classification of the other singularities we exhibited remains true when we consider dark energy interacting with scalar fields carrying a non-zero pressure, or when we introduce high-energy (branes) or long-distance corrections to general relativity for extending \cite{sergei1}, possibly accounting for the role of a massive graviton \cite{cai5,cai6}, bouncing cosmology \cite{cai2,cai4},  in quintom cosmology \cite{cai7}, in mimetic gravity \cite{sunny1}, or when we abandon the Copernican principle (e.g. by adopting a Bianchi or an inhomogeneous model). Also,  we will check whether our claim that the non-ideality of the fluid does not affect the values of the cosmological parameters at equilibrium, but for the age of the Universe, should be re-examined. Our results about the classification of the singularities depending on certain relationships between the parameters of the model are also important in  light of the studies of their quantum corrections (accounted for by the Wheeler-de Wit equation), as done for example in the case of dark energy models based on the non-ideal Shan-Chen fluid in which the cases of singularities characterized by a finite or diverging energy density where compared and contrasted \cite{dew}.
\section*{Acknowledgments}
MK is supported in part by a CAS President’s International Fellowship Initiative Grant (No. 2018PM0054) and by the NSFC (No. 11847226). DG acknowledges support from China Postdoctoral Science Foundation (grant
No.2019M661944).

\appendix
\section{Dynamical system theory: a basic review} \label{appendix I}
Dynamical system  techniques have been widely adopted in the literature for studying the asymptotic behavior of a system of differential equations, in particular in the context of classical Hamiltonian dynamics \cite{ham1,ham2,ham3}. In this section we shortly review their properties that we apply in this paper.
Let $x=x_{\alpha_1,...,\alpha_n}(t)$ and $y=y_{\alpha_1,...,\alpha_n}(t)$ be two functions of a {\it time} variable $t$ and of $n$ (real) $\alpha_1$,...,$\alpha_n$ constant parameters. Assume that the evolution of the system is governed by the two differential equations:
\beq
x':=\frac{dx}{dt}=f_{\alpha_1,...,\alpha_n}(x,y), \qquad y':=\frac{dy}{dt}=g_{\alpha_1,...,\alpha_n}(x,y),
\eeq
with $f$ and $g$ some arbitrary $C^1$ functions. This type of system is called {\it autonomous} because the right hand side of the evolution equations do not depend explicitly on the time variable $t$. Assume also that the dynamical variables are bounded, i.e. that they satisfy $a\leq x(t)\leq b$ and  $c\leq y(t)\leq d$ $\forall t$ with $a$, $b$, $c$, $d$ some appropriate (finite) constants. Then, dynamical system techniques can provide qualitative information about the evolution of the system after an initial time $t_0$. In particular, an equilibrium point of the system ($x_{\rm eq}$, $y_{\rm eq}$) is such that 
\beq
x_{\alpha_1,...,\alpha_n}(t)=x_{\rm eq}, \qquad y_{\alpha_1,...,\alpha_n}(t)=y_{\rm eq} \qquad \forall t
\eeq
if $\exists \bar t$ such that $x_{\alpha_1,...,\alpha_n}(\bar t)=x_{\rm eq}, \, y_{\alpha_1,...,\alpha_n}(\bar t)=y_{\rm eq}$. From a computational point of view, equilibrium points can be found as the roots of the following algebraic system
\beq
f_{\alpha_1,...,\alpha_n}(x_{\rm eq},y_{\rm eq})=0, \qquad g_{\alpha_1,...,\alpha_n}(x_{\rm eq},y_{\rm eq})=0\,,
\eeq
and in general they depend on the values of the free parameters $\alpha_1$,...,$\alpha_n$. It is possible that, when more than one equilibrium is found, a particular choice of the values of the free parameters can bring to the identification of two or more equilibria into a single point: in this case we say that we have a {\it bifurcation}. In general, the system is expected to tend to a certain equilibrium point at late times (thus we can speak also of attractors), whose value depend on the initial conditions chosen (if more than one equilibrium is allowed). Hartman-Grobman theorem provides the classification of the stability of equilibrium points in the linearized regime \cite{hart}. Let
\beq
J:=\begin{pmatrix} \frac{\partial f_{\alpha_1,...,\alpha_n}(x,y)}{\partial x } & \frac{\partial f_{\alpha_1,...,\alpha_n}(x,y)}{\partial y }   \\
	\frac{\partial g_{\alpha_1,...,\alpha_n}(x,y)}{\partial x } & \frac{\partial g_{\alpha_1,...,\alpha_n}(x,y)}{\partial y }
\end{pmatrix} _{x=x_{\rm eq},\, y=y_{\rm eq}}  \,,
\eeq
be the Jacobian matrix of the system evaluated at the equilibrium point of interest. Then, according to the nature of its eigenvalues, or equivalently of its determinant  $\rm DetJ$ and of its trace $\rm TrJ$, the stability of the equilibrium point can be classified as follows:
\begin{itemize}
	\item ${\rm Det J} <0$ $\Leftrightarrow$  the  eigenvalues  are  real  and  of  opposite  sign $\Leftrightarrow$ the
	phase portrait is a saddle (which is always unstable);
	\item   If $0<{\rm Det J}<\frac{({\rm Tr J})^2}{4}$ $\Leftrightarrow$ the eigenvalues are real, distinct, and of the same
	sign $\Leftrightarrow$  the phase portrait is a node, stable if
	${\rm Tr J}<0$, unstable if ${\rm Tr J}>0$;
	\item   If $0<\frac{({\rm Tr J})^2}{4}<{\rm Det J}$ $\Leftrightarrow$ the eigenvalues are neither real nor purely imaginary $\Leftrightarrow$ and the phase portrait is a spiral, stable if
	${\rm Tr J}<0$, unstable if ${\rm Tr J}>0$.
\end{itemize}
Essentially this classification describes whether the system returns back to the equilibrium point or depart from it after a small perturbation.

As already pointed out, the only equilibrium points relevant for a cosmological analysis must fulfill the following requirements, which will impose strict restrictions on the free parameters of the models under investigation: $0 \leq x_{\rm eq} \leq 1$, $0 \leq z_{\rm eq} \leq 1$, $-\frac{6}{5}=-1.2 \leq w_{\rm eq} =\frac{y_{\rm eq}}{x_{\rm eq}} < 0$, $q_{\rm eq}=\frac{1}{2}(1+ 3y_{\rm eq}) <0$, well defined real and non-zero $H_{\rm eq}$ (for satisfying the Friedmann equation), and the general requirements $-1 \leq b \leq 1$, $-1 \leq \gamma \leq 1$, $\alpha >0$, $\beta<0$.

In the following we derive the dynamical autonomous system for each interaction model that we consider:

\subsection{\texorpdfstring{$Q_1 = 3Hb\rho_{\rm de} + \gamma \dot \rho_{\rm de}$}{Lg}} 

The choice of the interaction term (\ref{int1}) implies the following laws for the conservation of the dark energy and dark matter:
\begin{eqnarray}
\dot \rho_{\rm de} &=& -\frac{3H}{\gamma-1} [(b-1)\rho_{\rm de} - p_{\rm de}]\label{RDM}\\
\dot \rho_{\rm dm} &=& -\frac{3H}{\gamma-1} [(\gamma-b)\rho_{\rm de}+(\gamma -1)\rho_{\rm dm} +\gamma p_{\rm de}].\label{RDE}
\end{eqnarray}
Consequently the dynamical system to consider becomes:
\begin{eqnarray} 
\label{eqx1}
x' &=& 3  \left[  \left(1+y -\frac{ b-1 }{ \gamma-1}  \right) x                 +\frac{y}{ \gamma-1}   \right] \\
y' &=& -\frac{3H}{2} (1+y)  \frac{\partial y}{\partial H}   + 3 \left[   \left(1+y -\frac{ b-1}{ \gamma-1}  \right) x                 +\frac{y}{ \gamma-1}              \right]   \frac{\partial y}{\partial x}.
\end{eqnarray}
The evolution equation for $y$ in our explicit cases (\ref{eos1}), (\ref{eos2}),  and (\ref{eos3})   reads as
\begin{eqnarray}
\label{eqy1a}
y' &=& 3 \frac{   \beta^2 (b-1)  x^3-(\beta+2 (1-\gamma)  y+b-\gamma)\beta  y  x^2-  y^2  (b-2\beta-1)  x+  y^3}{2 \beta x^2 (\gamma -1)}  \\
\label{eqy1b}
y' &= & 3y   \frac{\beta  (\gamma-1) (y+1) x^2+y (1-b) x + y^2}  {\beta x^2 (\gamma -1)}\\
\label{eqy1c}
y' &=&  \frac{ 3y \{  [x(b-1) -y] W^2 (\chi) +[((\gamma -1) y +4b +\gamma -5) x - 4y] W(\chi) +4 [(b-1)x -y]\} } {x (\gamma -1) W(\chi)}, \quad \chi=-\frac{2x \beta}{e^2 y};  \nonumber\\ 
\end{eqnarray}
respectively.

\subsection{\texorpdfstring{$Q_2 = 3Hb\rho_{\rm dm} + \gamma \dot \rho_{\rm dm}$}{Lg}}

The choice of the interaction term (\ref{int2})  implies the following laws for the conservation of the dark energy and dark matter:
\begin{eqnarray}
\dot \rho_{\rm de} &=& -\frac{3H}{\gamma+1} [(1+\gamma)\rho_{\rm de}+(\gamma -b)\rho_{\rm dm} +(1+\gamma) p_{\rm de}]\\
\dot \rho_{\rm dm} &=& -\frac{3H (1+b)}{\gamma+1} \rho_{\rm dm} .
\end{eqnarray}
Consequently the dynamical system to consider becomes:
\begin{eqnarray} 
\label{eqx2}
x' &=&  3 \left( \frac{ \gamma -b }{ \gamma+1} +y \right)  (x-1) \\
y' &=&  -\frac{3H}{2} (1+y)  \frac{\partial y}{\partial H}   +   3 \left( \frac{ \gamma -b }{ \gamma+1}+y   \right)  (x-1) \frac{\partial y}{\partial x}.\,,
\end{eqnarray}
where the latter in our explicit cases (\ref{eos1}), (\ref{eos2}), and (\ref{eos3}),  is given by:
\begin{eqnarray}
\label{eqy2a}
y' &=& 3  \frac{   [(1+b) x+(1+\gamma) y-b+\gamma]  x^2 \beta^2-2 (x-1) [b-\gamma - (1+\gamma) y ]  x y \beta-[(1+b) x+(1+\gamma) y-b+\gamma]y^2  }   { 2\beta x^2 (1+\gamma)}   \nonumber \\
&& \\
\label{eqy2b}
y' &= &  -3 y  \frac{  (1+\gamma) y^2+[-x^2 \beta  (1+\gamma)+(b+1) x-b+ \gamma] y-x^2\beta (1+\gamma) } { \beta x^2 (1+\gamma)}    \\
\label{eqy2c}
y' &=&  \frac{3y [ A W^2(\chi) +B W(\chi) +C] }{ x (\gamma +1) W(\chi)}  \,, \qquad \chi= -\frac{2 x\beta}{e^2 y}\,, \qquad A= (b+1) x +(y+1) \gamma -b +y= \frac{C}{4} \\
B &=& [(y+1)\gamma +4b +y +5] x +4[(y+1)\gamma -b +y] \nonumber\\ 
\end{eqnarray}
respectively.
\subsection{\texorpdfstring{$Q_3 = 3Hb (\rho_{\rm de}+\rho_{\rm dm}) + \gamma (\dot \rho_{\rm de} + \dot \rho_{\rm dm} )$}{Lg}}
The choice of the interaction term (\ref{int3})  implies the following laws for the conservation of the dark energy and dark matter:
\begin{eqnarray} 
\dot \rho_{\rm de} &=& -3H[  (\gamma+1-b)\rho_{\rm de}+(\gamma -b)\rho_{\rm dm} + (\gamma+1) p_{\rm de}   ]\label{A20} \\
\dot \rho_{\rm dm} &=& -3H  [ (b-\gamma+1)\rho_{\rm dm}+(b-\gamma)\rho_{\rm de}-\gamma p_{\rm de}   ].\label{A21}
\end{eqnarray}
Consequently the dynamical system to consider becomes:
\begin{eqnarray} 
\label{eqx3}
x' &=&  3 \Big[ (b-\gamma )(1-x) - (\gamma+1) y+x (y- \gamma + b  )  \Big]    \\
y'  &=&   -\frac{3H}{2} (1+y)  \frac{\partial y}{\partial H}   +   3 \Big[ (b-\gamma )(1-x) - (\gamma+1) y+x (y- \gamma + b  )  \Big]  \frac{\partial y}{\partial x}.
\end{eqnarray}
The evolution equation for $y$ in our explicit cases (\ref{eos1}), (\ref{eos2}), and (\ref{eos3}) reads as
\begin{eqnarray}
\label{eqy3a}
y' &=&  3 \frac{  x^2  [ x+(1+\gamma) y - b +\gamma] \beta^2 + 2 [x y -(1+\gamma)y+b-\gamma] x y\beta - y^2 [ x+(1+\gamma) y- b + \gamma]  }  {2\beta x^2}     \\
\label{eqy3b}
y' &= &  3y \frac{(\beta  x^2+b - \gamma  -x)  y+ \beta x^2 - (1+\gamma)  y^2  } { \beta x^2}     \\
\label{eqy3c}
y' &=&  \frac{3y [ A W^2(\chi) +B W(\chi) +C] }{ x W(\chi)}  \,, \qquad \chi= -\frac{2 x\beta}{e^2 y}\,, \qquad A= (1+\gamma ) y -b +x +\gamma=\frac{C}{4}  \\
B &=& (x+4\gamma +4) y -4b+5x +4 \gamma \nonumber\\ 
\end{eqnarray}
respectively.
\section{Phase Transitions} \label{appendix II}
\begin{figure}[H]	
	\centering
	\begin{subfigure}[b]{.24\linewidth}
		\centering
		\includegraphics[scale=0.3, angle=0]{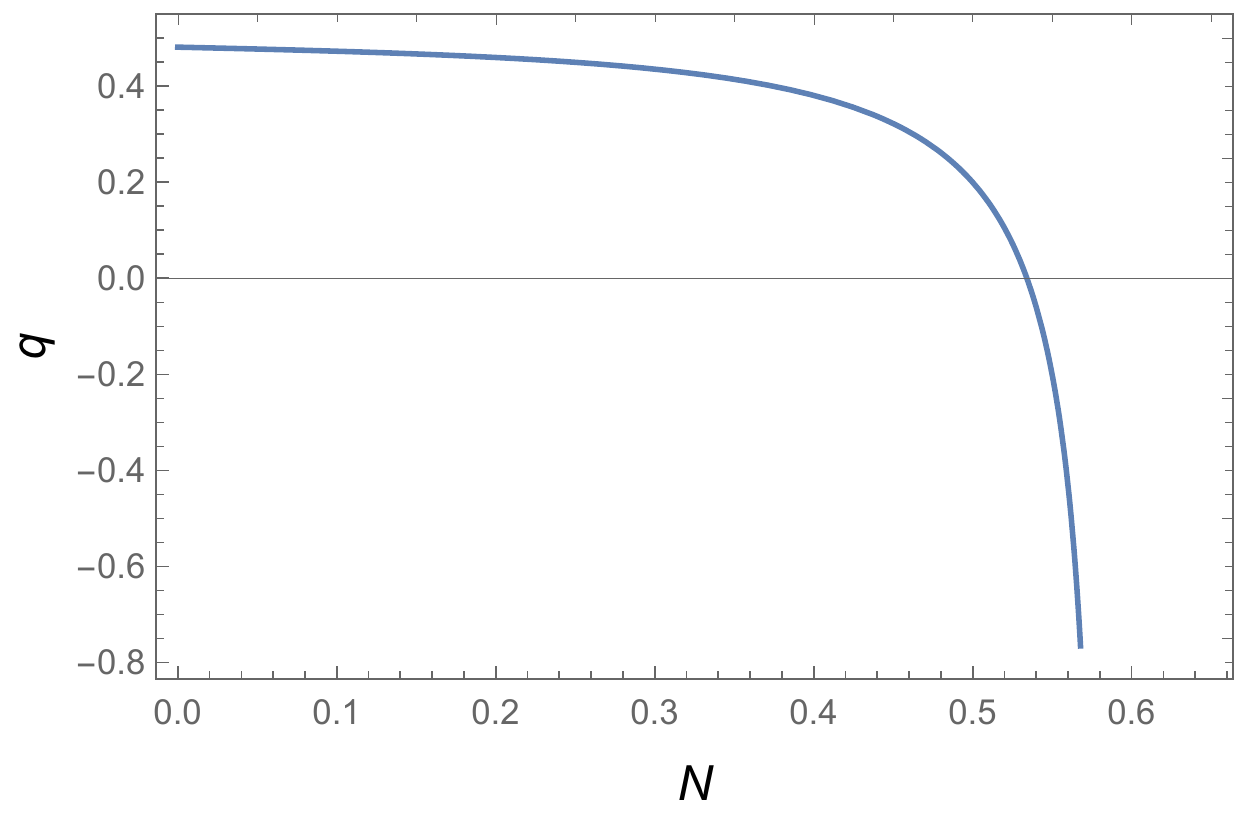}
		\caption{}
	\end{subfigure}
	\centering
	\begin{subfigure}[b]{.24\linewidth}
		\centering
		\includegraphics[scale=0.3, angle=0]{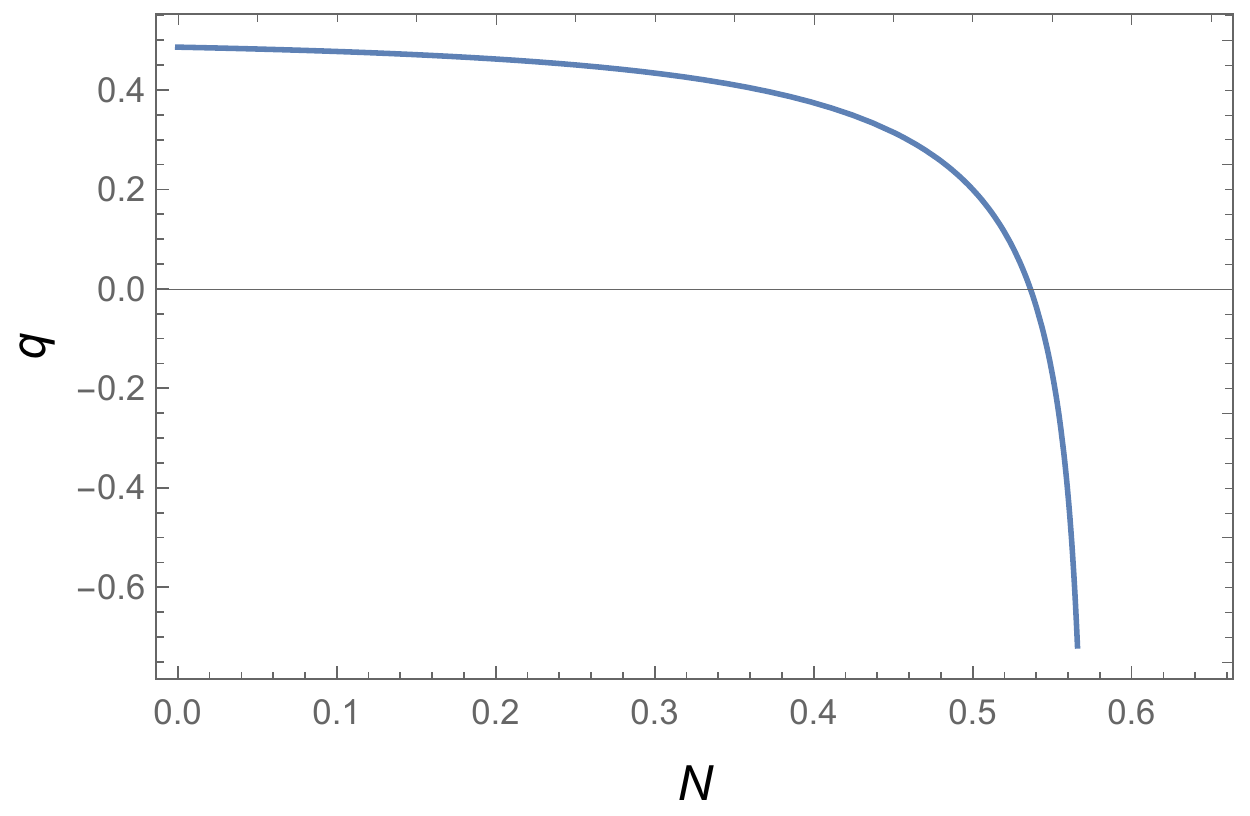}
		\caption{}
	\end{subfigure}
	\centering
	\begin{subfigure}[b]{.24\linewidth}
		\centering
		\includegraphics[scale=0.3]{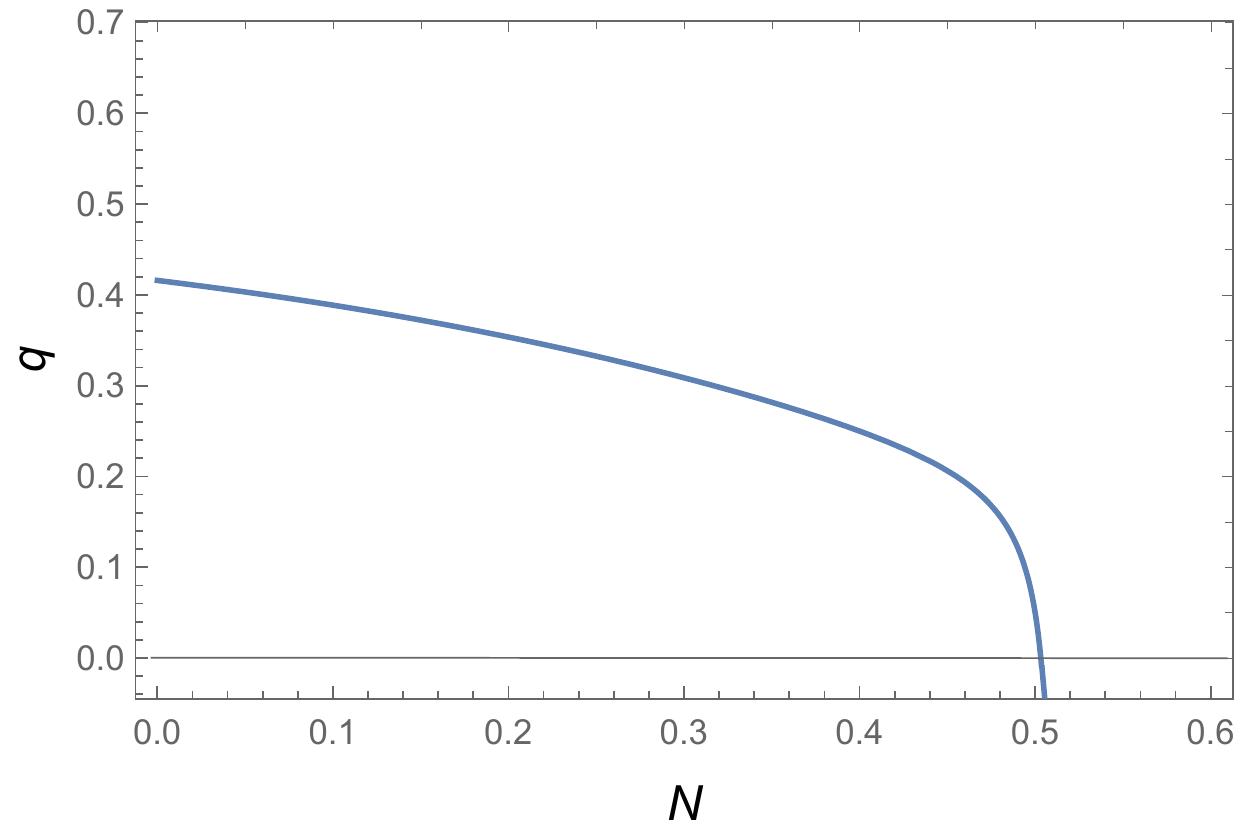}
		\caption{}
	\end{subfigure}\\
	\begin{subfigure}[b]{.24\linewidth}
		\centering
		\includegraphics[scale=0.3]{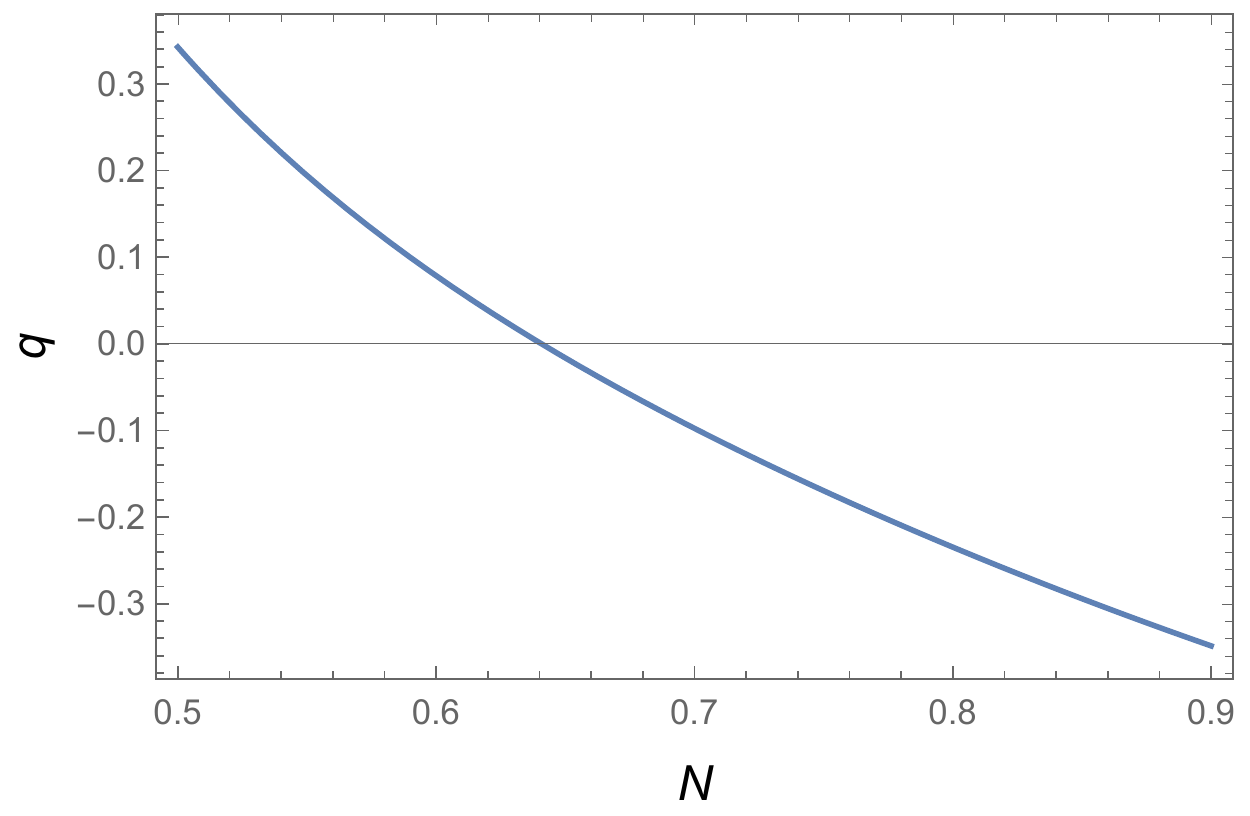}
		\caption{}
	\end{subfigure}
	\begin{subfigure}[b]{.24\linewidth}
		\centering
		\includegraphics[scale=0.3]{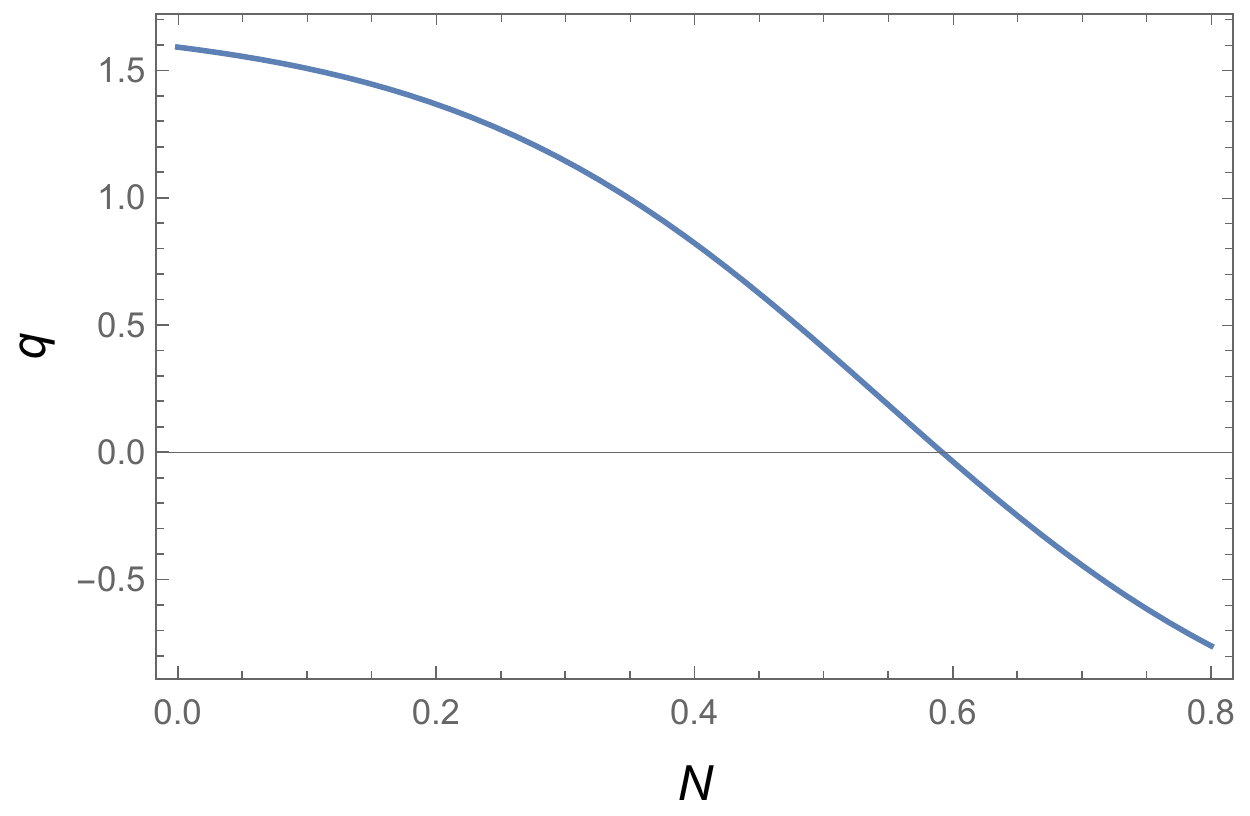}
		\caption{}
	\end{subfigure}
	\begin{subfigure}[b]{.24\linewidth}
		\centering
		\includegraphics[scale=0.3]{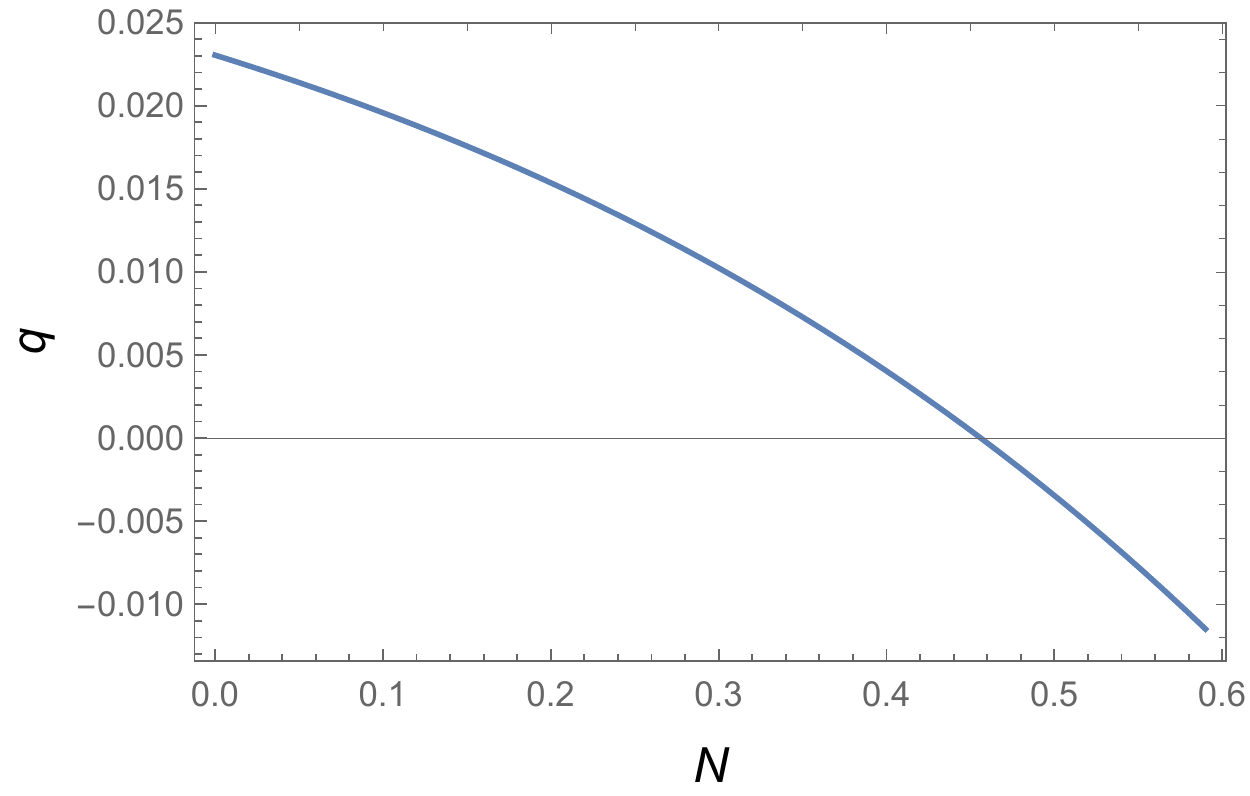}
		\caption{}
	\end{subfigure}
	
	\begin{subfigure}[b]{.24\linewidth}
		\centering
		\includegraphics[scale=0.3]{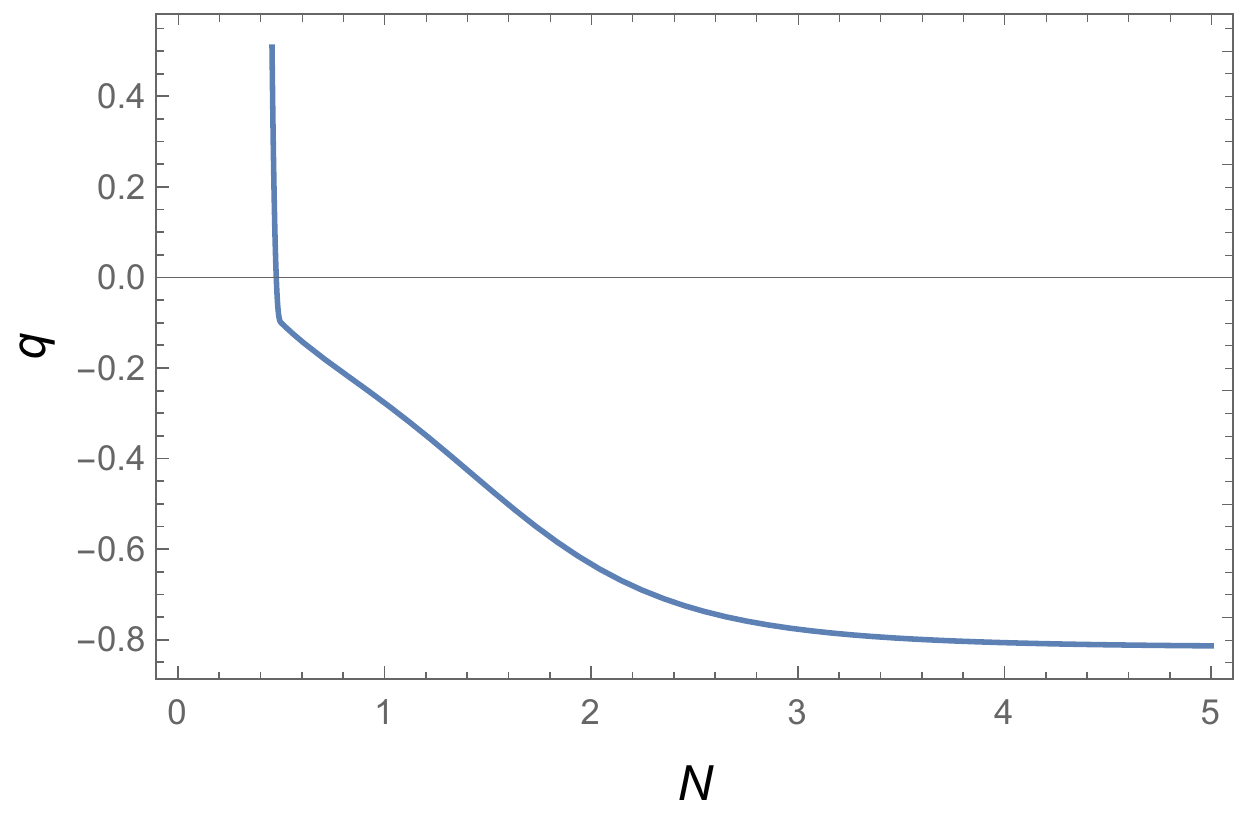}
		\caption{}
	\end{subfigure}
	\begin{subfigure}[b]{.24\linewidth}
		\centering
		\includegraphics[scale=0.3]{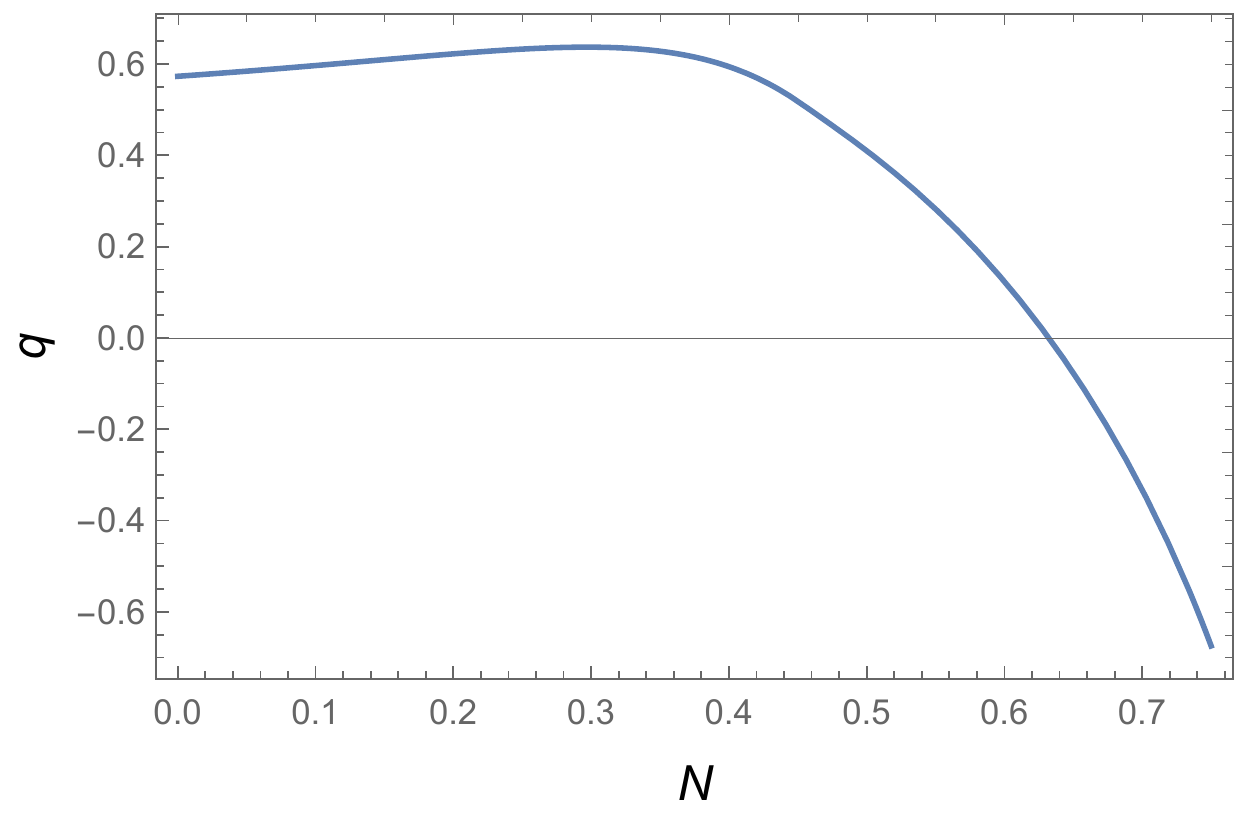}
		\caption{}
	\end{subfigure}
	\begin{subfigure}[b]{.24\linewidth}
		\centering
		\includegraphics[scale=0.3]{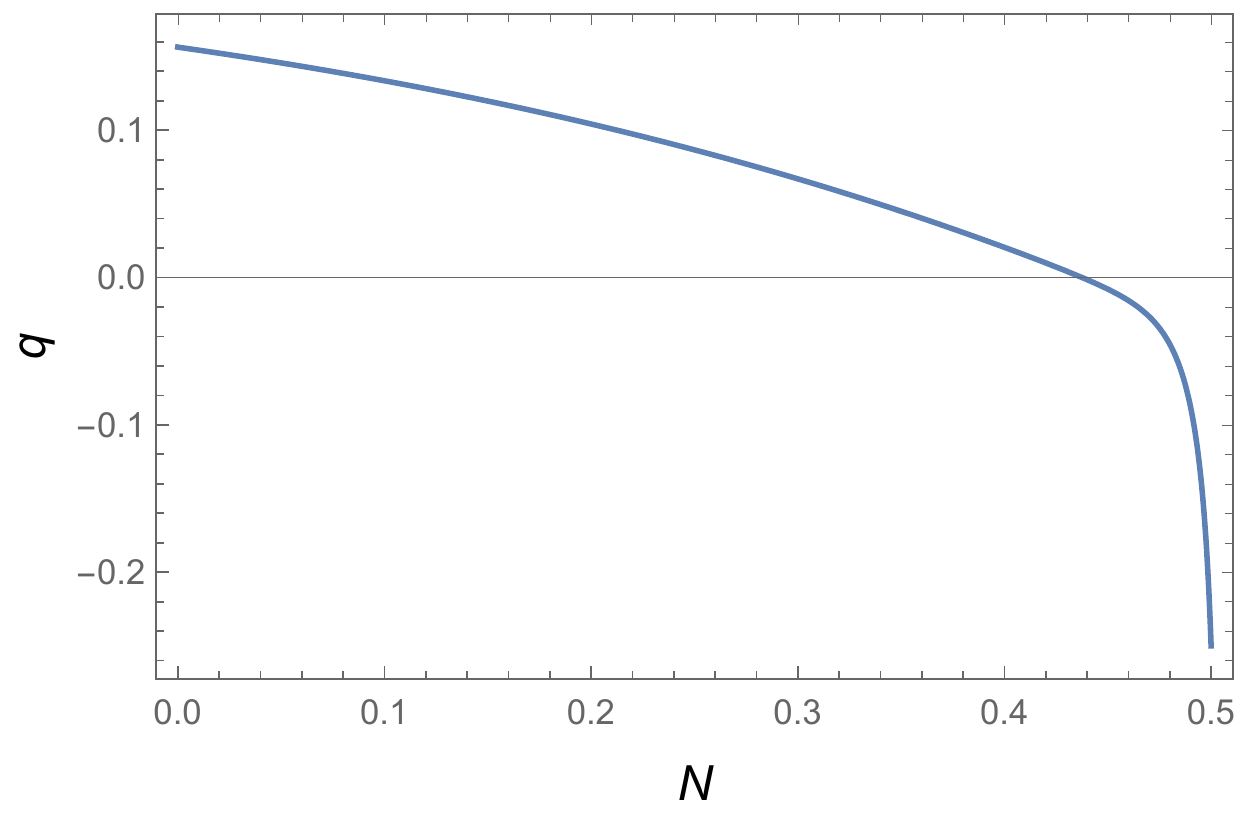}
		\caption{}
	\end{subfigure}
	\caption{This figure shows the phase transition between a decelerating to an accelerating epoch in each cosmological model which have been studied in this paper. Since the sign of $ q $  changes from positive to negative in each plot, it is clear that the universe experiences a transition from a decelerating phase in the early times into a late time accelerating phase. For the panels $(a), (b) $ and $ (c) $, which refer to the first interaction term for all three equations of state respectively, the value of the parameters  have been set to $ [b= -0.1, \gamma=0.5, \beta = -1.2, \alpha = 1.5] $. While for the panels $ (d) $, $ (e) $  which refer to the second interaction term  with first and second equations of state, the parameters has been fixed at $ [b= -0.1, \gamma=-0.5, \beta= -1.2, \alpha= 1.5] $. In the case of panel $ (f) $ which refers to  the third equation of state, the sign and value of $ \gamma $  is changed to  $\gamma=0.3$. Finally,  for the panels $ (g) $ and $ (h) $ which hold for the third interaction term with the first and second equation of state, the value of the parameters have been fixed at  $ [b=-0.1,\gamma =0.5,\beta =-1.2,\alpha =1.5] $. However for the third equation of state, these values are $  [b=-0.1,\gamma = 0.1,\beta =-1.2,\alpha =1.5]. $}
	\label{dec1}
\end{figure}

In this appendix we investigate the  phase transitions  between  early and late-time epochs of the universe in  each cosmological model which we have considered in our analysis. This can be done by looking at the evolution of the deceleration parameter respect to the e-folding number. The deceleration parameter is
\beq
q=\frac{1}{2}(1+3y),
\eeq
in which the evolution of $y$ is computed from the solution of the dynamical system. In  fig.(\ref{dec1}) we display the results for our nine cases.

{}


\begin{thebibliography}{66}

\bibitem{salvatelli}
Valentina Salvatelli, Najla Said, Marco Bruni, Alessandro Melchiorri, and David Wands,``Indications of a late-time interaction in the dark sector", \href{https://journals.aps.org/prl/abstract/10.1103/PhysRevLett.113.181301}{\color{blue}{Phys. Rev. Lett. \textbf{113} (2014)  181301}}, \href{https://arxiv.org/abs/1406.7297}{arXiv:1406.7297 [astro-ph.CO]}.

\bibitem{noint1}
Matteo Martinelli, Natalie B. Hogg, Simone Peirone, Marco Bruni, and David Wands, ``Constraints on the interacting vacuum -- geodesic CDM scenario", \href{https://academic.oup.com/mnras/article-abstract/488/3/3423/5532361?redirectedFrom=fulltext}{\color{blue}{Mon. Not. Roy. Acad. Soc. \textbf{488} (2019) 3423}}, \href{https://arxiv.org/abs/1902.10694}{arXiv:1902.10694 [astro-ph.CO]}.	

\bibitem{coin1}
Hermano E.S. Velten, Rodrigo vom Marttens, and  Winifried Zimdahl,``Aspects of the cosmological ``coincidence problem"", \href{https://link.springer.com/article/10.1140%2Fepjc%2Fs10052-014-3160-4}{\color{blue}{Eur. Phys. J. C \textbf{74} (2014)  3160}}, \href{https://arxiv.org/abs/arXiv:1410.2509}{arXiv:1410.2509 [astro-ph.CO]}.
	
\bibitem{coin2}
Paul J. Steinhardt, in {\it Critical Problems in Physics}, Edited by Val L. Fitch, Daniel R. Marlow, and Margit A.E. Dementi (Princeton University Press, Princeton, 1997).
	
\bibitem{coin3}
Edmund J. Copeland, M. Sami, and Shinji Tsujikawa, ``Dynamics of Dark Energy", \href{https://www.worldscientific.com/doi/abs/10.1142/S021827180600942X}{\color{blue}{Int. J. Mod. Phys. D \textbf{15}  (2006) 1753}}, \href{https://arxiv.org/abs/hep-th/0603057}{arXiv:0603057 [hep-th]}.
	
\bibitem{sunny2}
Eleonora Di Valentino, Alessandro Melchiorri, Olga Mena, and Sunny Vagnozzi, ``Interacting dark energy after the latest Planck, DES, and $H_0$ measurements: an excellent solution to the $H_0$ and cosmic shear tensions",  \href{https://arxiv.org/abs/1908.04281}{arXiv:1908.04281 [astro-ph.CO]}.

	
\bibitem{cai1}
Chunlong Li, Xin Ren, Martiros Khurshudyan, and Yi-Fu Cai, ``Implications of the possible 21-cm line excess at cosmic dawn on dynamics of interacting dark energy", \href{https://www.sciencedirect.com/science/article/pii/S0370269319308639?via%3Dihub}{\color{blue}{Phys. Lett. B  \textbf{801} (2020) 135141}}, \href{https://arxiv.org/abs/1904.02458}{arXiv:1904.02458 [astro-ph.CO]}.	
	
\bibitem{weinrev}
Steven Weinberg, ``The cosmological constant problem", \href{https://journals.aps.org/rmp/abstract/10.1103/RevModPhys.61.1}{\color{blue}{Rev. Mod. Phys. \textbf{61} (1989) 1}}.
	
\bibitem{bolejkosurv}
Krzysztof Bolejko, and Miko\l{}aj Korzy\'nski, ``Inhomogeneous cosmology and backreaction: current status and future prospects",
{\hypersetup{urlcolor=vividviolet}\href{https://www.worldscientific.com/doi/abs/10.1142/S0218271817300117}{Int. J. Mod. Phys. D  \textbf{26}  (2017)  1730011}}, \href{https://arxiv.org/abs/1612.08222}{arXiv:1612.08222 [gr-qc]}.
	
\bibitem{inhomo1}
Chris Clarkson, George Ellis, Julien Larena and Obinna Umeh, ``Does the growth of structure affect our dynamical models of the universe? The averaging, backreaction, and fitting problems in cosmology",
{\hypersetup{urlcolor=vividviolet}\href{http://iopscience.iop.org/article/10.1088/0034-4885/74/11/112901/meta}{Rept. Prog. Phys.  \textbf{74}  (2011)   112901}}, \href{https://arxiv.org/abs/1109.2314}{arXiv:1109.2314 [astro-ph.CO]}.
	
\bibitem{inhomo2}
Thomas Buchert, Mauro Carfora, G.F.R. Ellis, Edward W. Kolb, Malcolm MacCallum, Jan J. Ostrowski, Sysky R\"as\"anen,  Boudewijn F. Roukema, Lars Andersson, Alan Coley, and David L. Wiltshire, ``Is there proof that backreaction of inhomogeneities is irrelevant in cosmology?",
{\hypersetup{urlcolor=vividviolet}\href{http://iopscience.iop.org/article/10.1088/0264-9381/32/21/215021/meta}{Class. Quantum Grav.  \textbf{32}  (2015)   215021}}, \href{https://arxiv.org/abs/1505.07800}{arXiv:1505.07800 [gr-qc]}.

\bibitem{buch4}
Thomas Buchert, ``Dark Energy from structure: a status report", \href{https://link.springer.com/article/10.1007%2Fs10714-007-0554-8}{\color{blue}{Gen. Rel. Grav. \textbf{40} (2008) 467}}, \href{https://arxiv.org/abs/0707.2153}{arXiv:0707.2153 [gr-qc]}.	

\bibitem{buch3}
Thomas Buchert, and  Nathaniel Obadia, ``Effective inhomogeneous inflation: curvature inhomogeneities of the Einstein vacuum", \href{https://iopscience.iop.org/article/10.1088/0264-9381/28/16/162002}{\color{blue}{Class. Quantum Grav. \textbf{28} (2011) 162002}}, \href{https://arxiv.org/abs/1010.4512}{arXiv:1010.4512 [gr-qc]}.
	
\bibitem{nonideal1}
Maria da Conceicao Bento, Orfeu Bertolami, and Anjan A Sen, ``Generalized Chaplygin gas, accelerated expansion, and dark-energy-matter unification",
{\hypersetup{urlcolor=vividviolet}\href{https://journals.aps.org/prd/abstract/10.1103/PhysRevD.66.043507}{Phys.  Rev. D  \textbf{66}  (2002)  043507}}, \href{https://arxiv.org/abs/gr-qc/0202064}{arXiv:020206 [gr-qc]}.
	
\bibitem{nonideal2}
Kostas  Kleidis and Nicholas K. Spyrou, ``Polytropic dark matter flows illuminate dark energy and accelerated expansion", \href{https://www.aanda.org/10.1051/0004-6361/201424402}{\color{blue}{Astron. Astrophys. \textbf{576} (2015) A23}}, \href{https://arxiv.org/abs/1411.6789}{arXiv:1411.6789 [astro-ph.CO]}.
	
\bibitem{nonideal3}
Kayoomars Karami, Ahmad Sheykhi, Mubasher Jamil, Z. Azarmi, and M. M. Soltanzadeh, ``Interacting entropy-corrected new agegraphic dark energy in Brans-Dicke cosmology", \href{https://link.springer.com/article/10.1007%2Fs10714-010-1072-7}{\color{blue}{Gen. Rel. Grav. \textbf{43} (2011) 27}}, \href{https://arxiv.org/abs/1004.3607}{arXiv:1004.3607 [hep-th]}.
		
\bibitem{nonideal4}
Alexander Kamenshchik, Ugo Moschella, and VincentPasquier, ``An Alternative to quintessence", \href{https://linkinghub.elsevier.com/retrieve/pii/S0370269301005718}{\color{blue}{Phys. Lett. B \textbf{511} (2001) 265-268}}, \href{https://arxiv.org/abs/gr-qc/0103004}{arXiv:0103004 [gr-qc]}.
		
\bibitem{nonideal5}
Shin'ichi Nojiri, and Sergei D. Odintsov., ``Inhomogeneous equation of state of the universe: Phantom era, future singularity and crossing the phantom barrier", \href{https://journals.aps.org/prd/abstract/10.1103/PhysRevD.72.023003}{\color{blue}{Phys. Rev. D \textbf{72} (2005) 023003}}, \href{https://arxiv.org/abs/hep-th/0505215}{arXiv:0505215 [hep-th]}.
		
\bibitem{nonideal6}
Gilberto M. Kremer, ``Cosmological models described by a mixture of van der Waals fluid and dark energy", \href{https://journals.aps.org/prd/abstract/10.1103/PhysRevD.68.123507}{\color{blue}{Phys. Rev. D \textbf{68} (2003) 123507}}, \href{https://arxiv.org/abs/gr-qc/0309111}{arXiv:0309111 [gr-qc]}.
		
\bibitem{plank}
N. Aghanim et al. (The Planck Collaboration), ``Planck 2018 results. VI. Cosmological parameters", \href{https://arxiv.org/abs/1807.06209}{arXiv:1807.06209 [astro-ph.CO]}.
		
\bibitem{panteon}
Daniel Moshe Scolnic et al., ``The Complete Light-curve Sample of Spectroscopically Confirmed Type Ia Supernovae from Pan-STARRS1 and Cosmological Constraints from the Combined Pantheon Sample", \href{https://iopscience.iop.org/article/10.3847/1538-4357/aab9bb}{\color{blue}{Astrophys. J. \textbf{859} (2018) 101}}, \href{https://arxiv.org/abs/1710.00845v3}{arXiv:1710.00845 [astro-ph.CO]}.
		
\bibitem{bao}
Shadab Alam et al.,  ``The clustering of galaxies in the completed SDSS-III Baryon Oscillation Spectroscopic Survey: cosmological analysis of the DR12 galaxy sample", \href{https://academic.oup.com/mnras/article/470/3/2617/3091741}{\color{blue}{Mon. Not. Roy. Astr. Soc. \textbf{470} (2017) 2617}}, \href{https://arxiv.org/abs/1607.03155}{arXiv:1607.03155 [astro-ph.CO]}.
		
\bibitem{parametric1}
Joan Simon, Licia Verde, and Raul Jimenez, ``Constraints on the redshift dependence of the dark energy potential", \href{https://journals.aps.org/prd/abstract/10.1103/PhysRevD.71.123001}{\color{blue}{Phys. Rev. D \textbf{71} (2005) 123001}}, \href{https://arxiv.org/abs/astro-ph/0412269}{arXiv:0412269 [astro-ph]}.
		
\bibitem{parametric2}
Daniel Stern, Raul Jimenez, Licia Verde, Marc Kamionkowski, and  S. Adam Stanford, ``Cosmic chronometers: constraining the equation of state of dark energy. I: H(z) measurements", \href{https://iopscience.iop.org/article/10.1088/1475-7516/2010/02/008}{\color{blue}{Jour. Cosmol. Astropart. Phys.  \textbf{1002} (2010) 008}}, \href{https://arxiv.org/abs/0907.3149}{arXiv:0907.3149 [astro-ph.CO]}.
		
\bibitem{parametric3}
Varun Sahni, and Alexei Starobinsky, ``Reconstructing Dark Energy", \href{https://www.worldscientific.com/doi/abs/10.1142/S0218271806009704}{\color{blue}{Int. Jour. Mod. Phys. D \textbf{15} (2006) 2105}}, \href{https://arxiv.org/abs/astro-ph/0610026}{arXiv:0610026 [astro-phO]}.
		
\bibitem{qual1}
Somnath Bhattacharya, Pradip Mukherjee, Amit Singha Roy, and Anirban Saha, ``Non-minimally coupled quintessence dark energy model with a cubic galileon term: a dynamical system analysis", \href{https://link.springer.com/article/10.1140/epjc/s10052-018-5644-0#citeas}{\color{blue}{Eur. Phys. Jour. C \textbf{78} (2018) 201}}, \href{https://arxiv.org/abs/1512.03902}{arXiv:1512.03902 [gr-qc]}.
		
\bibitem{qual2}
Genly Leon, and Emmanuel N. Saridakis, ``Phase-space analysis of Horava-Lifshitz cosmology", \href{https://iopscience.iop.org/article/10.1088/1475-7516/2009/11/006}{\color{blue}{Jour. Cosmol. Astropart. Phys. \textbf{0911} (2009) 006}}, \href{https://arxiv.org/abs/0909.3571v1}{arXiv:0909.3571 [hep-th]}.
		
\bibitem{qual3}
Sujay Kr. Biswas, and Subenoy Chakraborty, ``Interacting Dark Energy in  $f(T)$  cosmology: A Dynamical System analysis", \href{https://www.worldscientific.com/doi/abs/10.1142/S0218271815500467}{\color{blue}{Int. Jour. Mod. Phys. D \textbf{24} (2015) 1550046}}, \href{https://arxiv.org/abs/1504.02431}{arXiv:1504.02431 [gr-qc]}.
		
\bibitem{qual4}
Robert J. Van Den Hoogen, Alan Coley, B. Alhulaimi, S. Mohandas, E. Knighton, and S. O'Neil, ``Kantowski-Sachs Einstein-Aether Scalar Field Cosmological Models", \href{https://iopscience.iop.org/article/10.1088/1475-7516/2018/11/017}{\color{blue}{Jour. Cosmol. Astropart. Phys. \textbf{1811} 2018 017}}, \href{https://arxiv.org/abs/1809.01458}{arXiv:1809.01458 [gr-qc]}.

\bibitem{cai3}
Yi-Fu Cai, Jinn-Ouk Gong, Shi Pi, Emmanuel N. Saridakis, and Shang-Yu Wu, ``On the possibility of blue tensor spectrum within single field inflation", \href{https://www.sciencedirect.com/science/article/pii/S0550321315003442?via%3Dihub}{\color{blue}{Nucl. Phys. B  \textbf{900} (2015) 517}}, \href{https://arxiv.org/abs/1412.7241}{arXiv:1412.7241 [hep-th]}.

\bibitem{buch1}
Xavier Roy, Thomas Buchert, Sante Carloni, and Nathaniel Obadia, ``Global gravitational instability of FLRW backgrounds - interpreting the dark sectors", \href{https://iopscience.iop.org/article/10.1088/0264-9381/28/16/165004}{\color{blue}{Class. Quantum Grav. \textbf{28} (2011) 165004}}, \href{https://arxiv.org/abs/1103.1146}{arXiv:1103.1146 [gr-qc]}.	
		
\bibitem{mar1}
Martiros Khurshudyan, ``A varying polytropic gas universe and phase space analysis", \href{https://www.worldscientific.com/doi/abs/10.1142/S0217732316500978}{\color{blue}{Mod. Phys. Lett. A \textbf{31} (2016) 1650097}}.
		
\bibitem{mar2}
Martiros Khurshudyan, and Ratbay Myrzakulov, ``Phase space analysis of some interacting Chaplygin gas models", \href{https://link.springer.com/article/10.1140/epjc/s10052-017-4634-y}{\color{blue}{Eur. Phys. Jour. C \textbf{77} (2017) 65}}, \href{https://arxiv.org/abs/1509.02263}{arXiv:1509.02263 [gr-qc]}.
		
\bibitem{mar3}
Martiros Khurshudyan, and Amalya Khurshudyan, ``Interacting varying Ghost Dark energy models in General Relativity", \href{https://link.springer.com/article/10.1007%2Fs10509-015-2341-4}{\color{blue}{ Astrophys. Space Sci. \textbf{357} (2015) 113}}, \href{https://arxiv.org/abs/1307.7859}{arXiv:1307.7859 [gr-qc]}.
			
\bibitem{mar4}
Jafar Sadeghi, Aram Movsisyan, Martiros Khurshudyan, and H.M. Farahani, ``Interacting Ghost Dark Energy Models with Variable $G$ and $\Lambda$", \href{https://iopscience.iop.org/article/10.1088/1475-7516/2013/12/031}{\color{blue}{Jour. Cosmol. Astropart. Phys. \textbf{1312} (2013) 031}}, \href{https://arxiv.org/abs/1308.3450}{arXiv:1308.3450 [gr-qc]}.
			
\bibitem{coley1}
Ed. by John Wainwright, and George F. R. Ellis, {\it  Dynamical systems in cosmology} (Cambridge University Press, Cambridge, 1997).
			
\bibitem{coley2}
Alan Coley, {\it Dynamical systems and cosmology} (Springer, The Netherlands, 2003).
			
\bibitem{coley3}
Vladimir Belinski, and  Marc Henneaux, {\it The cosmological singularity}  (Cambridge University Press, Cambridge, 2017).


\bibitem{sergei2}
Sergei D. Odintsov, Vasilis K. Oikonomou, and Petr V. Tretyakov, ``Phase space analysis of the accelerating multifluid Universe", \href{https://journals.aps.org/prd/abstract/10.1103/PhysRevD.96.044022}{\color{blue}{Phys. Rev. D  \textbf{96} (2017) 044022}}, \href{https://arxiv.org/abs/1707.08661}{arXiv:1707.08661 [gr-qc]}.
			
\bibitem{tsing1}
Amalkumar Raychaudhuri, ``Relativistic cosmology. I", \href{https://journals.aps.org/pr/abstract/10.1103/PhysRev.98.1123} {\color{blue}{Phys. Rev. \textbf{98} (1955) 1123}}.
			
			
\bibitem{tsing2}
J\"urgen Ehlers, ``Beitrage zur relativistischen Mechanik kontinuierlicher Medien", Akad. Wiss. Lit. Mainz, Abhandl. Math.–Nat. Kl. \textbf{11} (1961) 793837.
English translation: ``Contributions to the relativistic mechanics of continuous media", \href{https://link.springer.com/article/10.1007%2FBF00759031}{\color{blue}{Gen. Rel. Grav. \textbf{25} (1993) 1225}}.
				
\bibitem{tsing3}
George F. R. Ellis, ``Relativistic cosmology", \href{https://link.springer.com/article/10.1007%2Fs10714-009-0760-7}{\color{blue}{Gen. Rel. Grav. \textbf{41} (2009) 581}}.
					
\bibitem{tsing4}
George F. R. Ellis, {\it Carg\'ese Lectures in Physics}, Vol. 6, pp. 1-60  (Gordon and Breach. E. Schatzman, New York, 1973).
					
					
\bibitem{capo}
Vincenzo Fabrizio Cardone,   Crescenzo Tortora, Antonio Troisi, and Salvatore Capozziello, ``Beyond the perfect fluid hypothesis for the dark energy equation of state", {\hypersetup{urlcolor=vividviolet}\href{https://journals.aps.org/prd/abstract/10.1103/PhysRevD.73.043508}{Phys. Rev. D \textbf{73} (2006)   043508}}, \href{https://arxiv.org/abs/astro-ph/0511528}{arXiv:0511528 [astro-ph]}.
					
\bibitem{exact}
Hans Stephani, Dietrich Kramer, Malcolm MacCallum, Cornelius Hoenselaers, and Eduard Herlt,  {\it Exact Solutions of Einstein's Field Equations}  (Cambridge University Press, Cambridge, England, 2002).
					
\bibitem{peebles}
Phillip James Edwin Peebles,  {\it Principles of Physical Cosmology}   (Princeton University Press, Princeton, NJ, 1993).
					
\bibitem{gravitation}
Charles W. Misner, Kip S. Thorne, and John Archibald Wheeler,  {\it Gravitation}  (Freeman, New York, 1973).
					
\bibitem{linde}
Anderi Linde, {\it Particle physics and Inflationary Cosmology}  (Harwood, Chur, 1990). \href{https://journals.aps.org/prd/abstract/10.1103/PhysRevD.78.023505}{\color{blue}{Contemp. Concepts Phys. \textbf{5} (1990) 1}}, \href{https://arxiv.org/abs/hep-th/0503203}{arXiv:0503203 [hep-th]}. 
					
\bibitem{rint1}
Christian G. B\"ohmer, Gabriela Caldera-Cabral, Ruth Lazkoz, and Roy Maartens, ``Dynamics of dark energy with a coupling to dark matter", \href{https://journals.aps.org/prd/abstract/10.1103/PhysRevD.78.023505}{\color{blue}{Phys. Rev. D \textbf{78} (2008) 023505}}, \href{https://arxiv.org/abs/0801.1565}{arXiv:0801.1565 [gr-qc]}.
					
\bibitem{rint2}
Miguel Quartin, Mauricio O. Calvao, Sergio E. Joras, Ribamar R. R. Reis, and Ioav Waga, ``Dark Interactions and Cosmological Fine-Tuning", \href{https://iopscience.iop.org/article/10.1088/1475-7516/2008/05/007}{\color{blue}{Jour. Cosmol. Astropart. Phys. \textbf{0805} (2008) 007}}, \href{https://arxiv.org/abs/0802.0546}{ arXiv:0802.0546 [astro-ph]}.
					
\bibitem{rint3}
Gabriela Caldera-Cabral, Roy Maartens, and L. Arturo Urena-Lopez, ``Dynamics of interacting dark energy", \href{https://journals.aps.org/prd/abstract/10.1103/PhysRevD.79.063518}{\color{blue}{Phys. Rev. D \textbf{79} (2009) 063518}}, \href{https://arxiv.org/abs/0812.1827}{arXiv:0812.1827 [gr-qc]}.
					
\bibitem{rint4}
Luis P. Chimento, Monica Forte, and Gilberto M. Kremer, ``Cosmological model with interactions in the dark sector", \href{https://link.springer.com/article/10.1007%2Fs10714-008-0694-5}{\color{blue}{Gen. Rel. Grav. \textbf{41} (2009) 1125}}, \href{https://arxiv.org/abs/0711.2646}{arXiv:0711.2646 [astro-ph]}.
						
\bibitem{rint5}
Luis P. Chimento,``Linear and nonlinear interactions in the dark sector", \href{https://journals.aps.org/prd/abstract/10.1103/PhysRevD.81.043525}{\color{blue}{Phys. Rev. D \textbf{81} (2010) 043525}}, \href{https://arxiv.org/abs/0911.5687}{arXiv:0911.5687 [astro-ph.CO]}.
						
\bibitem{rint6}
Maria Belen Gavela Legazpi, Daniel Hernandez, Laura Lopez Honorez, Olga Mena Requejo, and  Rigolin Stefano, ``Dark coupling", \href{https://iopscience.iop.org/article/10.1088/1475-7516/2009/07/034}{\color{blue}{Jour. Cosmol. Astropart. Phys. \textbf{0907} (2009) 034}}; \href{https://iopscience.iop.org/article/10.1088/1475-7516/2010/05/E01}{\color{blue}{Erratum-ibid. \textbf{1005} (2010) E01}}, \href{https://arxiv.org/abs/0901.1611}{arXiv:0901.1611 [astro-ph.CO]}.
						
\bibitem{rint7}
Maria Belen Gavela Legazpi,  Laura Lopez Honorez, Olga Mena Requejo, and  Rigolin Stefano, ``Dark Coupling and Gauge Invariance", \href {https://iopscience.iop.org/article/10.1088/1475-7516/2010/11/044}{\color{blue}{Jour. Cosmol. Astropart. Phys. \textbf{1011} (2010) 044}}, \href{https://arxiv.org/abs/1005.0295}{arXiv:1005.0295 [astro-ph.CO]}.
						
\bibitem{binwang}
Bin Wang, Elcio Abdalla, Fernando Atrio-Barandela, and Diego Pavon, ``Dark Matter and Dark Energy Interactions: Theoretical Challenges, Cosmological Implications and Observational Signatures", \href{https://iopscience.iop.org/article/10.1088/0034-4885/79/9/096901}{\color{blue}{Rept. Prog. Phys. \textbf{79} (2016) 9}}, \href{https://arxiv.org/abs/1603.08299}{arXiv:1603.08299 [astro-ph.CO]}.
						
\bibitem{reos1}
Otto Redlich, and J.N.S.  Kwong, ``On the Thermodynamics of Solutions. V. An Equation of State. Fugacities of Gaseous Solutions.", {\hypersetup{urlcolor=vividviolet}\href{https://pubs.acs.org/doi/10.1021/cr60137a013}{Chem. Rev.  \textbf{44} (1949)   233244}}.
						
\bibitem{reos3}
D. Berthelot, in Travaux et Memoires du Bureau international des Poids et Mesures Tome XIII (Paris: Gauthier-Villars, 1907).
						
\bibitem{reos2}
C. Dieterici, ``Ueber den kritischen Zustand", {\hypersetup{urlcolor=vividviolet}\href{http://www3.interscience.wiley.com/cgi-bin/abstract/112504868/ABSTRACT}{Ann. Phys. \textbf{305} (1899)   11}}.
						
						
\bibitem{chemistry1}
Peter Atkins, {\it Atkins' Physical Chemistry} (Oxford University Press, USA,  2006).
						
						
\bibitem{chemistry2}
R. Stephen Berry, Stuart A. Rice, and John Ross, {\it  Physical Chemistry} (Oxford University Press, USA, 2000).


\bibitem{mcg2}
Ujjal Debnath, Asit Banerjee, and Subenoy Chakraborty, ``Role of Modified Chaplygin Gas in Accelerated Universe", \href{https://iopscience.iop.org/article/10.1088/0264-9381/21/23/019}{\color{blue}{Class. Quantum Grav.  \textbf{21} (2004) 5609}}, \href{https://arxiv.org/abs/gr-qc/0411015}{arXiv:0411015 [gr-qc]}.

\bibitem{mcg3}
Jose Beltran Jimenez, Diego Rubiera-Garcia, Diego Saez-Gomez, and Vincenzo Salzano, ``Cosmological future singularities in interacting dark energy models", \href{https://journals.aps.org/prd/abstract/10.1103/PhysRevD.94.123520}{\color{blue}{Phys. Rev. D  \textbf{94} (2016) 123520}}, \href{https://arxiv.org/abs/1607.06389}{arXiv:1607.06389 [gr-qc]}.
	
\bibitem{refstaro}
Alexei Starobinsky, ``Future and Origin of our Universe: Modern View", \href{https://ui.adsabs.harvard.edu/abs/2000GrCo....6..157S/abstract}{\color{blue}{Grav. Cosmol.  \textbf{6} (2000) 157}}, \href{https://arxiv.org/abs/astro-ph/9912054}{arXiv:9912054 [astro-ph]}.						
						
\bibitem{Sergei}
Shin'ichi Nojiri, Sergei D. Odintsov, and Shinji Tsujikawa, ``Properties of singularities in (phantom) dark energy universe", \href{https://journals.aps.org/prd/abstract/10.1103/PhysRevD.71.063004}{\color{blue}{Phys. Rev. D \textbf{71} (2005) 063004} }, \href{https://arxiv.org/abs/hep-th/0501025}{arXiv:0501025 [hep-th]}.
						
\bibitem{class1}
Robert R. Caldwell, Marc Kamionkowski, and Nevin N. Weinberg, ``Phantom energy and cosmic doomsday", \href {https://journals.aps.org/prl/abstract/10.1103/PhysRevLett.91.071301}{\color{blue}{Phys. Rev. Lett. \textbf{91} (2003) 071301}}, \href{https://arxiv.org/abs/astro-ph/0302506}{arXiv:0302506 [astro-ph]}.
						
\bibitem{class1a}
L. Fern\'andez-Jambrina, and  Ruth Lazko, ``Classification of cosmological milestones", \href{https://journals.aps.org/prd/abstract/10.1103/PhysRevD.74.064030}{\color{blue}{Phys. Rev. D \textbf{74} (2006) 064030}}, \href{https://arxiv.org/abs/gr-qc/0607073}{arXiv:0607073 [gr-qc]}.
						
\bibitem{class2}
John D. Barrow, Gregory J. Galloway, and Frank J. Tipler, ``The closed-universe recollapse conjecture", \href{http://adsabs.harvard.edu/full/1986MNRAS.223..835B}{\color{blue}{Mon. Not. Roy. Astron. Soc. \textbf{223} (1986) 835}}.
							

\bibitem{sergei4}
Shin'ichi Nojiri, and Sergei D.Odintsov, ``Quantum escape of sudden future singularity", \href{https://www.sciencedirect.com/science/article/pii/S0370269304009232?via%3Dihub}{\color{blue}{Phys. Lett. B  \textbf{595} (2004) 1}}, \href{https://arxiv.org/abs/hep-th/0405078}{arXiv:0405078 [hep-th]}.
					
						
\bibitem{class3}
John D. Barrow, ``Sudden future singularities", \href{https://iopscience.iop.org/article/10.1088/0264-9381/21/11/L03}{\color{blue}{Class. Quant. Grav. \textbf{21} (2004) L79}}, \href{https://arxiv.org/abs/gr-qc/0403084}{arXiv:0403084 [gr-qc]}.
							
\bibitem{class3a}
John D. Barrow, ``More general sudden singularities", \href{https://iopscience.iop.org/article/10.1088/0264-9381/21/23/020}{\color{blue}{Class. Quant. Grav. \textbf{21} (2004) 5619}}, \href{https://arxiv.org/abs/gr-qc/0409062}{arXiv:0409062 [gr-qc]}.
							
\bibitem{class4}
Mariam Bouhmadi-Lopez, Pedro F. Gonzalez-Diaz, and Prado Martin-Moruno, ``Worse than a big rip?", \href{https://linkinghub.elsevier.com/retrieve/pii/S0370269307013457}{\color{blue}{Phys. Lett. B \textbf{659} (2008) 1}}, \href{https://arxiv.org/abs/gr-qc/0612135}{arXiv:0612135 [gr-qc]}.
							
							
\bibitem{sergei3}
Shin'ichi Nojiri, and Sergei D.Odintsov, ``The Final state and thermodynamics of dark energy universe", \href{https://journals.aps.org/prd/abstract/10.1103/PhysRevD.70.103522}{\color{blue}{Phys. Rev. D  \textbf{70} (2004) 103522}}, \href{https://arxiv.org/abs/hep-th/0408170}{arXiv:0408170 [hep-th]}.						
							
							
\bibitem{class5}
Mariusz P. D\c{a}browski, Konrad Marosek, and  Adam Balcerzak,  ``Standard and exotic singularities regularized by varying constants",  \href{http://sait.oat.ts.astro.it/MmSAI/85/PDF/44.pdf}{\color{blue}{Mem. Soc. Ast. It. \textbf{85} (2014) 44}}, \href{https://arxiv.org/abs/1308.5462}{arXiv:1308.5462 [astro-ph.CO]}.
							
							
\bibitem{class6}
Mariusz P. D\c{a}browski, and Tomasz Denkiewicz, ``Barotropic index $ w $-singularities in cosmology", \href{https://journals.aps.org/prd/abstract/10.1103/PhysRevD.79.063521}{\color{blue}{Phys. Rev. D \textbf{79} (2009) 063521}}, \href{https://arxiv.org/abs/0902.3107}{arXiv:0902.3107 [gr-qc]}.
							
\bibitem{class6a}
L. Fern\'andez-Jambrina, ``Hidden past of dark energy cosmological models", \href{https://www.sciencedirect.com/science/article/abs/pii/S0370269307011550?via%3Dihub}{\color{blue}{Phys. Lett. B \textbf{656} (2007) 9}}, \href{https://arxiv.org/abs/0704.3936}{arXiv:0704.3936 [gr-qc]}.
	
\bibitem{lead1}
Kazuharu Bamba, Salvatore Capozziello, Shin'ichi Nojiri, and Sergei D. Odintsov, ``Dark energy cosmology: the equivalent description via different theoretical models and cosmography tests", \href{https://link.springer.com/article/10.1007%2Fs10509-012-1181-8}{\color{blue}{ Astrophys. Space. Sci. \textbf{342} (2012) 155}}, \href{https://arxiv.org/abs/1205.3421}{arXiv:1205.3421 [gr-qc]}.
								
\bibitem{lead2}
Spiros Cotsakis, and John D. Barrow, 	``The dominant balance at cosmological singularities",  \href{https://iopscience.iop.org/article/10.1088/1742-6596/68/1/012004}{\color{blue}{Jour. Phys. Conf. Ser. \textbf{68} (2007)  012004}}, \href{https://arxiv.org/abs/gr-qc/0608137}{arXiv:0608137 [gr-qc]}.
								
								
								
\bibitem{chmod}
Maye Elmardi, and  Amare Abebe,  ``Cosmological Chaplygin gas as modified gravity", \href{https://www.iopscience.iop.org/article/10.1088/1742-6596/905/1/012015}{\color{blue}{Jour. Phys. Conf. Ser. \textbf{905} (2017) 012015}}. 

\bibitem{buch2}
Xavier Roy, and Thomas Buchert, ``Chaplygin gas and effective description of inhomogeneous universe models in general relativity", \href{https://iopscience.iop.org/article/10.1088/0264-9381/27/17/175013}{\color{blue}{Class. Quantum Grav. \textbf{27} (2010) 175013}}, \href{https://arxiv.org/abs/0909.4155}{arXiv:0909.4155 [gr-qc]}.	
								
\bibitem{nloc}
Friedrich W.Hehl, and Bahram Mashhoon, ``Nonlocal Gravity Simulates Dark Matter", \href{https://www.sciencedirect.com/science/article/pii/S0370269309002111}{\color{blue}{Phys. Lett. B \textbf{673} (2009) 279}}, \href{https://arxiv.org/abs/0812.1059}{arXiv:0812.1059 [gr-qc]}.

\bibitem{dege1}
Saulo Carneiro, and H. A. Borges, ``On dark degeneracy and interacting models", \href{https://iopscience.iop.org/article/10.1088/1475-7516/2014/06/010}{\color{blue}{Jour. Cosmol. Astropart. Phys. \textbf{1406} (2014) 010}}, \href{https://arxiv.org/abs/1402.2316}{arXiv:1402.2316 [astro-ph.CO]}.
								
\bibitem{dege2}
George Efstathiou, and J. Richard Bond, ``Cosmic confusion: degeneracies among cosmological parameters derived from measurements of microwave background anisotropies, \href{https://academic.oup.com/mnras/article/304/1/75/972361}{\color{blue}{Mon. Not. Roy. Astron. Soc. \textbf{304} (1999) 75}}, \href{https://arxiv.org/abs/astro-ph/9807103}{arXiv:9807103 [astro-ph]}.
								
\bibitem{dege3}
Steffen Hagstotz, Max Gronke, David Mota, and Marco Baldi, ``Breaking cosmic degeneracies: Disentangling neutrinos and modified gravity with kinematic information", \href{https://www.aanda.org/articles/aa/abs/2019/09/aa35213-19/aa35213-19.html}{\color{blue}{ Astron. Astrophys.  \textbf{629}  (2019) A46}}, \href{https://arxiv.org/abs/1902.01868}{arXiv:1902.01868 [astro-ph.CO]}.

\bibitem{sergei1}
Sergei D. Odintsov, and Vasilis K. Oikonomou,  ``Dynamical Systems Perspective of Cosmological Finite-time Singularities in $f(R)$ Gravity and Interacting Multifluid Cosmology", \href{https://journals.aps.org/prd/abstract/10.1103/PhysRevD.98.024013}{\color{blue}{Phys. Rev. D  \textbf{98} (2018) 024013}}, \href{https://arxiv.org/abs/1806.07295}{arXiv:1806.07295 [gr-qc]}.





\bibitem{cai5}
Yi-Fu Cai, and Emmanuel N. Saridakis, ``Cosmology of $F(R)$ nonlinear massive gravity", \href{https://journals.aps.org/prd/abstract/10.1103/PhysRevD.90.063528}{\color{blue}{Phys. Rev. D  \textbf{90} (2014) 063528}}, \href{https://arxiv.org/abs/1401.4418}{arXiv:1401.4418 [astro-ph.CO]}.

\bibitem{cai6}
Yi-Fu Cai, Francis Duplessis, Emmanuel N. Saridakis, ``$F(R)$ nonlinear massive theories of gravity and their cosmological implications", \href{https://journals.aps.org/prd/abstract/10.1103/PhysRevD.90.064051}{\color{blue}{Phys. Rev. D  \textbf{90} (2014)   064051}}, \href{https://arxiv.org/abs/1307.7150}{arXiv:1307.7150 [hep-th]}.

		
\bibitem{cai2}
Yi-Fu Cai, Antonino Marciano, Dong-Gang Wang, and Edward Wilson-Ewing, ``Bouncing cosmologies with dark matter and dark energy", \href{https://www.mdpi.com/2218-1997/3/1/1}{\color{blue}{Univ.  \textbf{3} (2017) 1}}, \href{https://arxiv.org/abs/1610.00938}{arXiv:1610.00938 [astro-ph]}.
		
			
\bibitem{cai4}
Yi-Fu Cai, Francis Duplessis, Damien A. Easson, and Dong-Gang Wang, ``Searching for a matter bounce cosmology with low redshift observations", \href{https://journals.aps.org/prd/abstract/10.1103/PhysRevD.93.043546}{\color{blue}{Phys. Rev. D  \textbf{93} (2016)  043546}}, \href{https://arxiv.org/abs/1512.08979}{arXiv:1512.08979 [astro-ph.CO]}.

\bibitem{cai7}
Yi-Fu Cai, Emmanuel N. Saridakis, Mohammad R. Setare, and Jun-Qing Xia, ``Quintom Cosmology: Theoretical implications and observations", \href{https://www.sciencedirect.com/science/article/abs/pii/S0370157310000943?via%3Dihub}{\color{blue}{Phys. Rept.  \textbf{493} (2010) 1}}, \href{https://arxiv.org/abs/0909.2776}{arXiv:0909.2776 [hep-th]}.
	


\bibitem{sunny1}
Jibitesh Dutta, Wompherdeiki Khyllep, Emmanuel N. Saridakis, Nicola Tamanini, and Sunny Vagnozzi, ``Cosmological dynamics of mimetic gravity", \href{https://iopscience.iop.org/article/10.1088/1475-7516/2018/02/041}{\color{blue}{Jour. Cosmo. Astropart. Phys.  \textbf{1802} (2018) 041}}, \href{https://arxiv.org/abs/1711.07290}{arXiv:1711.07290 [gr-qc]}.
								
\bibitem{dew}
Donato Bini, Giampiero Esposito, and Andrea Geralico, ``Late-time evolution of cosmological models with fluids obeying a Shan-Chen-like equation of state", \href{https://journals.aps.org/prd/abstract/10.1103/PhysRevD.93.023511}{\color{blue}{Phys. Rev. D \textbf{93} (2016) 023511}}, \href{https://arxiv.org/abs/1601.04177}{arXiv:1601.04177 [gr-qc]}.
																
\bibitem{ham1}
Steven Strogatz, {\it Nonlinear dynamics and chaos: with applications in to physics, biology, chemistry and engineering} (CRC Press, 1994).
								
\bibitem{ham2}
Anatole Katok,  and  Boris Hasselblatt , {\it Introduction to the modern theory of dynamical systems} (Cambridge University Press, Cambridge, 1995).

\bibitem{ham3}
Morris W. Hirsch, Robert L. Devaney, and  Stephen Smale, {\it Differential equations, dynamical systems, and linear algebra} (AcademicPress, London, 1974).
								
\bibitem{hart}
Philip Hartman, {\it Ordinary Differential Equations} (Birkhauser, Boston-BaselStuttgart, 1982).
								
							

								
\end{thebibliography}
\end{document}